\documentclass[preprint]{aastex}
\usepackage{epsfig}
\usepackage{amssymb}
\usepackage{makeidx}
\usepackage{ifthen}
\usepackage{verbatim}
\begin{document}
\title{Eclipsing binary stars in the Large and Small Magellanic Clouds 
from the MACHO project: The Sample}

\author{Lorenzo Faccioli \altaffilmark{1}}
\affil{Department of Physics and Astronomy, University of Pennsylvania,
	Philadelphia, PA 19104, USA\newline
	and
	Lawrence Berkeley National Laboratory,
	Berkeley, CA 94720, USA}

\author{Charles Alcock \altaffilmark{2}} 
\affil{Harvard-Smithsonian Center for Astrophysics, 
	Cambridge, MA 02138, USA}

\author{Kem Cook \altaffilmark{3}}
\affil{Lawrence Livermore National Laboratory, 
	Livermore, CA 94550, USA}

\author{Gabriel E. Prochter \altaffilmark{4}}
\affil{Department of Astronomy and Astrophysics, University of
	California Santa Cruz, Santa Cruz, CA 95064, USA}

\author{Pavlos Protopapas \altaffilmark{5}}
\affil{Harvard-Smithsonian Center for Astrophysics, 
	Cambridge, MA 02138, USA}

\and

\author{David Syphers \altaffilmark{6}}
\affil{Department of Physics, University of Washington, 
	Seattle, WA 98195, USA}

\altaffiltext{1}{\textsl{LFaccioli@lbl.gov}}
\altaffiltext{2}{\textsl{calcock@cfa.harvard.edu}}
\altaffiltext{3}{\textsl{kcook@igpp.ucllnl.org}}
\altaffiltext{4}{\textsl{prochter@astro.ucsc.edu}}
\altaffiltext{5}{\textsl{pprotopapas@cfa.harvard.edu}}
\altaffiltext{6}{\textsl{dsyphers@u.washington.edu}}
\begin{abstract}
We present a new sample of $4634$ eclipsing binary stars in the Large 
Magellanic Cloud (LMC), expanding on a previous sample of 611 objects and 
a new sample of $1509$ eclipsing binary stars in the Small Magellanic Cloud 
(SMC), that were identified in the light curve database of the MACHO 
project.
We perform a cross correlation with the OGLE-II LMC sample, finding 1236 
matches.
A cross correlation with the OGLE-II SMC sample finds 698 matches.
We then compare the LMC subsamples corresponding to center and the 
periphery of the LMC and find only minor differences between the two 
populations.
These samples are sufficiently large and complete that statistical studies 
of the binary star populations are possible.
\end{abstract}
\section{Introduction}
Eclipsing binary stars (EBs) are important for astrophysical research in 
many ways.
They may be used to obtain accurate estimates of star masses and radii 
\cite[and references therein]{andersen91}.
Precise determination of stellar parameters can in turn be used to put 
theories of stellar structure and evolution to a stringent test by 
comparing measured parameters with theoretical predictions 
\citep[and references therein]{lvg02,lvg03}.
\par
EBs may also be used for distance determination and this use goes back 
several decades; its history is reviewed by \citet{kruszewski99}.
Since \citet{stebbins11} used an estimate of the parallax to $\beta$ 
Aurig\ae~to infer the surface brightness of both its components, it has been known
that a good photometric light curve plus a double line spectroscopic orbit 
admits a simple geometric relationship between the surface brightnesses 
of the stars and the distance to the EB; \citet{stebbins11} however
had no way at the time to make the reverse ``surface brightness to distance'' 
inference and his paper does no mention this possibility.
After \citet{stebbins11}, other early papers
\citep{gaposchkin33,woolley34,pilowski36,kopal39,gaposchkin38,gaposchkin40}
used parallaxes obtained independently to estimate surface brightnesses, 
but, as remarked by \citet{kruszewski99}, these pioneers surely 
knew of the potential of this technique to estimate distances.
Modern analyses of EBs have usually focussed on this technique 
(e.g. \cite{andersen91}).
The method affords high precision due to its purely geometrical nature and 
has been applied by a number of authors to determine the distance to the 
Large Magellanic Cloud (LMC) using HV2274 \citep{ud98a,guinan98,nelson99,
groen01}, HV982 \citep{fitzp02}, EROS 1044 \citep{ribas02} and HV 5936 
\citep{fitzp03}; an attempt to use EBs to determine the distance to M31 is 
currently under way \citep{ribas03} and the DIRECT project is attempting 
to measure the distance to M31 and M33 via EBs and Cepheids
\citep{kaluzny98,bonanos03,bonanos05}; other recent examples include
\citet{michalska05} who present a sample of detached binaries in the LMC 
for distance determination and \citet{ribas05} who present the first
determination of the distance and properties of an EB in M31; 
\citet{north05} presents a sample of EBs with total eclipses in the LMC 
suitable for spectroscopic studies.
In general it is important that distances be determined using a large 
sample of EBs to minimize the impact of systematic errors.
A recent collection of references on extragalactic binaries can be found 
in \citet{ribas04}.
\par
Large-scale surveys to detect gravitational microlensing events have 
identified and collected light curves for large numbers of variable stars 
in the bulge of the Milky Way and in the Magellanic Clouds. 
Eclipsing binary stars comprise a significant fraction of these collections.
The MACHO collaboration\footnote{\url{http://www.macho.mcmaster.ca/}} has 
presented a sample of 611 EBs in the LMC with preliminary analyses of their
orbits \citep{lacy97}.
A catalogue of $3031$ EBs in the LMC found in the MACHO database has been just
published by \citet{derekas07}; this catalogue was compiled by analyzing a list
of $6835$ stars classified as possible EBs in the MACHO database; a cross correlation
between these $6835$ stars and our sample finds just $1987$ matches, thus
at least about $2700$ EBs in our catalogue are new identifications.
The $6835$ classified as possible EBs were found in regions of parameter space such 
as color, magnitude, and period, where one does not expect to find pulsating 
variables and therefore the detected variability of these stars was 
\emph{tentatively} ascribed to eclipses.
Regions where pulsating variables could exist were not considered while 
making this preliminary classification and EBs there were therefore not 
included in the list.
In our search we did not rely primarily on cuts in parameter space and and we did
not exclude a priori regions of this space where pulsating variables are present; 
therefore we were able to classify many EBs in these regions, that were not included in the
preliminary classification.
The OGLE collaboration\footnote{\url{http://sirius.astrouw.edu.pl/\~{}ogle/}} 
has introduced a sample of $2580$ EBs in the LMC \citep{wyr03} and of $1351$ 
EBs in the Small Magellanic Cloud \citep[SMC:][]{wyr04}.
Both samples were selected from their catalogue of variable stars in the 
Magellanic Clouds \citep{zebrun01a} compiled from observations taken during
the second part of the project \citep[OGLE II:][]{udalski97} and reduced 
via Difference Image Analysis \citep[DIA:][]{zebrun01b}.
An earlier sample of $79$ EBs in the bar of the LMC was presented by the 
EROS collaboration\footnote{\url{http://eros.in2p3.fr/}}\citep{gri95}.
Other large variable star data sets are being produced by surveys not 
specifically designed to detect gravitational microlensing, such as the
All Sky Automated Survey 
\citep[ASAS:][]{pojmanski97}
\footnote{\url{http://www.astrouw.edu.pl/\~{}gp/asas/asas.html}}.
\par
The availability of large samples of EBs (and the even larger ones that 
can be found by future surveys such as 
Pan-STARRS\footnote{\url{http://pan-starrs.ifa.hawaii.edu/public/}} and 
LSST\footnote{\url{http://www.lsst.org/lsst\_home.shtml/}}) can have an
important impact on stellar astrophysics.
This impact can arise in two qualitatively different approaches.
First, a large catalogue allows the discerning researcher to select 
carefully a few EBs for detailed follow-up study; the distance 
estimation described above is an example of this.
Second, statistical analyses of an entire population become possible when 
a large collection is assembled; such analyses of EBs have not previously 
been possible.
To fulfill this promise there are challenges to overcome, including 
finding EBs in large data sets and automating their analysis.
With regard to the first task, the \textit{discovery} problem is 
complicated by the fact that EBs do not have clear relationships between 
their parameters (period, luminosity, colors) as do the major classes of 
pulsating variables.
This makes it difficult to find them via simple and well understood cuts 
in parameter space.
The first step toward automated discovery is thus to have a large sample 
of data on which to experiment with search techniques. 
This non-trivial exercise in mining large data sets can be useful for 
future surveys that are not necessarily aimed at binary star research.
An example is given by \citet{wyr03} and \citet{wyr04} who employ an 
artificial neural network to identify EBs in the OGLE-II LMC and SMC samples,
but more needs to be done.
With regard to \textit{analysis} of EBs, the traditional approach has been 
to carefully analyze individual systems with the help of dedicated 
computer codes such as the Wilson-Devinney code
\citep[WD:][]{wilson71,wilson79}. 
This becomes impracticable when many thousands of stars are involved and an
automated approach is required.
The light curves in a previous sample of 1459 EBs in the SMC found by 
OGLE-II \citep{ud98b} were systematically solved by 
\citet{wilson01,wilson02} using an automated version of the WD code;
the ASAS collaboration has developed an automated classification algorithm 
for variable stars based on Fourier decomposition \citep{pojmanski02};
\citet{devor05} found and analyzed 10000 Bulge EBs from OGLE-II using 
DEBiL\footnote{\url{http://www.cfa.harvard.edu/\~{}jdevor/DEBiL.html}}, 
an EB analysis code that allows automated solutions of large EB 
data sets and works best for detached EBs; a genetic algorithm based
approach to finding good initial parameters for WD is described in 
\citet{metcalfe99}.
\par
This paper is the first of a series of papers aimed at describing the EB 
samples in the MACHO database and is organized as follows: Section 
\ref{sec:sample} introduces the LMC and SMC samples; Section 
\ref{sec:diagrams} describes the Color Magnitude Diagram (CMD) and the
Color Period Diagram, pointing out significant features in them, 
Section \ref{sec:comparison} compares the LMC and SMC samples, 
Section \ref{sec:crossogle} describes the results of the 
cross correlation with the OGLE LMC and SMC samples, and
Section \ref{sec:online} reports where and in what form 
the data presented in the paper can be accessed on line.
\section{The Samples}
\label{sec:sample}
\subsection{The MACHO Project}
The MACHO Project was an astronomical survey whose primary aim was to 
detect gravitational microlensing events of background sources by compact 
objects in the halo of the Milky Way. 
The gravitational background sources were located in the LMC, SMC and the 
bulge of the Milky Way; more details on the detection of microlensing 
events can be found in \citet{alcock00a} and references therein.
Observations were carried out from July 1992 to December 1999 with the 
dedicated $1.27\mathrm{m}$ telescope of Mount Stromlo, Australia, using a $2\times 2$ 
mosaic of $2048\times 2048$ CCD in two bandpasses simultaneously. 
These are called MACHO ``blue", hereafter indicated with $V_{\mathrm{MACHO}}$, with 
a bandpass of $\sim 440-590\mathrm{nm}$ and MACHO ``red", hereafter 
indicated with $R_{\mathrm{MACHO}}$, with a bandpass of $\sim 590-780\mathrm{nm}$;
these widths are between the half-response points as estimated from
Figure $1$ of \citep{alcock99}.
The bandpasses and the transformations to standard Johnson $V$ and Cousins 
$R$ bands are described in detail in \citep{alcock99}; see in particular
their Figure $1$ for the instrumental throughput of the two MACHO bands.
Each MACHO object is identified by its field number ($1-82$ for the LMC, 
$201-213$ for the SMC), its tile number (which can overlap more than one 
field), and its sequence number in the tile.
These form the so called MACHO Field.Tile.Sequence (FTS), which is used in 
this paper to label EBs.
Note that, since some overlap exists between fields, one star may have two
or more FTS identifiers.
\subsection{The Large Magellanic Cloud Sample}
\begin{figure}
\footnotesize
\plotone{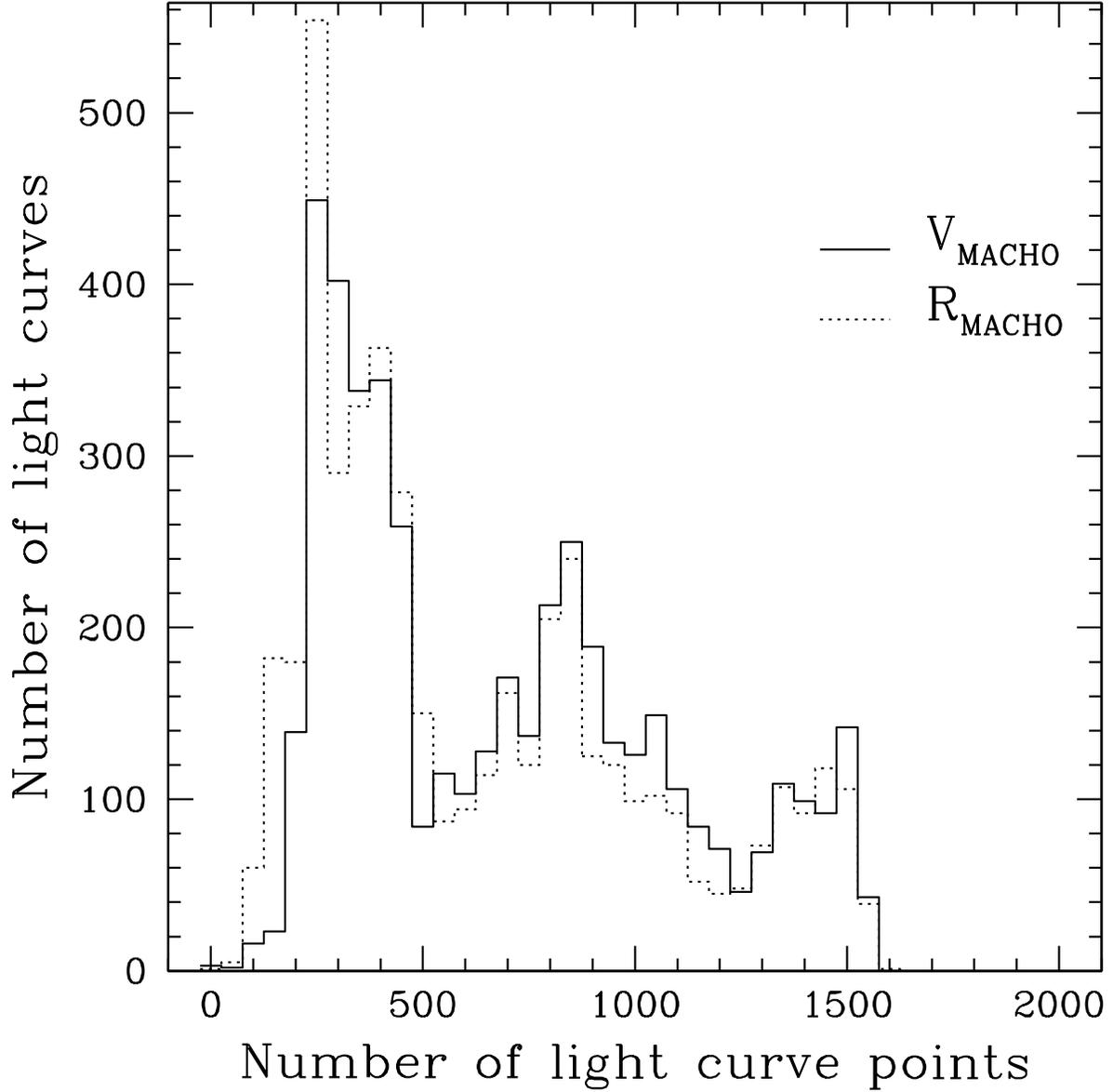}
\caption{Histograms of the number of light curve points for both bands for 
the LMC sample.
The first peak corresponds to objects in the LMC periphery which was 
observed less often than the center, the other two peaks to the regions in
and near the central bar.}
\label{fig:pointhist}
\normalsize 
\end{figure}

The LMC sample we present comprises $4634$ EBs selected by a variety of methods which we describe in this section; the sample includes the 611 EBs described in \citet{lacy97}.
The LMC magnitudes quoted in this paper have been obtained by using the 
following transformation:
\begin{eqnarray}\label{eq:machocal}
V&=&V_{\mathrm{MACHO}}+24.22~\mathrm{mag}-0.18(V_{\mathrm{MACHO}}-R_{\mathrm{MACHO}})
\nonumber \\
R&=&R_{\mathrm{MACHO}}+23.98~\mathrm{mag}+0.18(V_{\mathrm{MACHO}}-R_{\mathrm{MACHO}}).
\end{eqnarray}
From now on we will use the symbols $V$, $R$, and $\vr$ to refer to
standard magnitudes obtained from $V_{\mathrm{MACHO}}$ and $R_{\mathrm{MACHO}}$ via Eq.
\ref{eq:machocal} for the LMC and Eq. \ref{eq:smcmachocal} for the
SMC, and \emph{not corrected for reddening};
we will also use 
$V_{\mathrm{MACHO}}$ and $R_{\mathrm{MACHO}}$ to indicate instrumental magnitudes.
The observations number in several hundreds in both bandpasses for most 
light curves; Figure \ref{fig:pointhist} shows histograms of the number of 
light curve points of the EBs in both  bands; the $V_{\mathrm{MACHO}}$ band has on 
average more observations than the $R_{\mathrm{MACHO}}$ band  because one half of 
one of the red CCDs was out of commission during part of the project.
The central fields of the LMC were observed more  often and the periphery 
less often as shown by the three peaks in the distribution where the first 
peak corresponds to the LMC periphery and the other two correspond to the 
center.
\subsection{Identifying Eclipsing Binary Stars in the MACHO database}
This Section describes the techniques employed to identify EBs 
both in the LMC and in the SMC; the results we quote are relative to the 
LMC.
From now on we will always use the term \emph{unfolded light curve} to 
indicate a set of time ordered observations and will reserve the term 
\emph{light curve} to indicate a set of time ordered observations 
\emph{folded} (or \emph{phased}) around a period, omitting for brevity the 
adjectives ``folded'' and ``phased''; we will also use the terms
EB, system, and object interchangeably.
\par
All sources in the survey were subjected to a test for variability 
\citep{cook95} and a large number of variable sources were identified 
\citep{machovar1,machovar2,machovar3,machovar4}.
This first test starts by first eliminating the $20\%$ most extreme 
photometric data points; the resulting unfolded light curve is fitted to 
constant brightness and its $\chi^2$/dof is calculated.
The elimination of the most extreme points is expected to reduce the 
influence of noise 
and yield a good fit for a constant source, but not for a truly variable 
one.
We consider the source variable if the $\chi^2$/dof thus computed can occur
by chance with probability $1\%$ or less. 
Sources that were flagged for variability were tested for periodicity.
Periods were found using the Supersmoother algorithm \citep[][first 
published by \citet{fri84}]{reimann94}.
The algorithm folds the unfolded light curve around trial periods
and selects those periods in which the smoothed light 
curve matches the data best in a statistical sense; we have selected the 
best 15 possible periods ranked by the smoothness of the light curve.
Periods were found separately for the red and blue unfolded light curves.
The period selected as the best one by the program turned out to be 
``correct'' in $88\%$ of the EBs for at least one band; for $10\%$ of the EBs 
the second best period turned out to be correct for at least one band 
and only for $<0.5\%$ of the EBs did the procedure fail to find a good 
period.
In these cases ``correctness'' was determined by direct visual inspection.
\tabletypesize{\footnotesize}
\begin{deluxetable}{lccccc}
\tablecolumns{6}
\tablewidth{0pc}
\tablecaption{Period determination via Supersmoother.
\label{tab:period}}
\tablehead{
\colhead{Color} 
&
\colhead{Number of EBs} 
&
\colhead{$P$\tablenotemark{\dagger}$=P_{\mathrm{SS}}$\tablenotemark{\ddagger}} 
& 
\colhead{$P$\tablenotemark{\dagger}$=2P_{\mathrm{SS}}$\tablenotemark{\ddagger}} 
& 
\colhead{$P$\tablenotemark{\dagger}$=P_{\mathrm{SS}}/2$\tablenotemark{\ddagger}} 
& 
\colhead{Other}
}
\startdata
Red &
$4634$ &
$3432$ &
$604$ &
$103$ &
$495$ \\
&
&
$74\%$ &
$13\%$ &
$2\%$ &
$11\%$ \\
Blue &
$4634$ &
$3460$ &
$583$ &
$108$ &
$483$ \\
&
&
$75\%$ &
$13\%$ &
$2\%$ &
$10\%$ \\
\enddata
\tablenotetext{\dagger}{Real period.}
\tablenotetext{\ddagger}{Best period found by Supersmoother.}
\end{deluxetable}
The Supersmoother program can fail in two manners when fitting EBs.
First when one eclipse (the secondary) is very shallow, Supersmoother may 
not recognize it and yield a period twice the correct one.
Second, when the two eclipses have nearly equal depth Supersmoother may 
confuse the secondary and the primary eclipses yielding a period half the 
correct one.
The first failure happened, for one or both bands, in about $2\%$ of EBs,
whereas the second happened in about $13\%$ of EBs.
These cases are easy to correct upon visual inspection.
For $\sim 100$ EBs Supersmoother gave a period which was some other 
multiple of the correct one for at least one band; these were fixed 
upon visual inspection.
For the remaining EBs we tried folding the light curves around the other 
periods selected by Supersmoother and managed to identify the correct 
period for most of them.
In $27$ cases in which there were OGLE-II counterparts we adopted OGLE 
periods since, though differing in some cases by less than $1\%$ from the 
periods found by Supersmoother, they gave a much better light curve.
We found $51$ stars in which the secondary eclipse was not evident, either 
because it was shallow or because the light curve was noisy, but with an 
OGLE-II counterpart in which it was clearly visible; these stars have not 
been included in the catalogue.
The periods in the two bands differ on average by $0.02\%$.
These results are summarized in Table \ref{tab:period}.
\par
The search for variable objects in the LMC gave $\sim 207,000$ objects
of which $\sim 66,000$ were found to be periodic.
To find EBs in this sample we considered a variety of properties of
light curves.
The techniques we employed are:
\begin{enumerate}
\item
Look at the number of photometric excursions (``dips'') in the light curve.
An EB is expected to show two ``dips'' in an entire period corresponding to 
the two eclipses as opposed to a Cepheid or an RRLyr\ae~star for which only
one is expected.
The number of photometric excursions was calculated by Supersmoother
by counting the number of times the smoothed light curve crosses the mean:
we selected stars with two excursions in both bands.
We additionally imposed a cut on light curve amplitudes:
calling $\mathrm{Ampl_V}$ and $\mathrm{Ampl_R}$ the amplitudes of the
blue and red light curves respectively, as computed by Supersmoother, we imposed
$\mathrm{Ampl_V}/\mathrm{Ampl_R}< 1.2$.
This amplitude cut was imposed to help in eliminating RRLyr{\ae}s from the
sample, since, considering a population of several thousands probable RRLyr{\ae}s 
found in the MACHO database, we found that, on average, 
$\mathrm{Ampl_V}/\mathrm{Ampl_R}\sim 1.27$; imposing a cut 
$\mathrm{Ampl_V}/\mathrm{Ampl_R}< 1.2$ should therefore filter out many RRLyr{\ae}.
In our sample just $2421$ EBs ($\sim 52\%$) pass this cut; we then removed 
the amplitude cut and look at the number of photometric excursions alone
we found that $3039$ EBs ($\sim 66\%$) pass this relaxed cut.
\item
Look at the ratio of number of points five standard deviations 
\textit{away} ($\mathrm{s}5$) from the median to the number of points five standard 
deviations \textit{below} the median ($\mathrm{s}5\mathrm{d}$).
This ratio is expected to be $\sim 2$ for a typical single variable star.
For an EB we expect this ratio to approach $\sim 1$ as the signal to 
noise in the photometry increases, as most ``outlier'' points are due to 
eclipses.
We imposed a cut $\mathrm{s}5/\mathrm{s}5\mathrm{d}<1.2$ in both bands and 
found that $3417$ EBs pass it ($\sim 74\%$), whereas for the overall variable 
star dataset the figure is $\sim 19,600$ out of $66,000$ or $\sim 30\%$.
\item
Use a decision tree.
We applied the decision tree program described in \citet{murthy94}, which 
was run on all the variable objects on the catalogue and gave for each the 
probability that it was an EB, an RRLyr\ae, a Cepheid, a long period 
variable or an unknown object. 
In all $\sim 17000$ objects were found most likely by the decision tree to be EBs 
but only $3281$ were found to be real on visual inspection and included in the sample.
\item
Use a similarity technique.
We finally tested the sample with a technique described in 
\citet{protopapas06} aimed at finding ``outliers'' in large data sets of variable star light curves.
The technique aims at finding objects whose light curves are most dissimilar, in a statistical
sense, from an ``average'' light curve built out of all the light curve in the data set.
This is accomplished by looking at all the pairs of light curves to find their mutual 
similarity as defined in \citet{protopapas06} and, for each light curve, by then 
combining these measures, to find its overall similarity to the rest of the sample.
Light curves with low measure of similarity are flagged as outliers.
This approach was useful in finding misfolded lightcurves since it
found many objects for which the period for one band gave a badly folded light curve but the period for the other band gave a good folding.
This happened for $198$ EBs, despite these periods differing on average by 
just $0.18\%$; in this case we selected the period that gave the good light 
curve for both bands.
\end{enumerate}
\par
Since all the techniques we used give some false positives, each candidate was also visually inspected before inclusion in the sample; we paid closer attention to those stars which could more easily be classified as EBs without being so, 
like ellipsoidal variables (see Subsection \ref{subsec:ell}) and Cepheids and RRLyr\ae~mistakenly folded around a period twice their real one.  
\tabletypesize{\footnotesize}
\begin{deluxetable}{cccccc}
\tablecolumns{6}
\tablewidth{0pc}
\tablecaption{Summary of cuts applied to define the LMC sample.
\label{tab:cuts}}
\tablehead{
\colhead{} &
\colhead{Number of EBs} &
\colhead{$2$ dips} & 
\colhead{$2$ dips and $\mathrm{Ampl_V}<1.2\mathrm{Ampl_R}$} & 
\colhead{$\mathrm{s}5/\mathrm{s}5\mathrm{d}<1.2$} & 
\colhead{Decision tree}
}
\startdata
EB &
$4634$&
$3039$ &
$2421$ &
$3417$ &
$3281$ \\
&
&
$66\%$ &
$52\%$ &
$74\%$ &
$71\%$ \\
All variable sources &
$\sim 66,000$ & 
$\sim 8000$ & 
$\sim 7300$ &
$\sim 19,600$ & 
$\sim 7900$ \\
& 
& 
$\sim 12\%$ &
$\sim 11\%$ &
$\sim 30\%$ & 
$\sim 12\%$ \\
\enddata
\end{deluxetable}
\tabletypesize{\footnotesize}
\begin{deluxetable}{lc}
\tablecolumns{2}
\tablewidth{0pc}
\tablecaption{Summary of EBs passing more than one cut.
\label{tab:morethanonecut}}
\tablehead{
\colhead{Cuts} &
\colhead{Number of EBs}
}
\startdata
$2$ dips and $\mathrm{s}5/\mathrm{s}5\mathrm{d}<1.2$ &
$2061$ \\
$2$ dips, $\mathrm{Ampl_V}<1.2\mathrm{Ampl_R}$, and $\mathrm{s}5/\mathrm{s}5\mathrm{d}<1.2$ &
$1686$ \\
$2$ dips and decision tree &
$2031$ \\
$\mathrm{s}5/\mathrm{s}5\mathrm{d}<1.2$, and decision tree &
$2574$ \\
$2$ dips, $\mathrm{Ampl_V}<1.2\mathrm{Ampl_R}$, and decision tree &
$1761$ \\
$2$ dips, $\mathrm{s}5/\mathrm{s}5\mathrm{d}<1.2$, and decision tree &
$1488$ \\
$2$ dips, $\mathrm{Ampl_V}<1.2\mathrm{Ampl_R}$, $\mathrm{s}5/\mathrm{s}5\mathrm{d}<1.2$, and decision tree &
$1280$ \\
\enddata
\end{deluxetable}
These results of our search techniques are summarized in Table 
\ref{tab:cuts}, and Table \ref{tab:morethanonecut} shows the number of 
EBs that pass more than one cut.
As the numbers show these heuristic tests are far from perfect and tend to 
give too many candidates; however we feel that the large dimension of our 
sample can allow the determination of more stringent tests, an absolute 
necessity for the analysis of future surveys.
\par
Our search gave $266$ objects observed in more than one field: these 
duplicates have different MACHO field numbers, but typically the same tile 
number.
In this case we summed the numbers of observations in both bands for each 
field and chose the one which had the highest total number of observations:
the object is identified by that corresponding FTS only.
\par
\begin{figure}
\footnotesize
\plotone{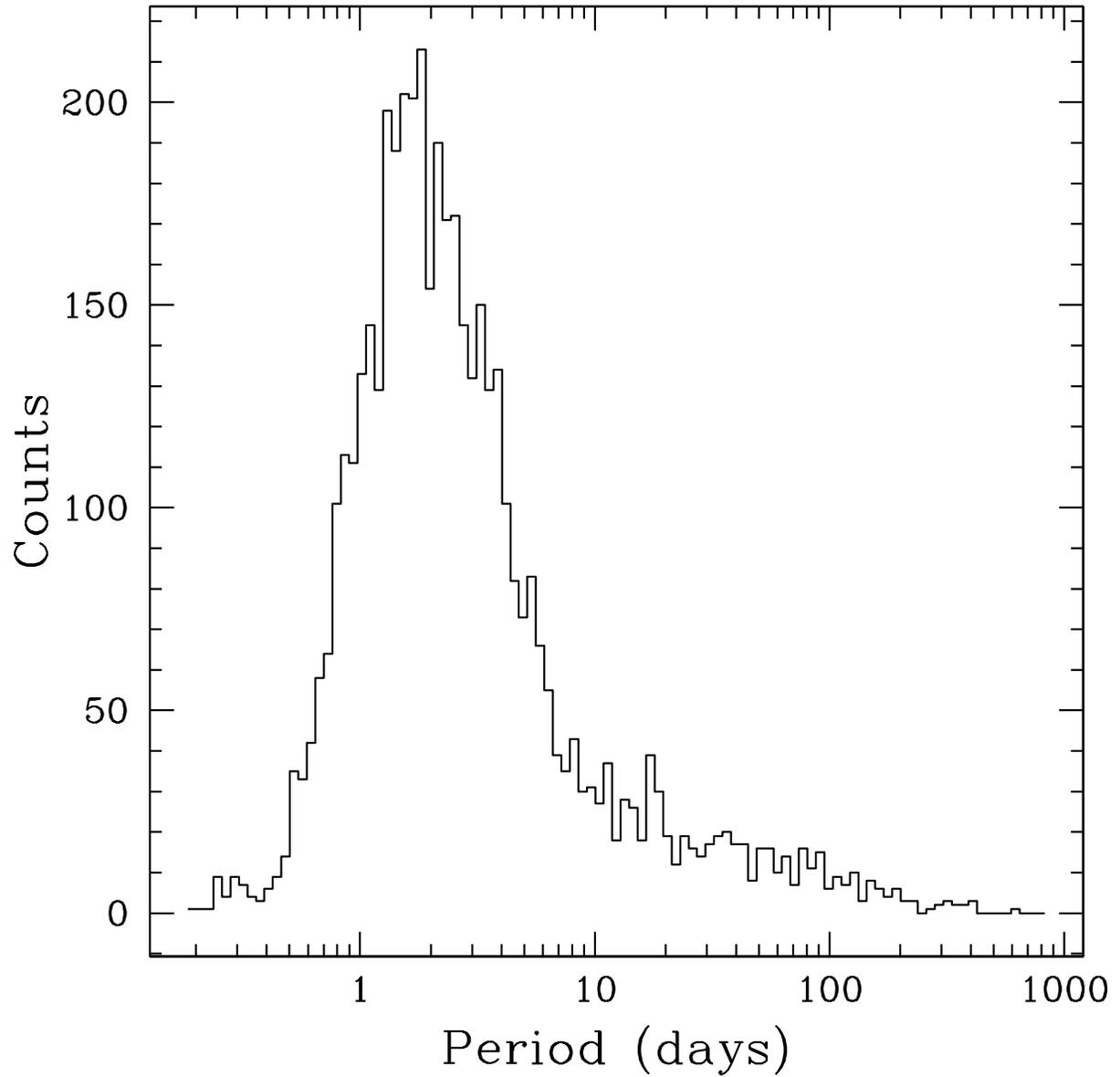}
\caption{Period histogram for $4634$ EBs in the LMC sample; the distribution 
peaks strongly in the $0.8-4\mathrm{d}$ range and has a tail in the 
$10-100\mathrm{d}$ range.
The size of the bins is $\sim 1/100$ of the span of the logarithms of the
periods.}
\label{fig:perhist}
\normalsize
\end{figure}
\begin{figure}
\footnotesize
\plotone{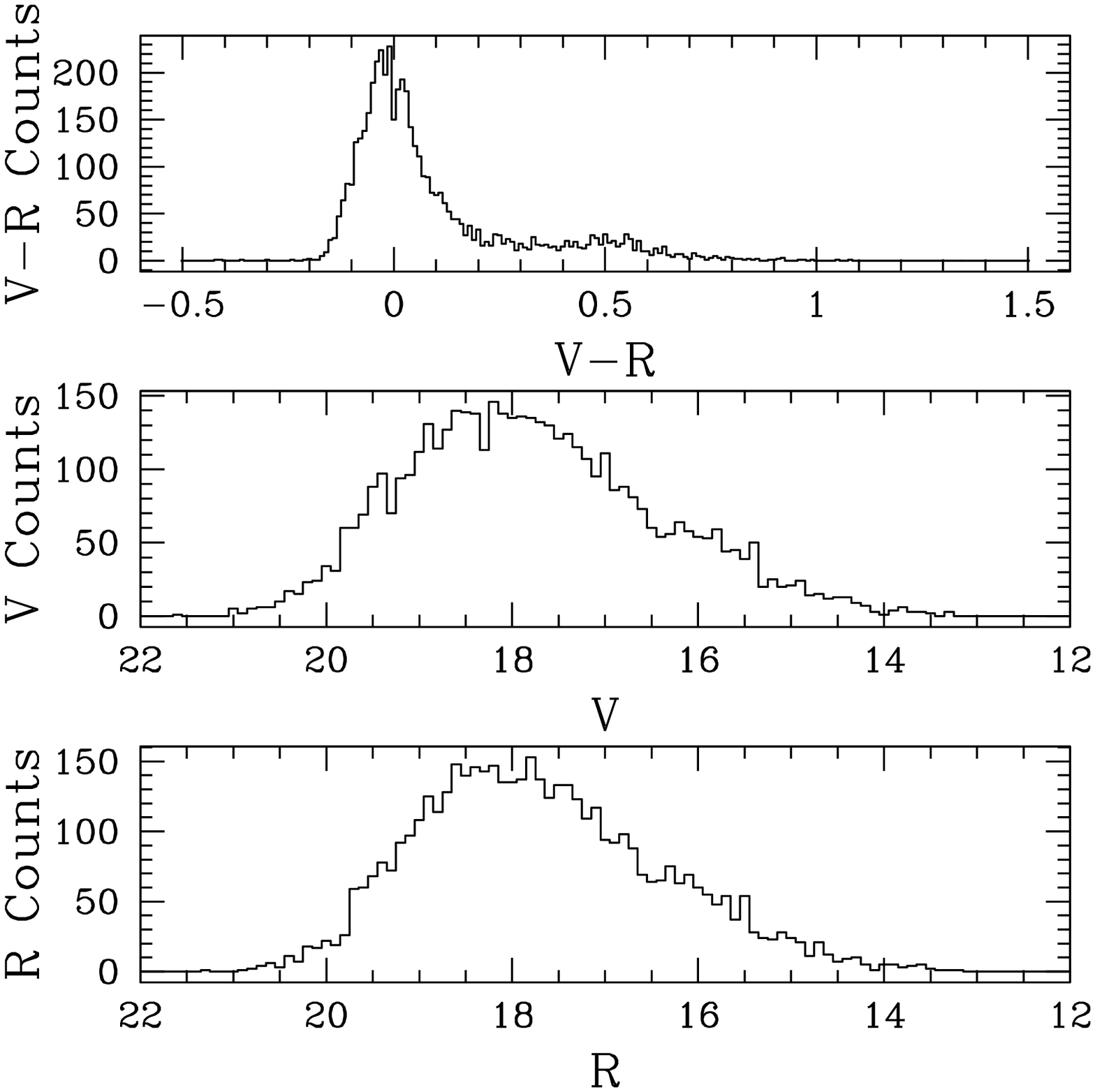}
\caption{Standard $V$ and $R$ magnitudes and $\vr$ histograms for the LMC 
sample; the bin size is $0.1~\mathrm{mag}$ for the $V$ and $R$ histograms and 
$0.01~\mathrm{mag}$ for the $\vr$ one.
The $\vr$ histogram is strongly peaked around $\vr\sim 0~\mathrm{mag}$, 
showing a majority of unevolved EBs, but the longer tail, with the smaller bump 
around $\vr\sim 0.5~\mathrm{mag}$ shows a sizeable minority of fairly evolved 
systems as also shown by the higher $V$ histogram values around $16~\mathrm{mag}$ 
with respect to the $R$ histogram.}
\label{fig:maghist}
\normalsize
\end{figure}
\begin{figure}
\footnotesize
\begin{center}
\plotone{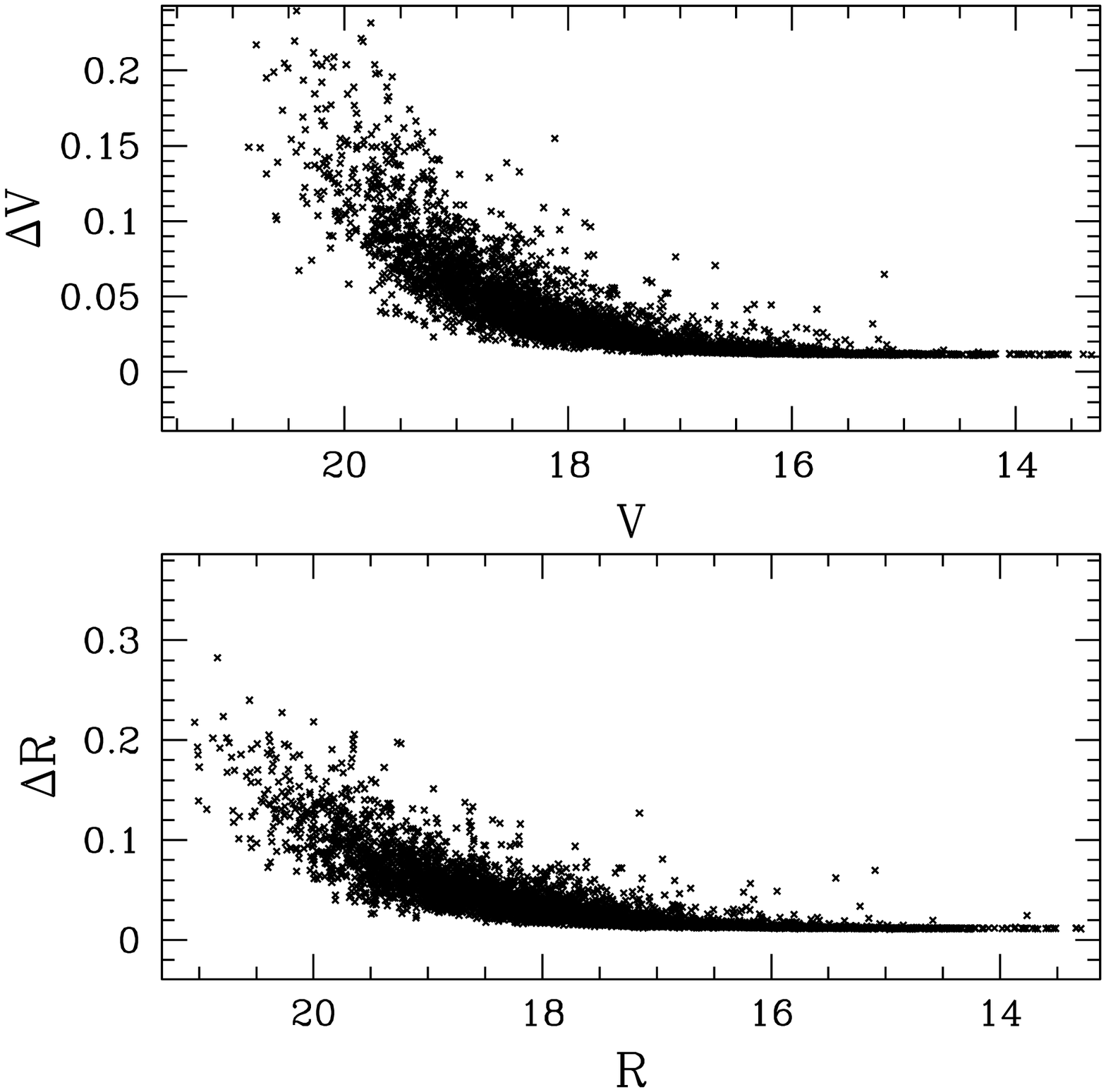}
\end{center}
\caption{Errors as a function of median magnitude for the LMC 
sample.} 
\label{fig:magerr}
\normalsize
\end{figure}

Figure \ref{fig:perhist} shows a logarithmic histogram of the period 
distribution.
Note that the periods range from a fraction of a day to several hundreds 
of days.
Figure \ref{fig:maghist} shows a histogram of the distribution of median 
$V$, $R$ and $\vr$.
Note that the magnitudes range in values from $\sim 20~\mathrm{mag}$ to 
$\sim 14~\mathrm{mag}$ both in $V$ and $R$ with a peak around $18~\mathrm{mag}$. 
The average photometric error for the LMC is $\sim 0.05~\mathrm{mag}$ in both 
instrumental bands; the average error for a light curve as a function of 
median relative magnitude is in standard magnitudes is shown in Figure 
\ref{fig:magerr}.
\subsection{Root Mean Square of residuals for the LMC sample}
\label{subsec:rms}
\begin{figure}
\footnotesize
\begin{center}
\plotone{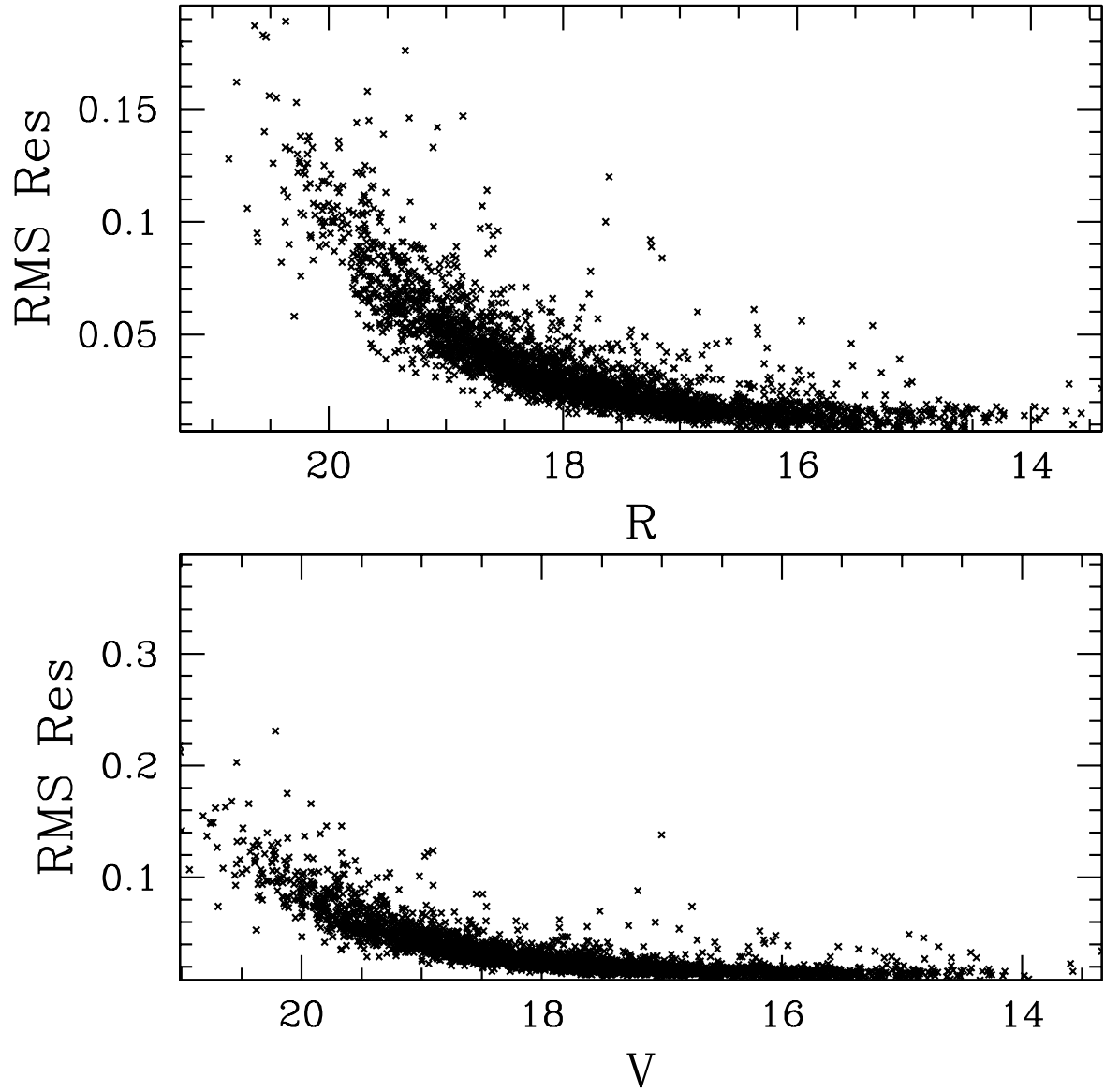}
\end{center}
\caption{RMS of residuals as a function of median relative magnitude for 
the LMC sample.} 
\label{fig:magrms}
\normalsize
\end{figure}

For the LMC sample we estimated the distribution the Root Mean Square (RMS) of the residuals $O_i-C_i$ of the observations around a theoretical light curve as a function of median relative magnitude where $O_i$ is the value of the observed magnitude orbital phase $\phi_i$, and $C_i$ is the theoretical value at the same phase.
Theoretical values were obtained by fitting the light curves using the 
JKTEBOP\footnote{\url{http://www.astro.keele.ac.uk/\~{}jkt/codes/jktebop.html}}
code \citep{southworth04a,southworth04b}.
The JKTEBOP code is based on the EBOP code \citep{etzel81,popper81}, which implements the model by \citet{nelson72} with some modifications; JKTEBOP in turn adds several modifications and extensions to the original EBOP code that make it easier to use, especially when fitting a large number of light curves.
Before fitting we eliminated outlying points by taking averages of all points in boxes containing from $\sim 10$ to $\sim 20$ points along a light curve and discarding the points more than $2$ standard deviations away from these averages.
We obtained starting values for the model parameters by running the DEBiL code
\citep{devor05} and using the values it computed; the limb darkening values
for the $V$ and $R$ bands were taken from \citep{cox00}.
We fixed the ratio of the masses, $q$, to $1$, $0.1$ and $10$ and did the fit 
in each case taking in the end the best result.
Finally we selected light curves with $\chi^2/\mathrm{dof}<2$ to show in Figure \ref{fig:magrms}.
Out of $4636$ EBs in the LMC the program converged in $4090$ cases in the $R$
band and in $4312$ cases in the $V$ band; we found $3067$ fits with $\chi^2/\mathrm{dof}<2$ in the $R$ band and $3198$ in the $V$ band.
We point out that those fits were made only with the aim of obtaining a good theoretical light curve for as many observed light curves as possible in a fast and automated manner, so that a residuals distribution could be calculated.
In particular we did \emph{not} attempt to accurately determine astrophysical parameters for our EBs.
This is also the reason why we discarded points at just $2$ standard deviations away from the 
moving averages, which could result in eliminating potentially interesting information
for some EBs; such objects are obviously deserving of more in depth study which
we did not attempt here.
While in general our fits were good in the case of largely separated, undistorted systems, they were often bad for close, strongly distorted ones, which is to be expected since JKTEBOP is not meant to be used for such systems; also in several cases our fits were bad because the 
scatter of the observed values was larger than the observational errors, which suggests that
other physical phenomena, such as pulsation of one or both components, are present.
The RMS distributions of the residuals vs. median magnitudes $V$ and $R$ are 
shown in Figure \ref{fig:magrms}.
\subsection{Examples of light curves}
\begin{figure}
\footnotesize
\begin{center}
\plottwo{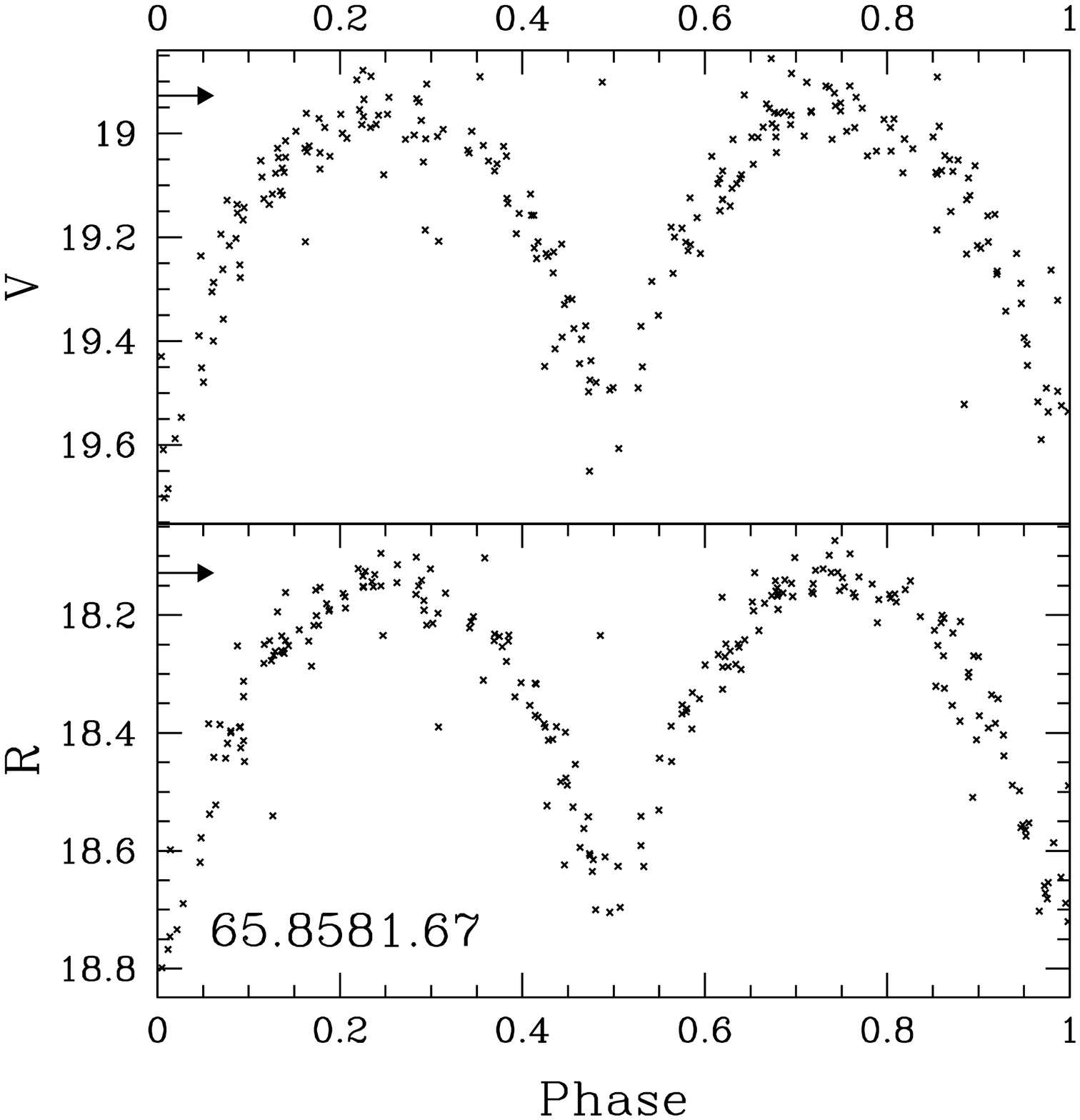}{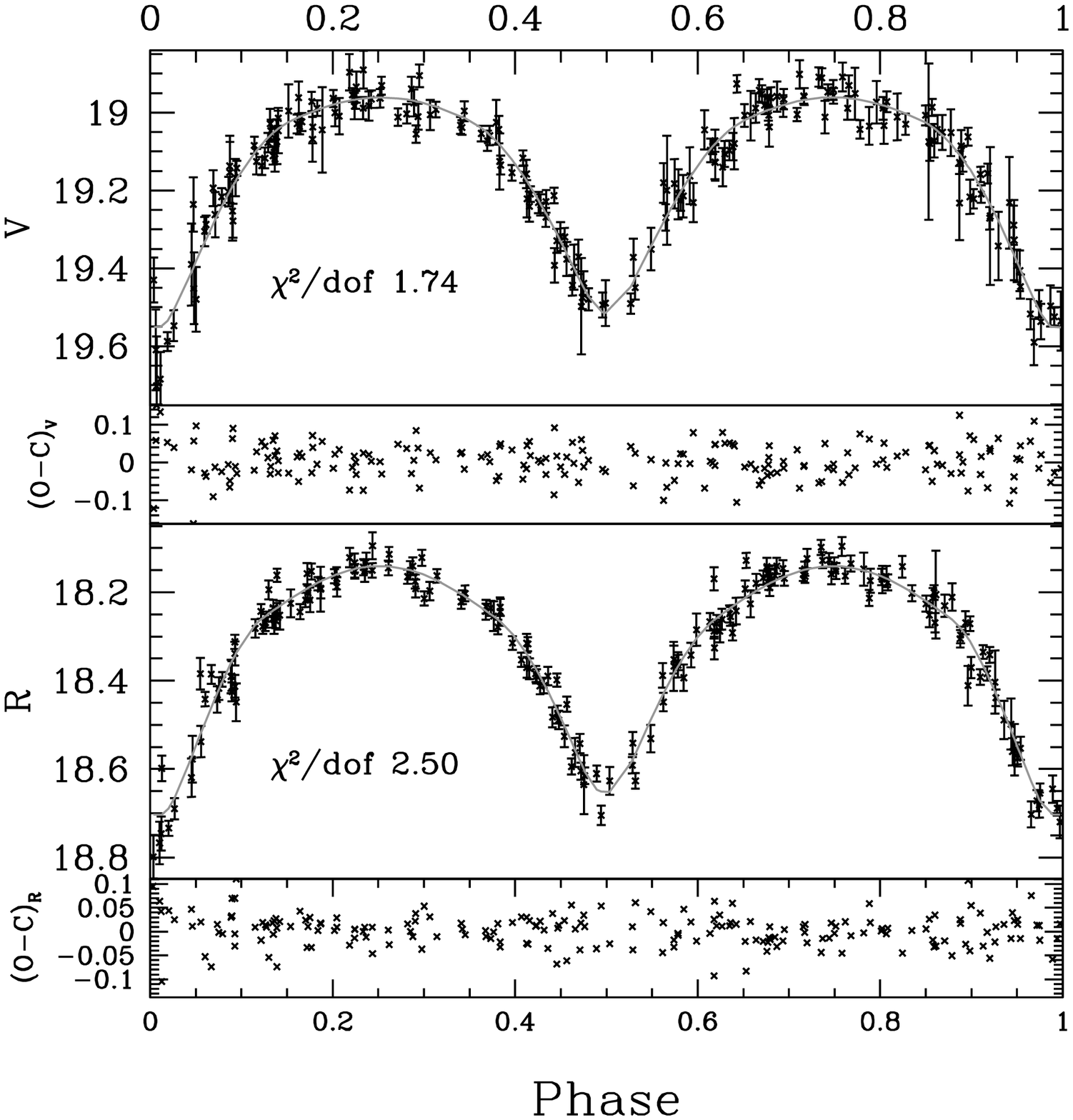}
\plottwo{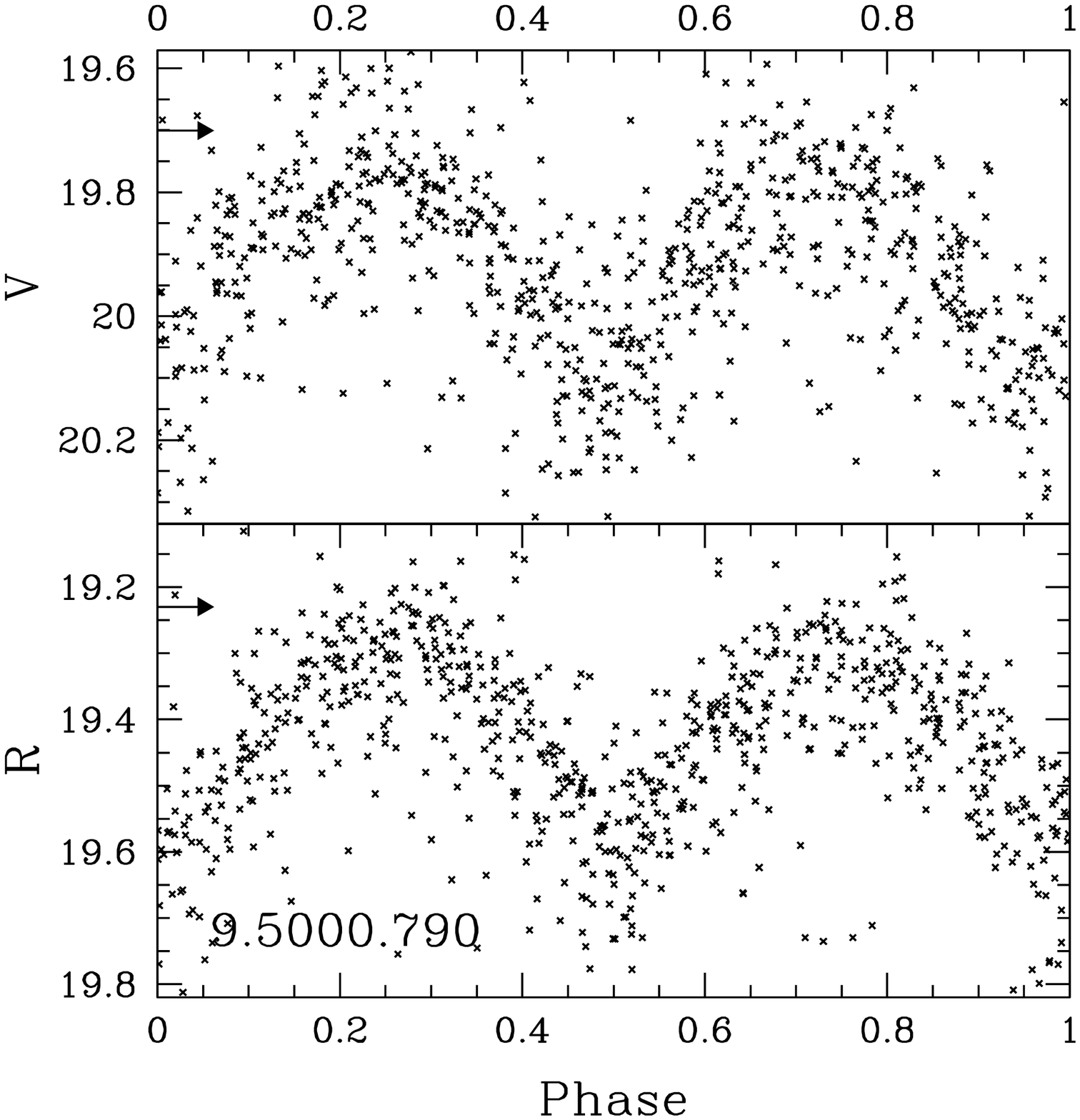}{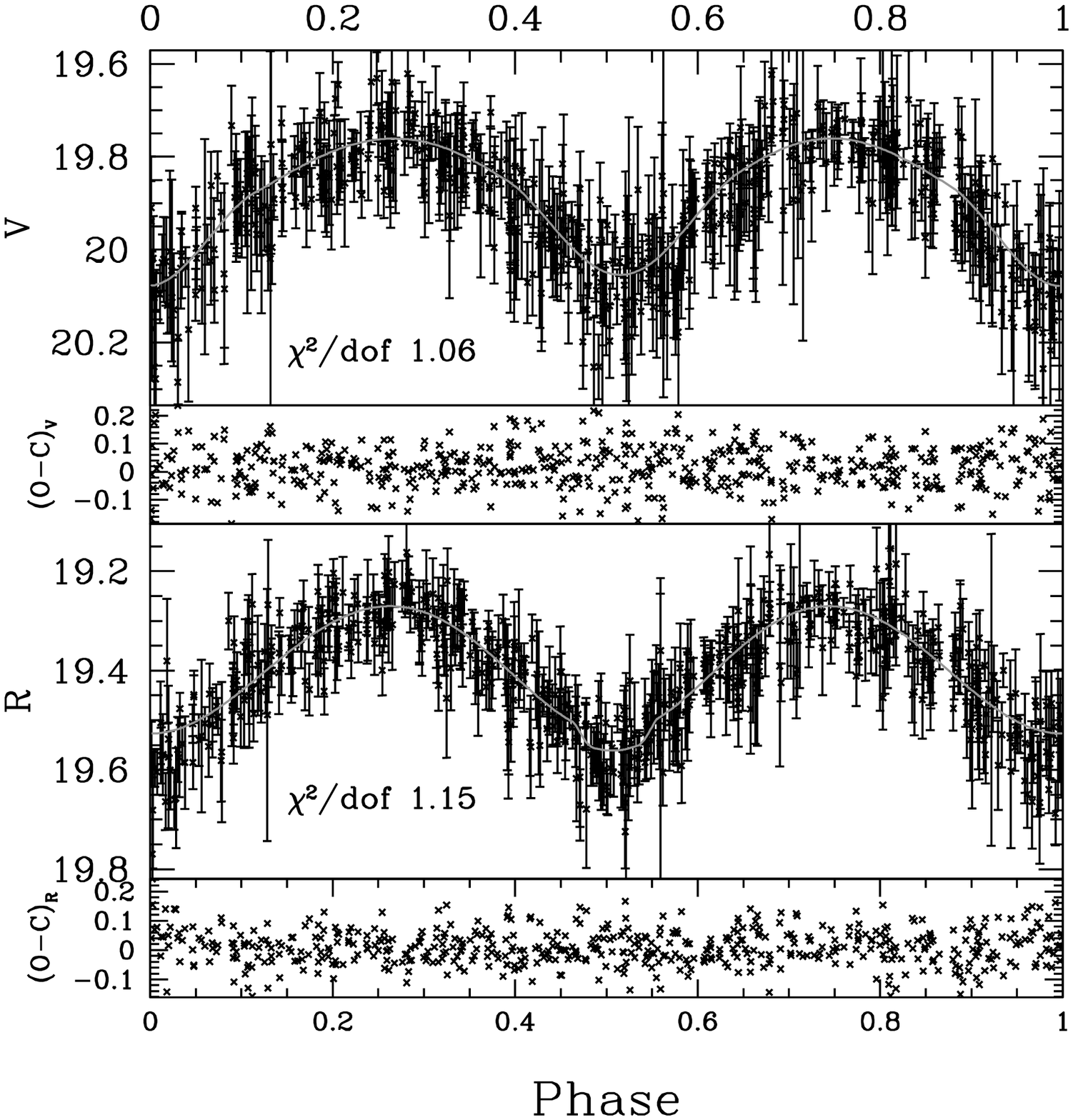}
\end{center}
\caption{Examples of LMC EBs light curves, arranged by ascending period; for basic data see Table \ref{tab:lcs}.
Left: observed light curves with all data points. 
The arrows show the baseline as defined in Table \ref{tab:lcs}.
Right: observed light curves with outlying points removed, theoretical
light curves from the fit and residuals.}
\label{fig:lcs1}
\normalsize
\end{figure}

\begin{figure}
\footnotesize
\begin{center}
\plottwo{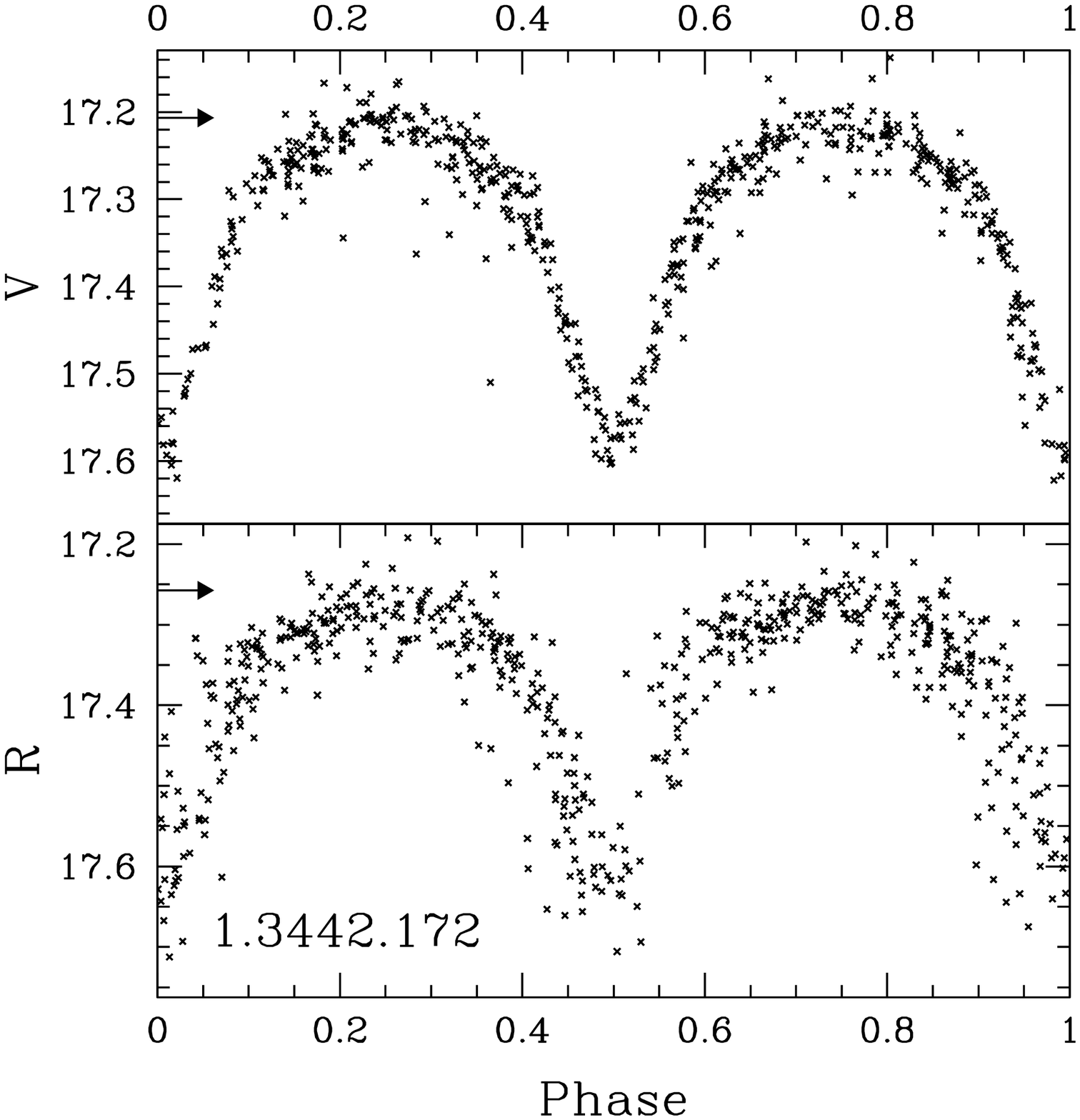}{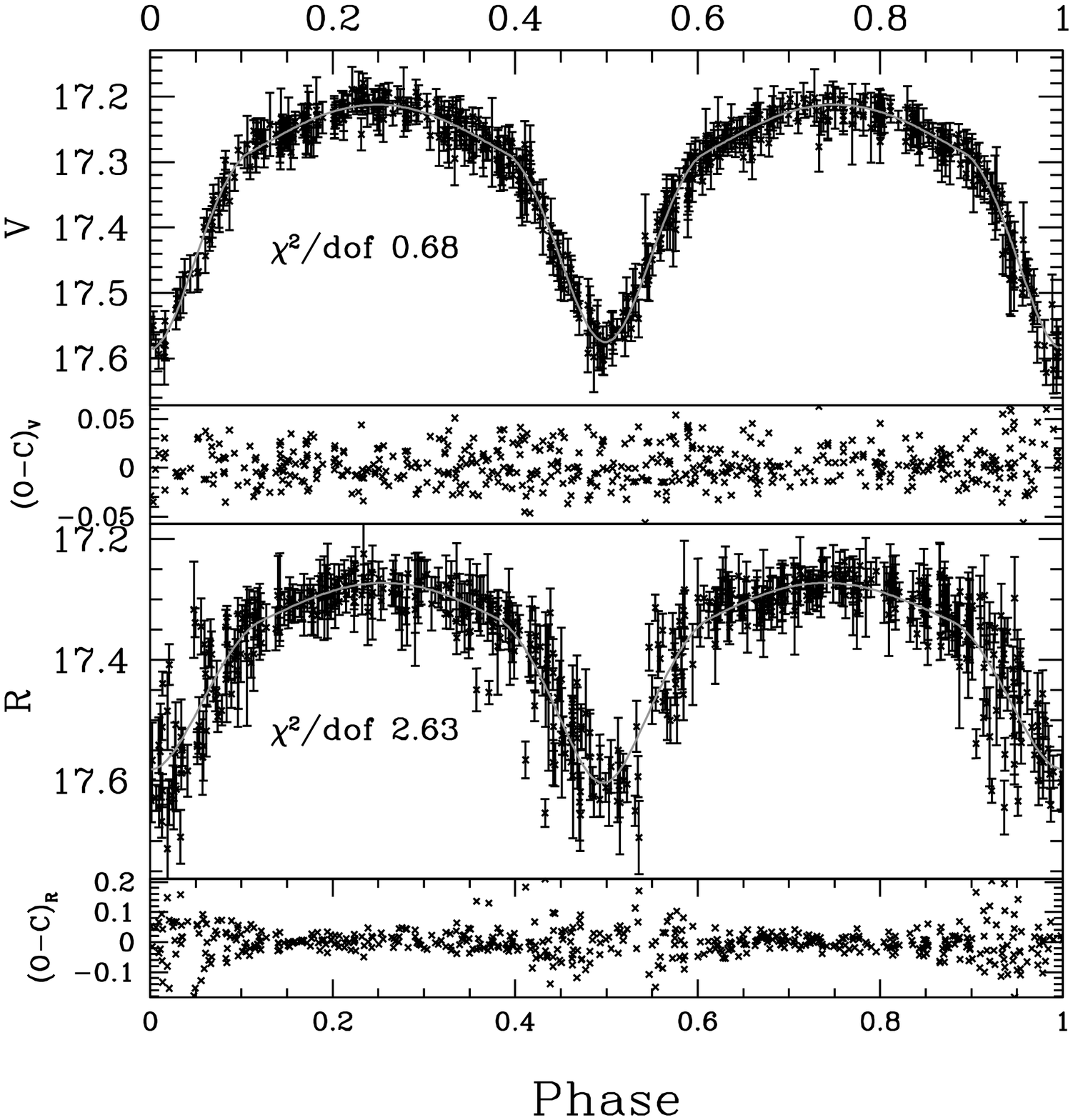}
\plottwo{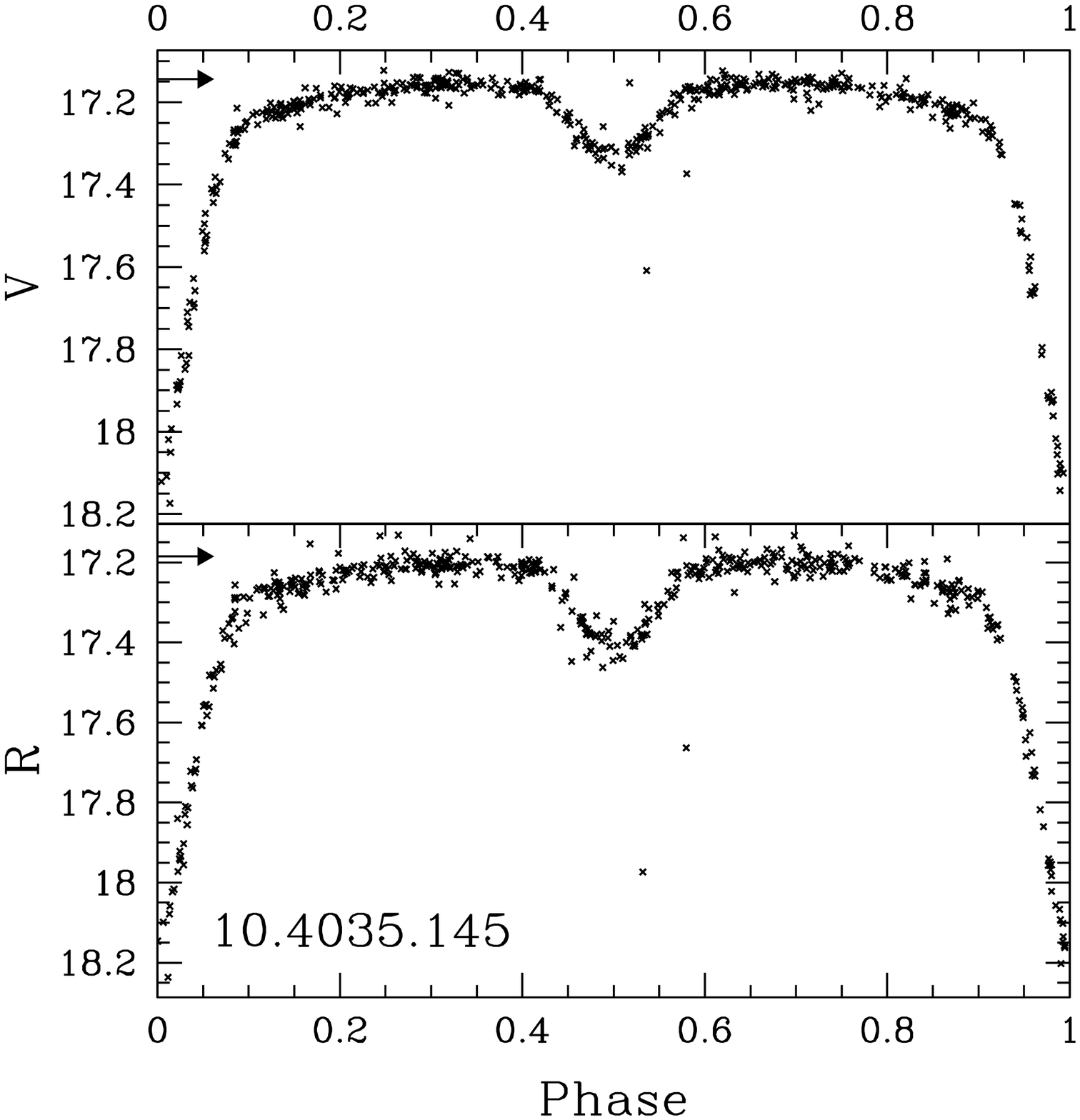}{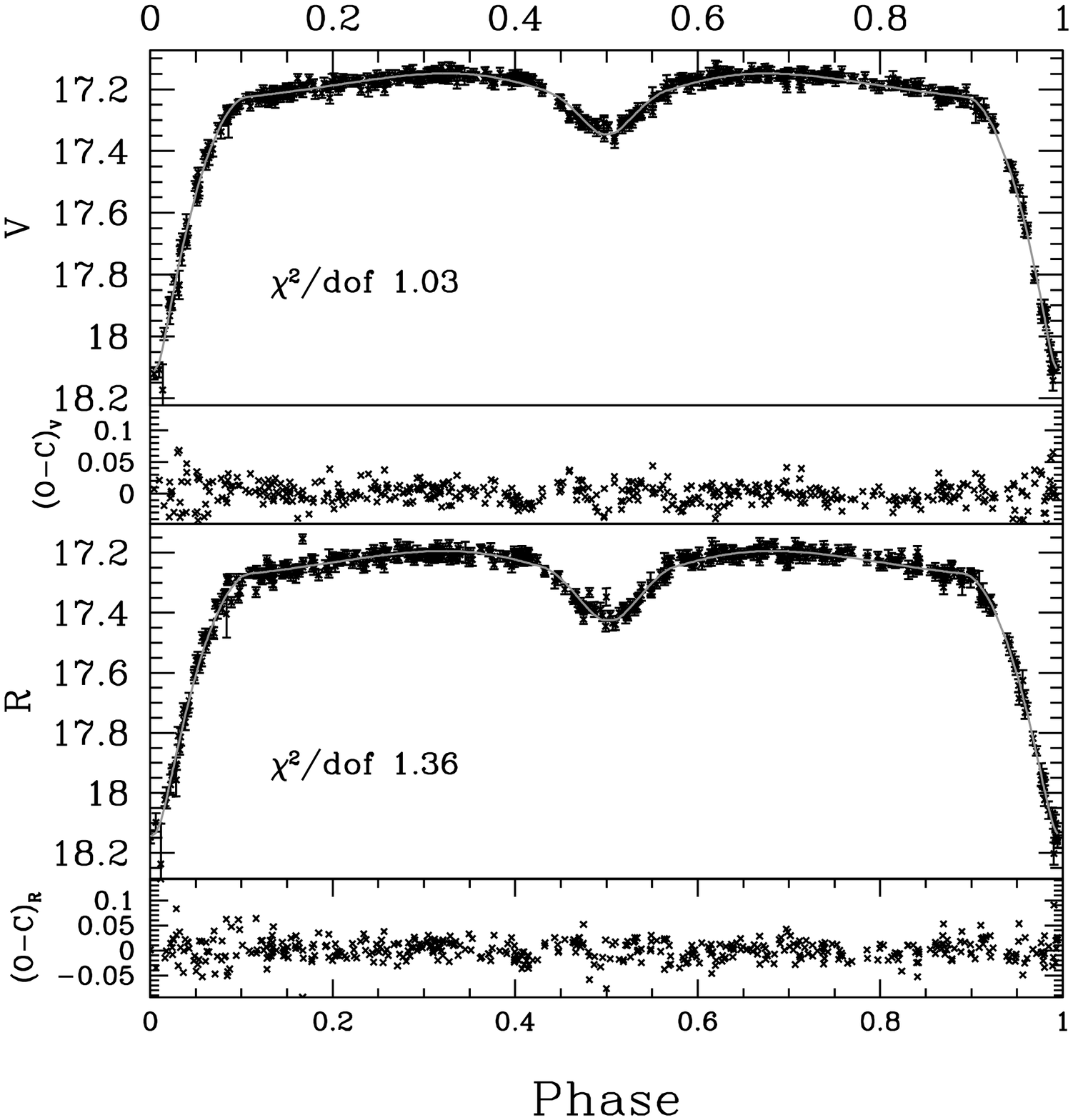}
\end{center}
\caption{Examples of LMC EBs light curves, arranged by ascending period; for basic data see Table \ref{tab:lcs}.
Left: observed light curves with all data points. 
The arrows show the baseline as defined in Table \ref{tab:lcs}.
Right: observed light curves with outlying points removed, theoretical
light curves from the fit and residuals.} 
\label{fig:lcs2}
\normalsize
\end{figure}

\begin{figure}
\footnotesize
\begin{center}
\plottwo{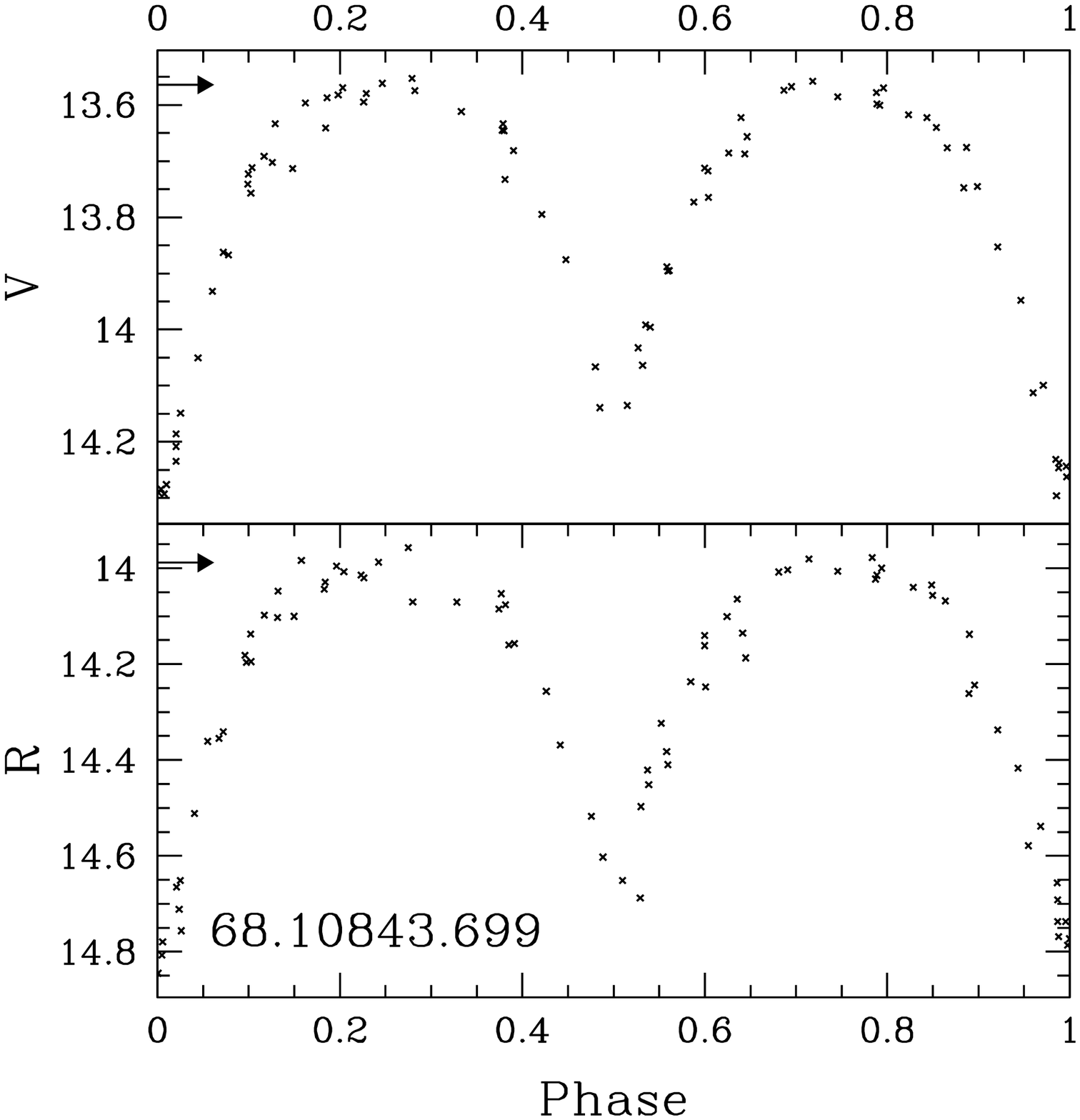}{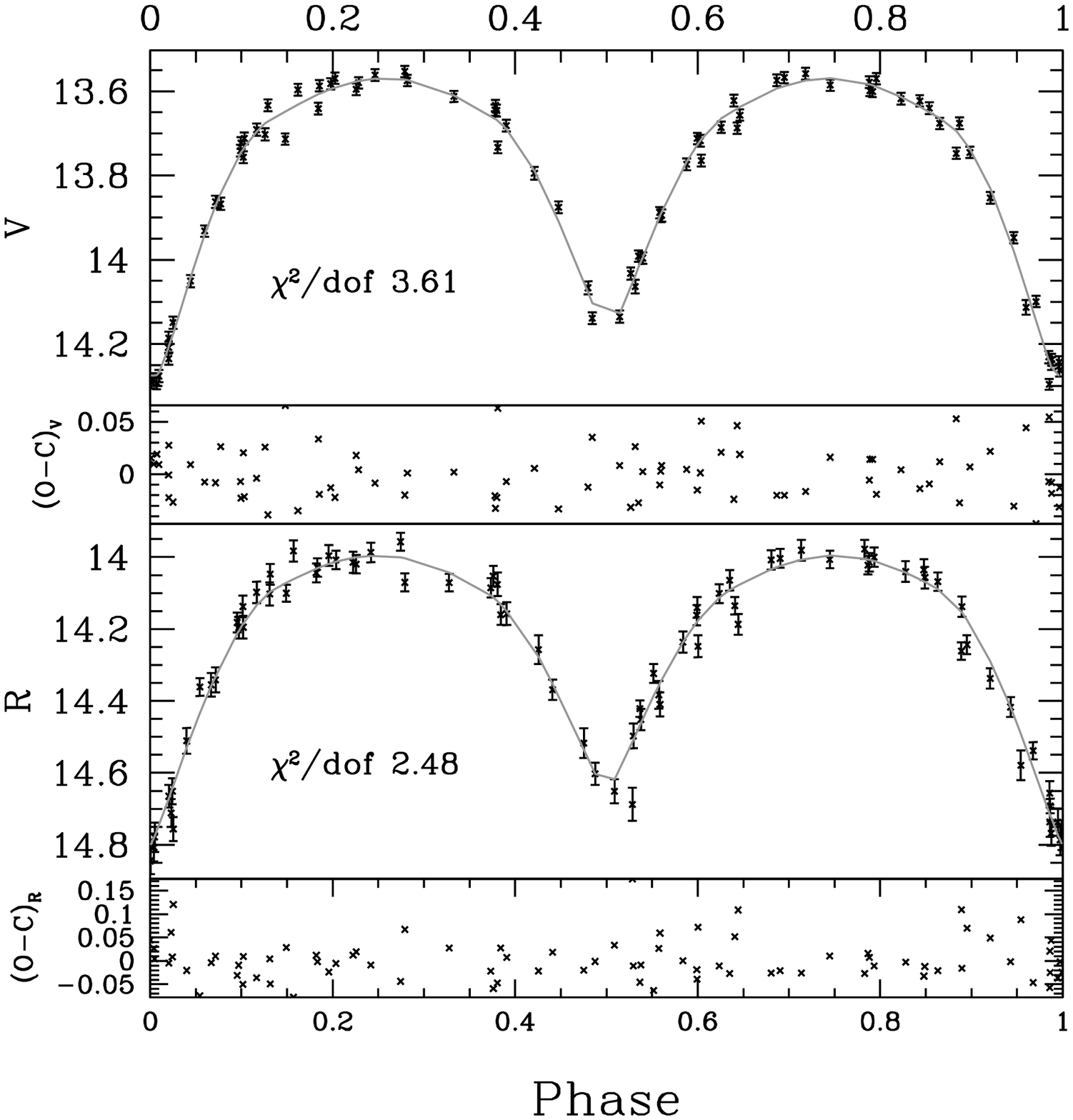}
\plottwo{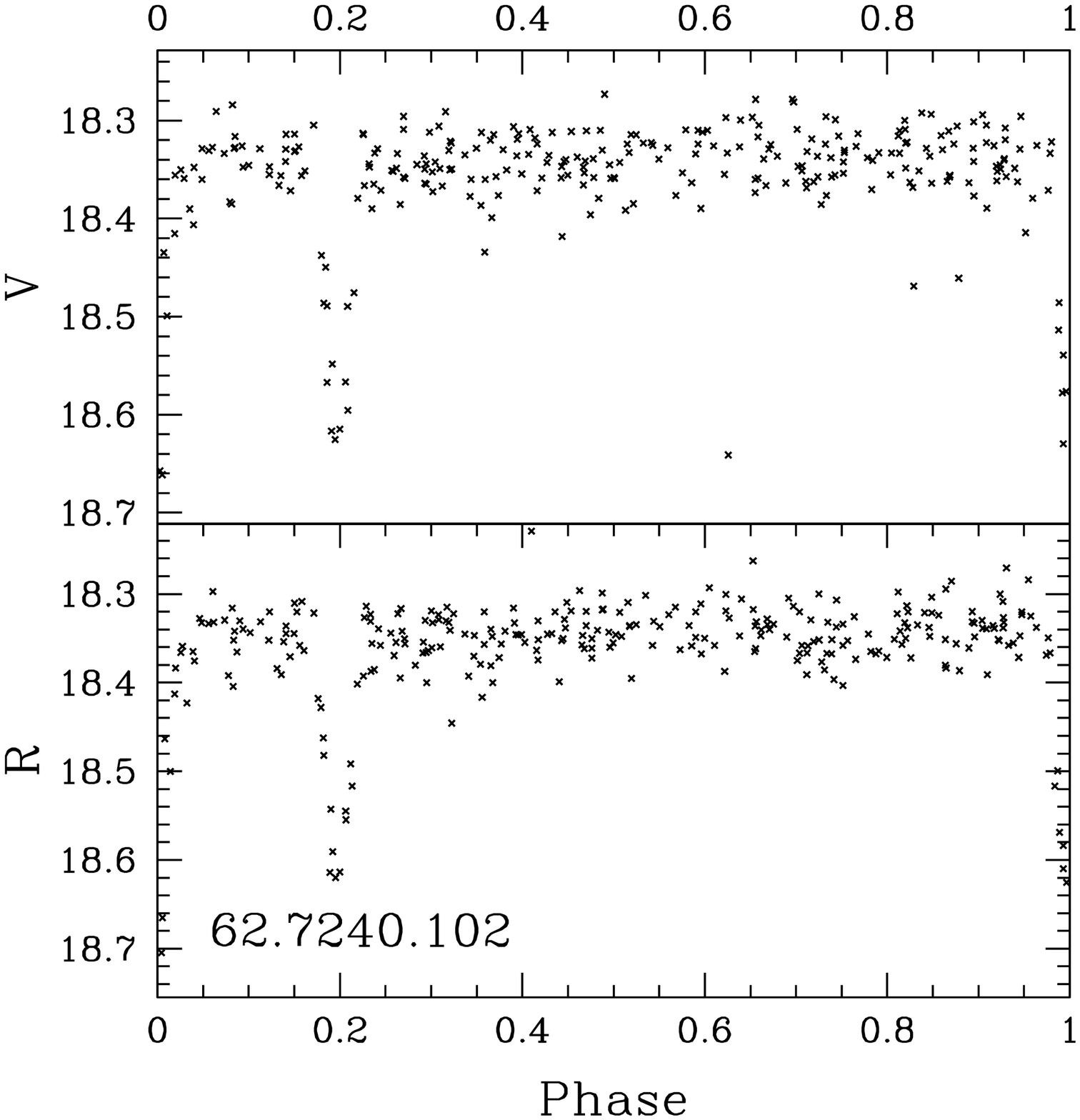}{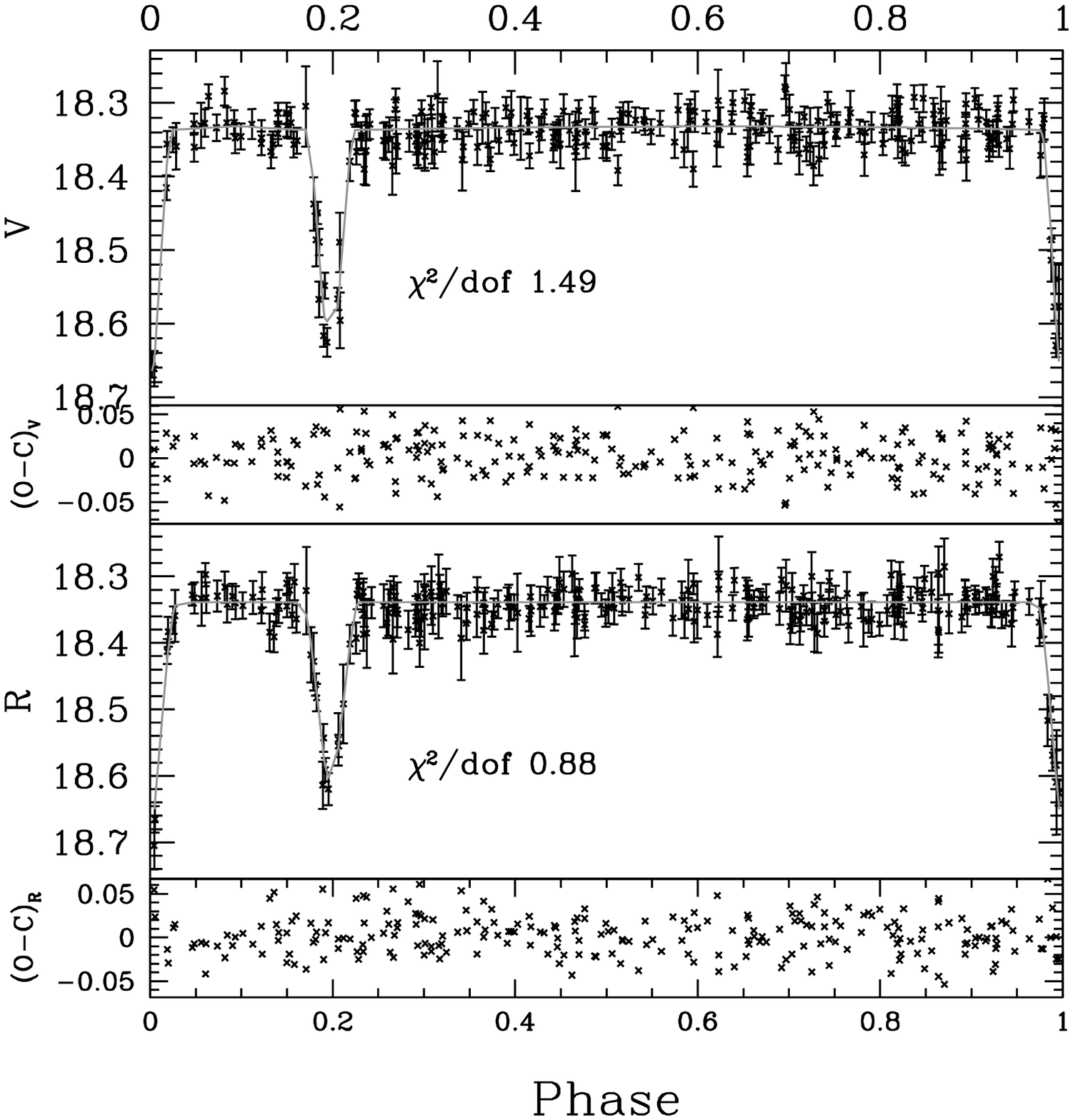}
\end{center}
\caption{Examples of LMC EBs light curves, arranged by ascending period; for basic data see Table \ref{tab:lcs}.
Left: observed light curves with all data points. 
The arrows show the baseline as defined in Table \ref{tab:lcs}.
Right: observed light curves with outlying points removed, theoretical
light curves from the fit and residuals.}
\label{fig:lcs3}
\normalsize
\end{figure}

\begin{figure}
\footnotesize
\begin{center}
\plottwo{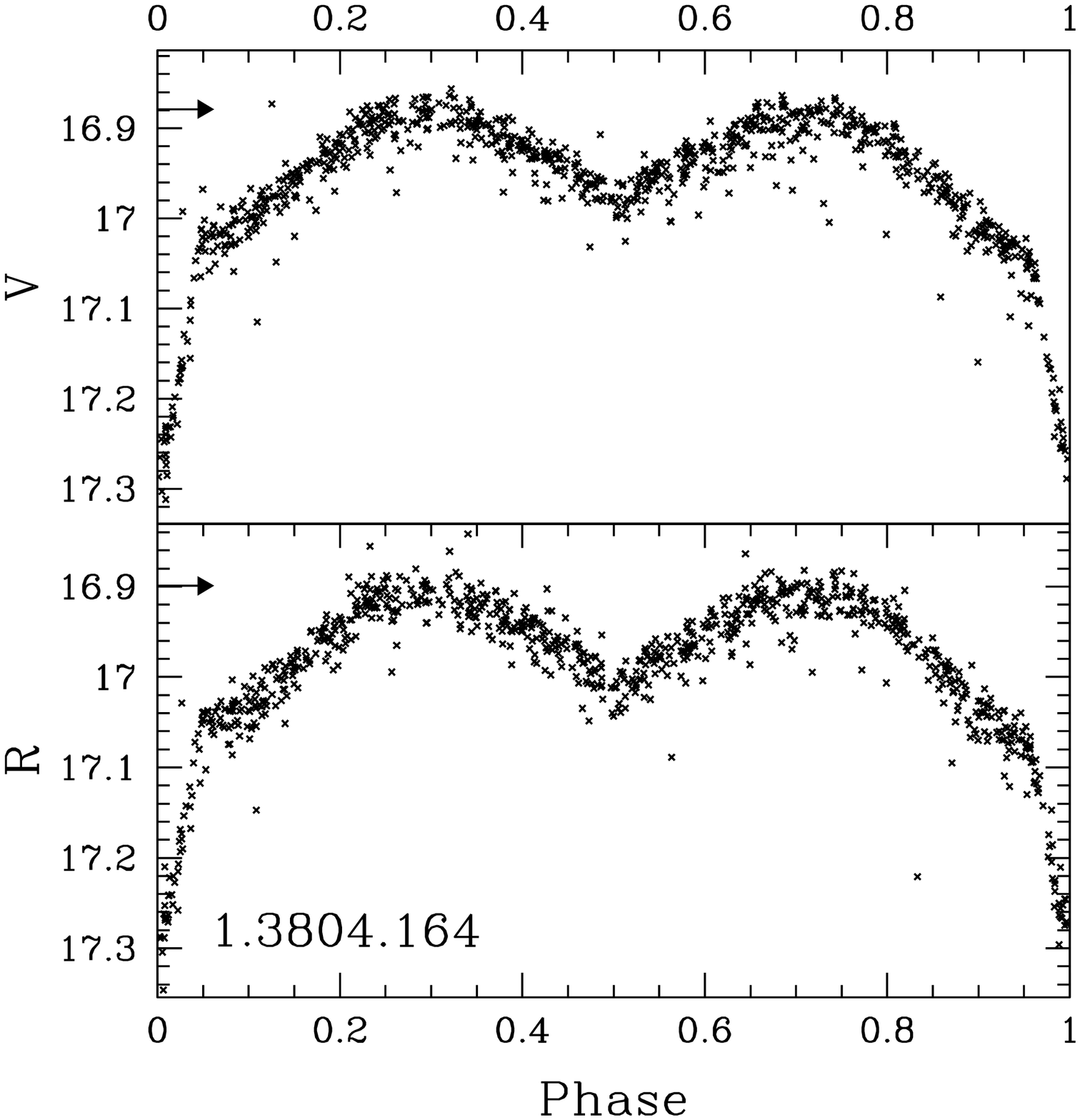}{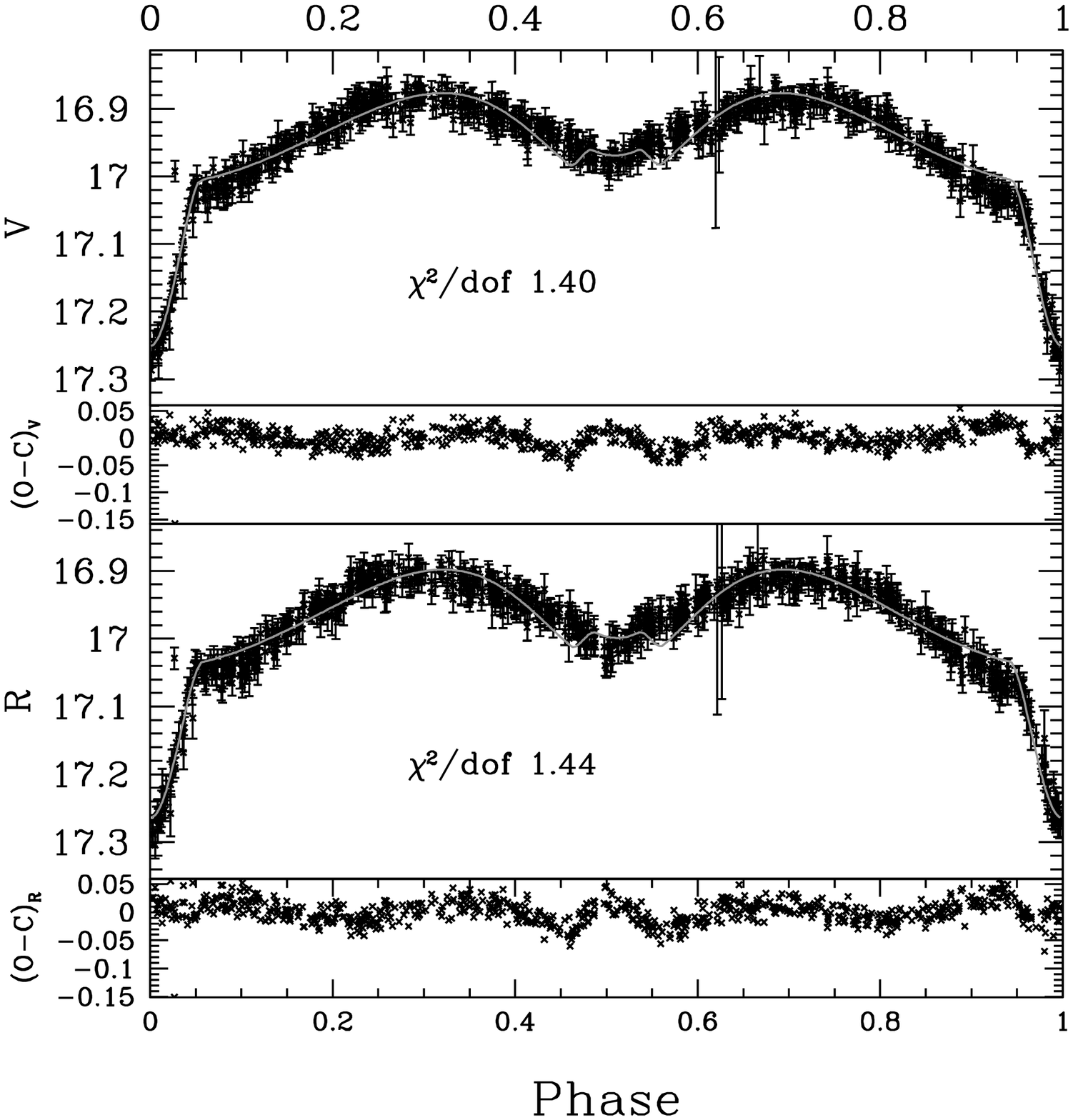}
\plottwo{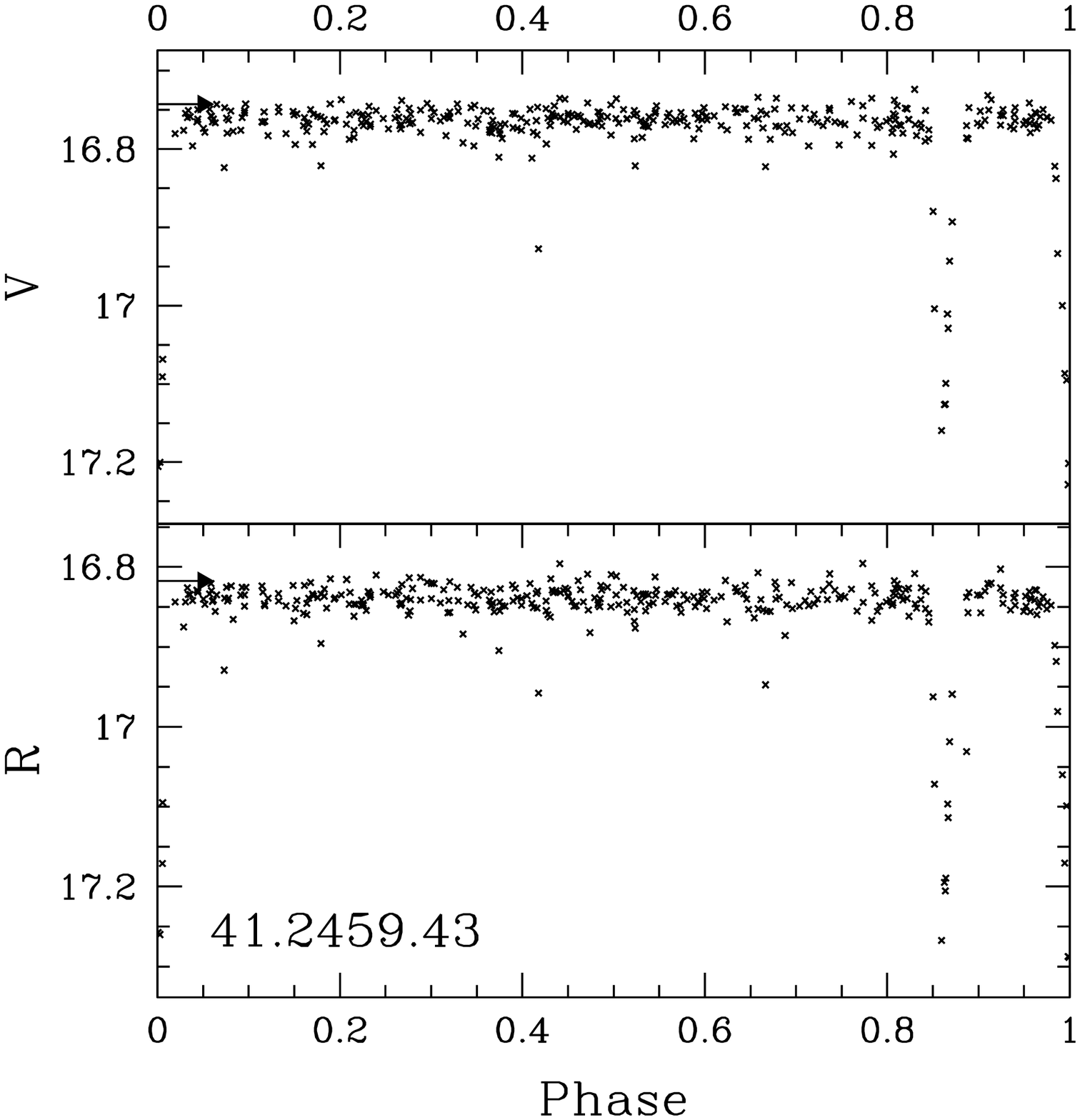}{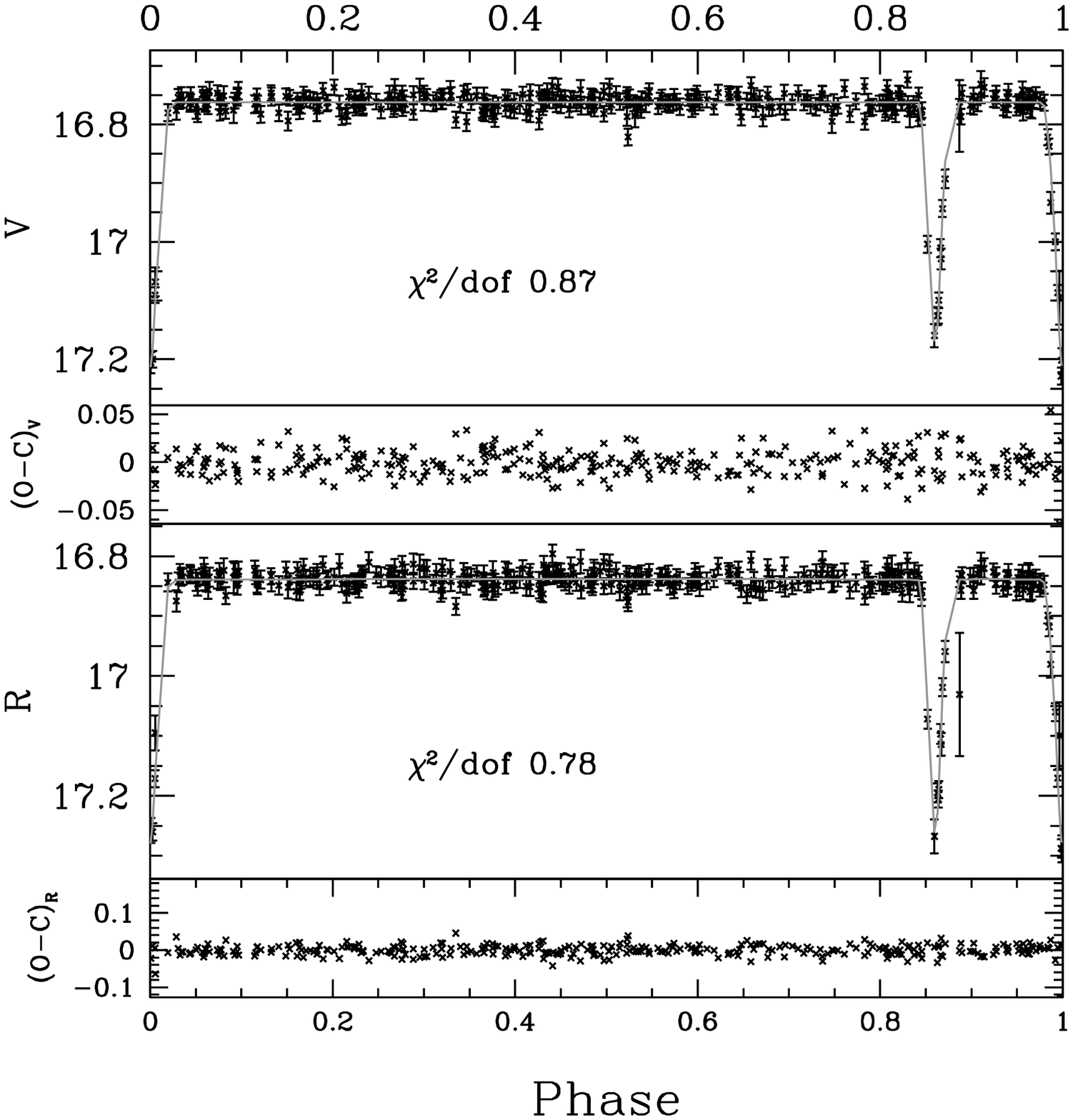}
\end{center}
\caption{Examples of LMC EBs light curves, arranged by ascending period; for basic data see Table \ref{tab:lcs}.
Left: observed light curves with all data points. 
The arrows show the baseline as defined in Table \ref{tab:lcs}.
Right: observed light curves with outlying points removed, theoretical
light curves from the fit and residuals.}
\label{fig:lcs4}
\normalsize
\end{figure}

\begin{figure}
\footnotesize
\begin{center}
\plottwo{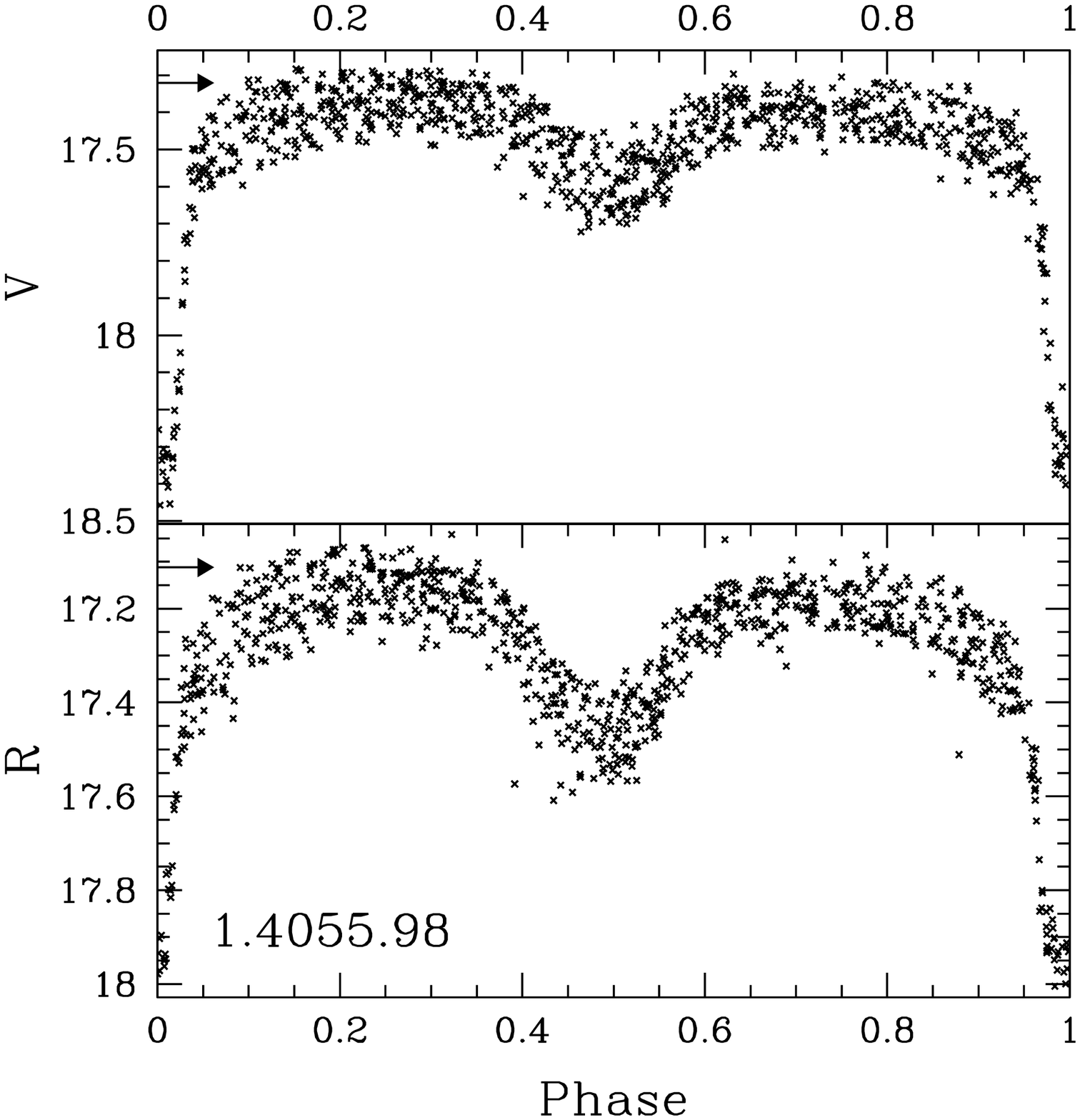}{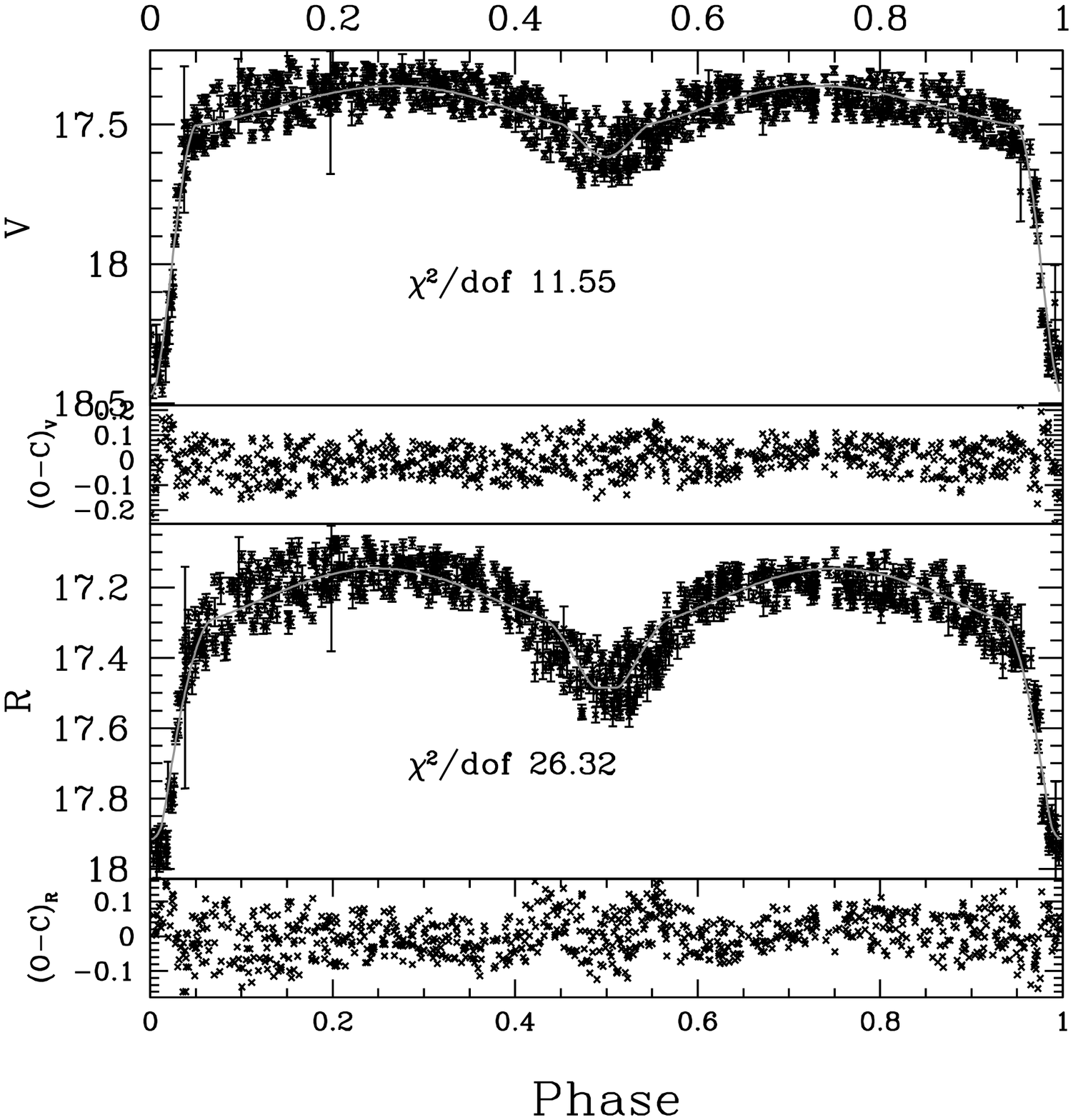}
\plottwo{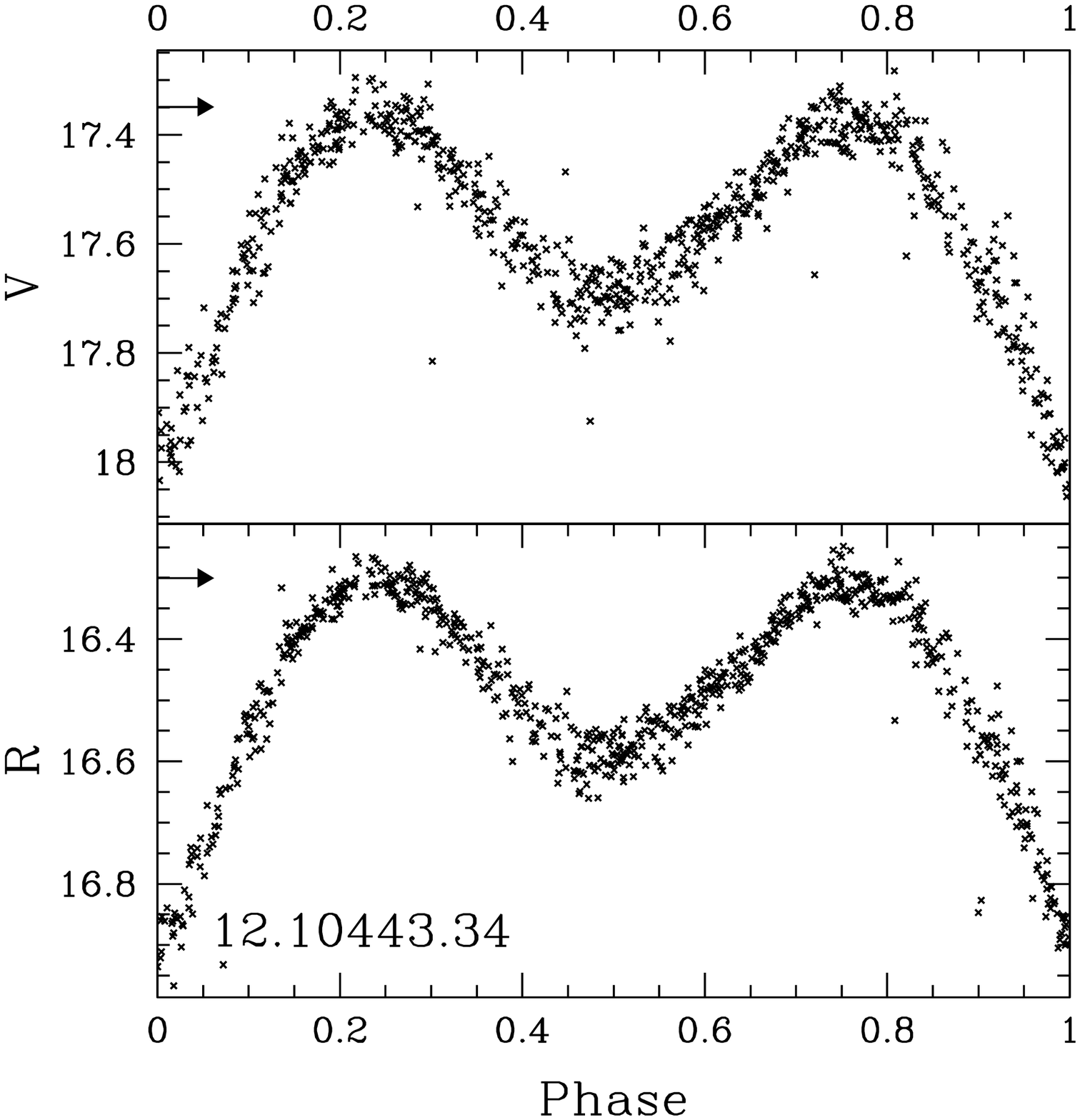}{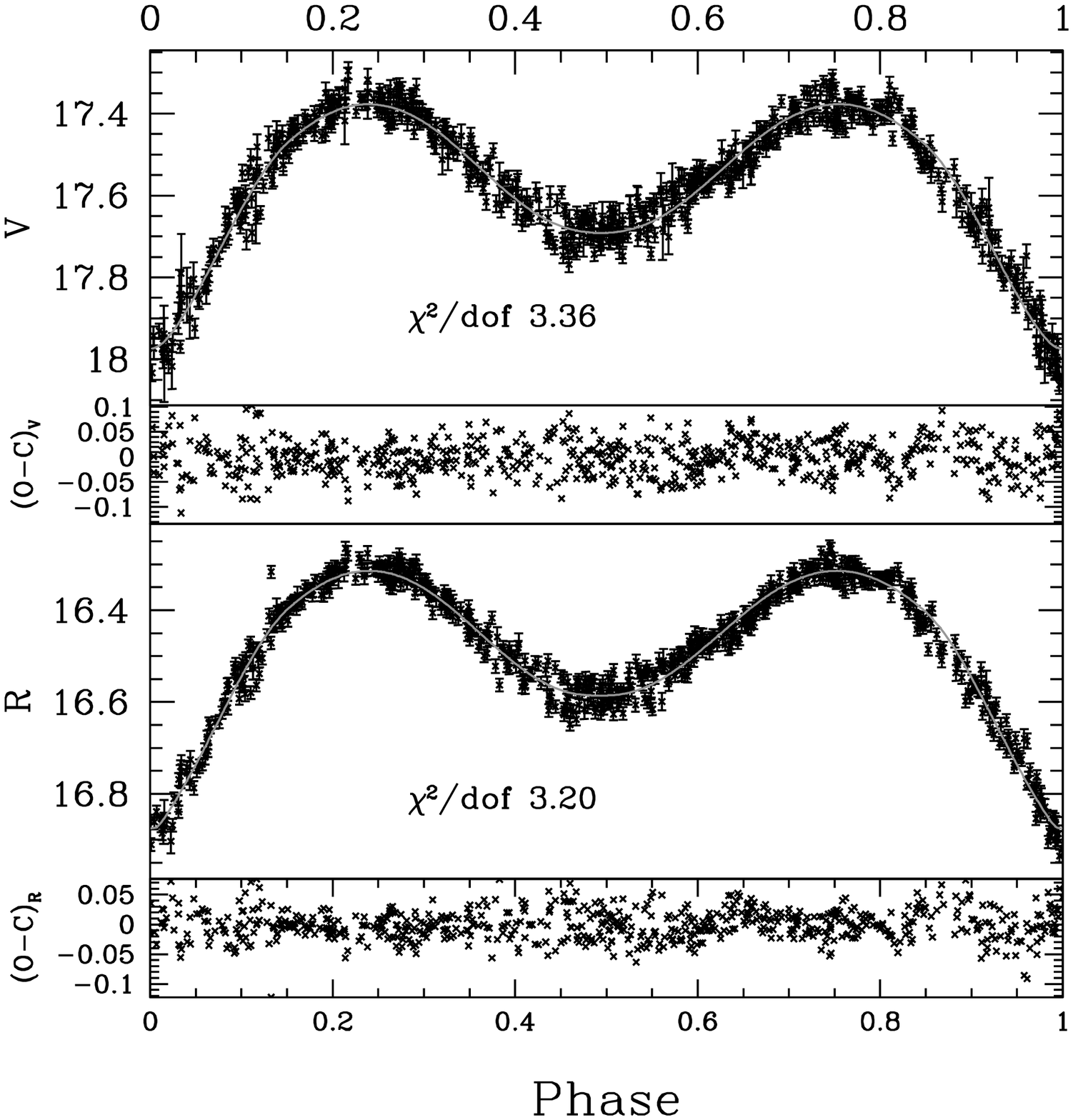}
\end{center}
\caption{Examples of LMC EBs light curves, arranged by ascending period; for basic data see Table \ref{tab:lcs}.
Left: observed light curves with all data points. 
The arrows show the baseline as defined in Table \ref{tab:lcs}.
Right: observed light curves with outlying points removed, theoretical
light curves from the fit and residuals.}
\label{fig:lcs5}
\normalsize
\end{figure}

\begin{figure}
\footnotesize
\begin{center}
\plottwo{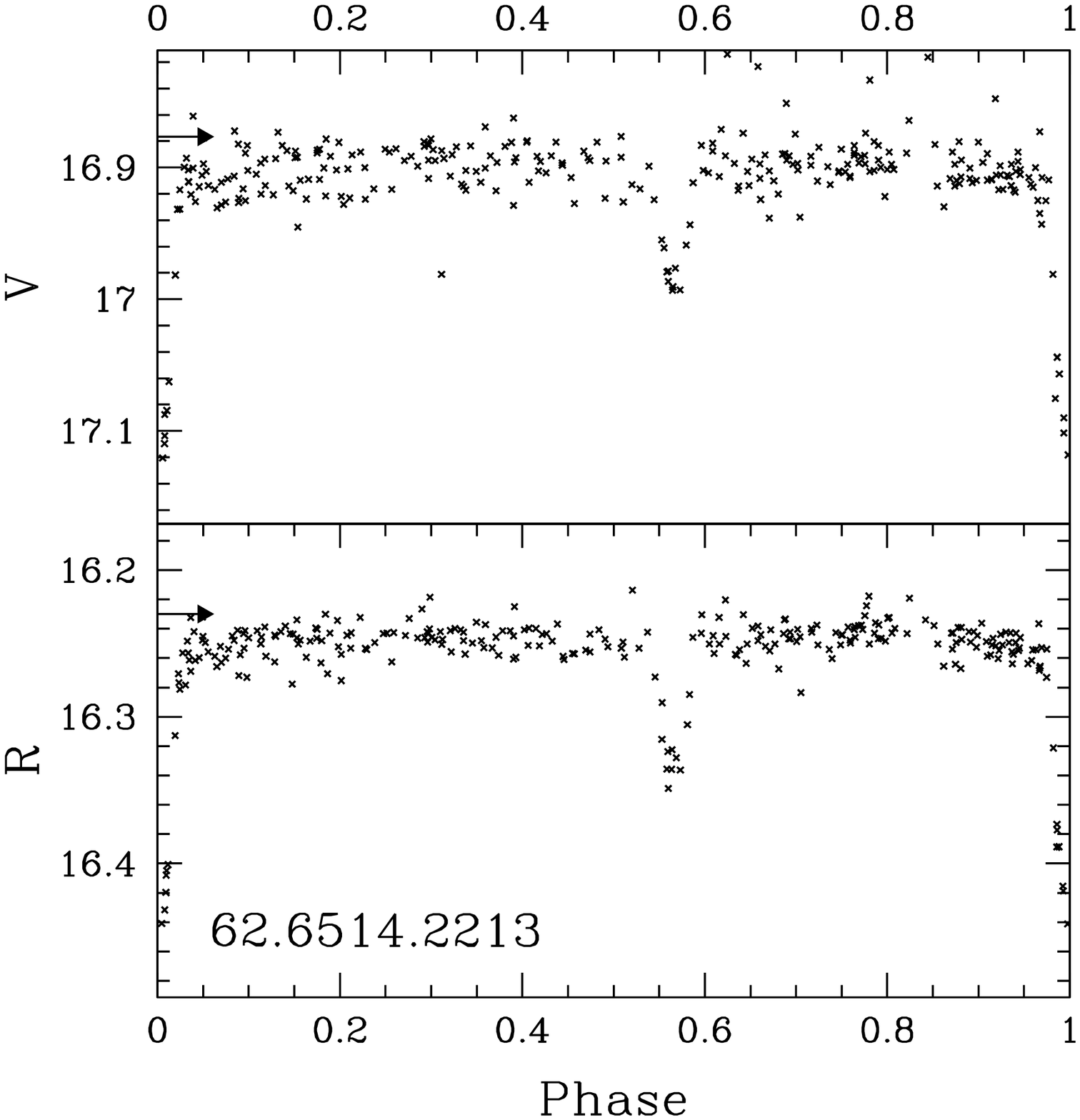}{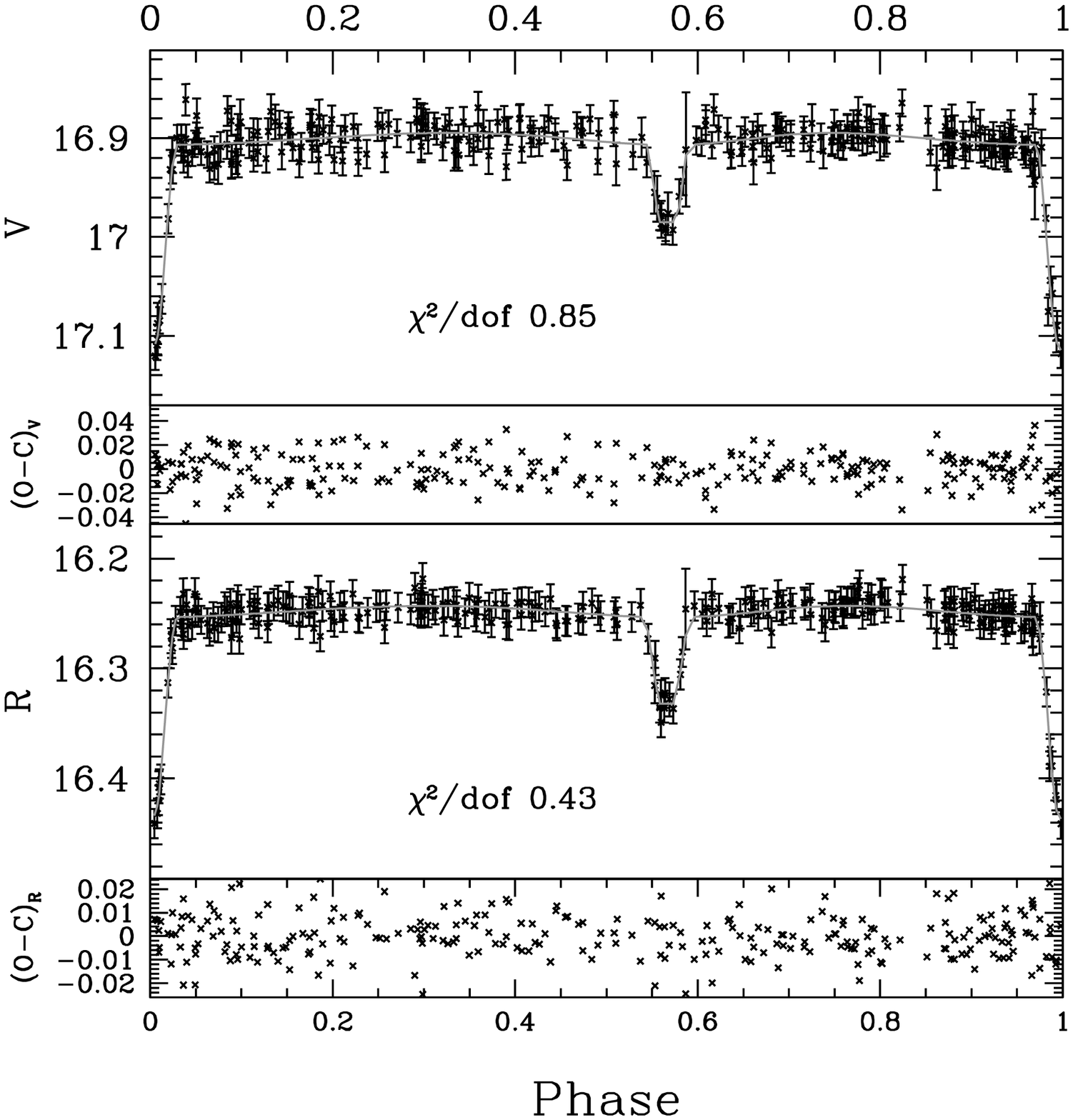}
\plottwo{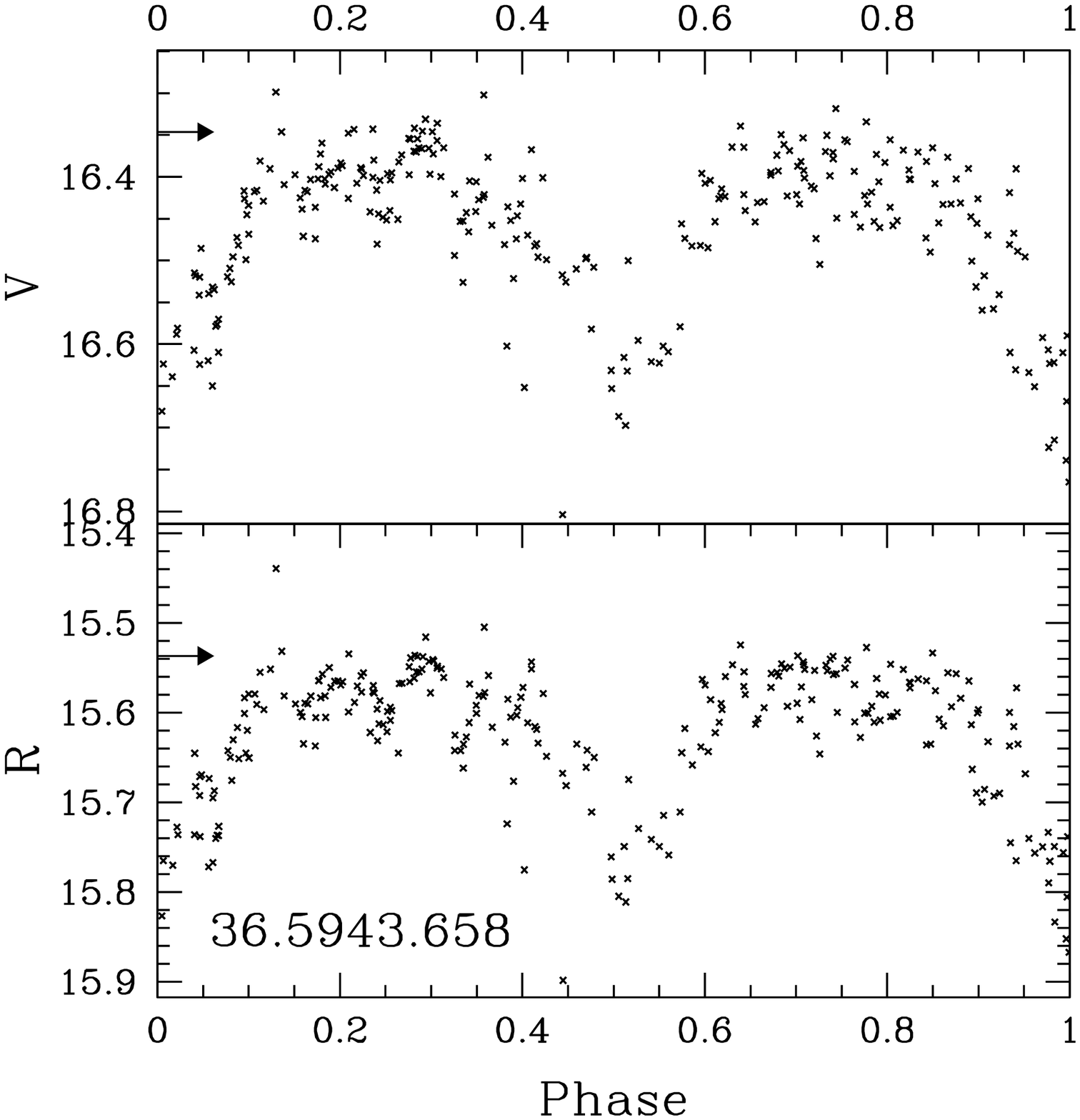}{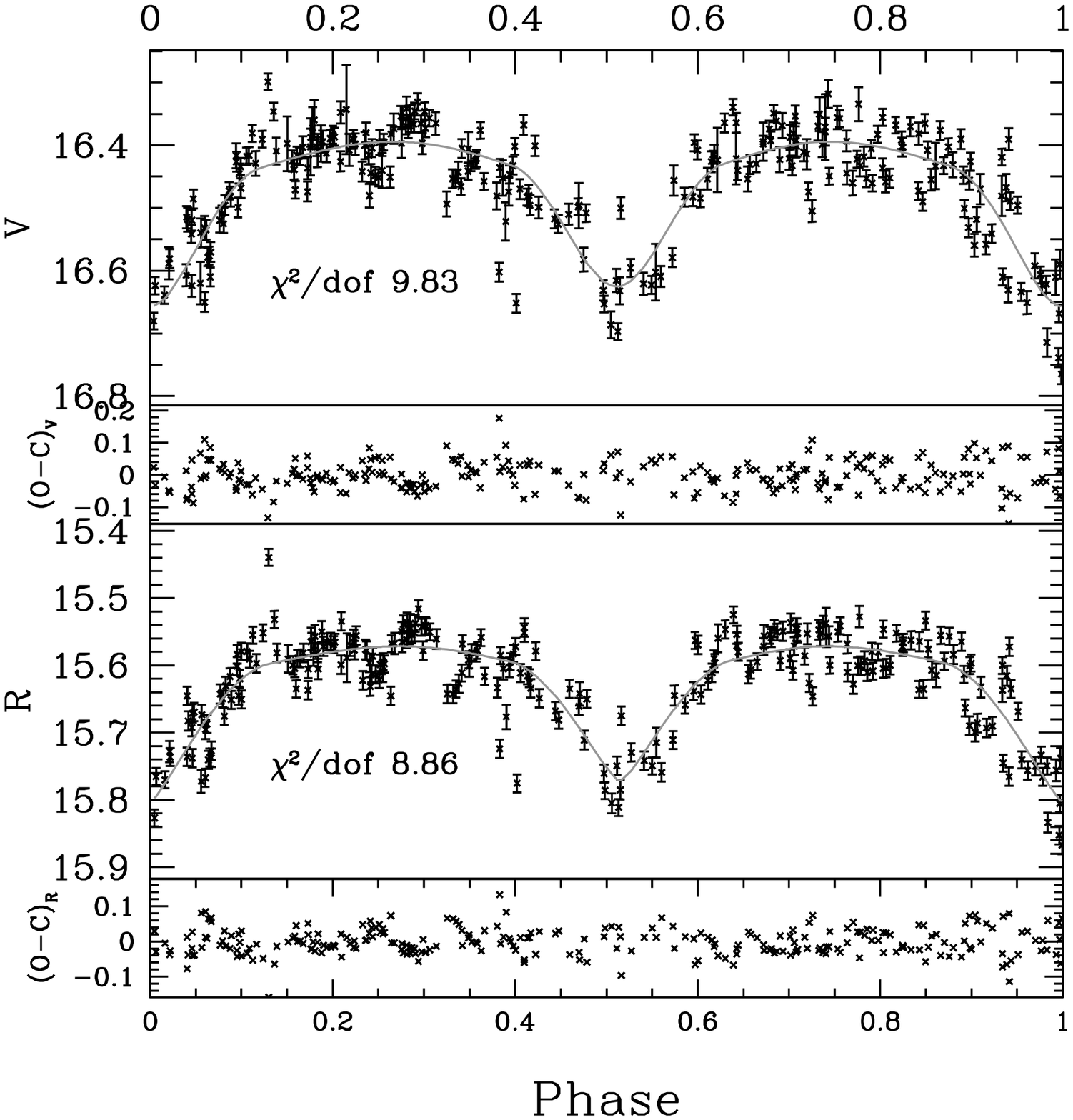}
\end{center}
\caption{Examples of LMC EBs light curves, arranged by ascending period; for basic data see Table \ref{tab:lcs}.
Left: observed light curves with all data points. 
The arrows show the baseline as defined in Table \ref{tab:lcs}.
Right: observed light curves with outlying points removed, theoretical
light curves from the fit and residuals.}
\label{fig:lcs6}
\normalsize
\end{figure}

Figures \ref{fig:lcs1}, \ref{fig:lcs2}, \ref{fig:lcs3}, \ref{fig:lcs4}, \ref{fig:lcs5}, and \ref{fig:lcs6} show some examples of light curves.
The panels on the left show the original light curves, those on the right
show the light curves with the outlying points removed, the error bars, the theoretical light curves from the fit and the residuals
The EBs shown are meant to be representative of the sample, this is why some examples of bad fits are included.
Comparing the panels on the left with the panels on the right for these figures gives
an idea of the effect of removing outlying points.
In particular the figures suggest that for those objects with good fits the procedure
resulted in the elimination of truly outlying points; these EBs are mostly detached
with undistorted components.
For objects with bad fits, which mostly comprise EBs with close and strongly distorted components, the situation is less clear.
For example the system labelled 1.3442.172 shown in Figure \ref{fig:lcs2} exhibits some
points in the $R$ band, at secondary eclipse around phase $\sim 0.4$, 
that run almost parallel to the main light curve but at a higher magnitude.
Such points may or may not be physically significant; some of these are removed
by our procedure, but the fit is nevertheless bad.  
The systems labelled 1.3804.164 in Figure \ref{fig:lcs4} and 1.4055.98 in Figure \ref{fig:lcs5} show a large and step-like scatter band, the reason of which is, we think, intrinsic variability of the secondary component\footnote{The component eclipsed at primary eclipse.},
as suggested by the fact that the band becomes much narrower at primary eclipse but not at secondary eclipse; this interpretation is also suggested by the fact that the residuals show an oscillating behavior as a function of phase.
In both 1.4055.98 and 36.5943.658 in Figure \ref{fig:lcs6} the scatter band is much larger than the observational error which explains their bad fits.
These examples show that the samples contain many EBs which could be deserving of more careful study which we did not attempt here.
The properties of these EBs are summarized in Table \ref{tab:lcs}; of the $12$ EBs shown $8$ have two photometric excursions in both bands and $\mathrm{Ampl_V}/\mathrm{Ampl_R}<1.2$, $7$ are found by the decision tree, $5$ have $\mathrm{s}5/\mathrm{s}5\mathrm{d}<1.2$, and
$2$ have a counterpart in the OGLE-II sample: 1.3442.172 
(counterpart OGLE050149.20-691945.5 in field LMC$\_$SC15) and 
1.4055.98 (counterpart OGLE050542.06-684732.8 in field LMC$\_$SC13).

\tabletypesize{\footnotesize}
\begin{deluxetable}{cccccccl}
\tablecolumns{8}
\tablewidth{0pc}
\tablecaption{Basic data for the LMC EBs shown in Figs. \ref{fig:lcs1},
\ref{fig:lcs2}, and \ref{fig:lcs3}.
The EBs are arranged by ascending period.
\label{tab:lcs}}
\tablehead{
\colhead{MACHO ID} & 
\colhead{RA(J2000)} & 
\colhead{DEC(J2000)} & 
\colhead{Period\tablenotemark{a}($\mathrm{d}$)} & 
\colhead{$V$ baseline\tablenotemark{b~d}} & 
\colhead{$R$ baseline\tablenotemark{b~d}} & 
\colhead{$\vr$\tablenotemark{c~d}} &
\colhead{Comment}
}
\startdata
65.8581.67 & 05:33:09.458 & -65:30:29.13 & 0.19 & 18.93 & 18.13 & 0.8 & 
Shortest period in sample\\
9.5000.790 & 05:11:20.548 & -70:21:00.16 & 0.24 & 19.70 & 19.23 & 0.47 & 
Very short period\\
1.3442.172 & 05:01:49.005 & -69:19:45.60 & 1.02 & 17.21 & 17.26 & -0.05 & 
Fairly typical EB\\
10.4035.145 & 05:05:02.233 & -70:06:13.27 & 2.53 & 17.14 & 17.19 & -0.05 &
Fairly typical EB\\
68.10843.699 & 05:47:10.738 & -67:58:55.74 & 3.07 & 13.56 & 13.99 & -0.43 &
Bluest in sample\\
62.7240.102 & 05:24:55.090 & -66:11:55.28 & 4.17 & 18.30 & 18.30 & 0.00 & 
Very high eccentric orbit\\
1.3804.164 & 05:03:36.536 & -69:23:32.27 & 4.19 & 16.88 & 16.90 & -0.02 & 
Algol type\\
41.2459.43 & 04:55:43.321 & -70:18:00.44 & 13.18 & 16.74 & 16.82 & -0.08 & 
Highest eccentric orbit\\
1.4055.98 & 05:05:42.201 & -68:47:33.44 & 24.51 & 17.32 & 17.11 & 0.21 & 
Fairly typical EB\\
12.10443.34 & 05:44:47.185 & -70:27:26.65 & 319.37 & 17.35 & 16.30 & 1.05 & 
Reddest in sample\\
62.6514.2213 & 05:20:32.499 & -66:13:17.92 & 417.60 & 16.88 & 16.23 & 0.65 &
Long Period\\
36.5943.658 & 05:17:15.478  & -71:57:45.69 & 633.70 & 16.35 & 15.54 & 0.81&
Longest Period in sample
\enddata
The information in Table \ref{tab:lcs} is also available in its entirety via the link to the machine-readable version above. The EBs are arranged by ascending period. Units of right ascension are hours, minutes, and seconds, and units of declination are degrees, arcminutes, and arcseconds. 
\tablenotetext{a}{Supersmoother provides different periods for the $V$
and $R$ unfolded light curves, but their difference is usually smaller than
the precision to which we report their values in this table.
On line summary tables provide both periods to $5$ significant digits.}
\tablenotetext{b}{The baseline is calculated in the following way.
First the outlying points are eliminated by dividing the light curve in 
boxes of $\sim 50$ data points and eliminating the points which are more 
than $2$ standard deviations away from the mean in each box.
Then the median of the $10\%$ most luminous points is taken.
This value is not the median of the whole light curve that is shown in the 
figures.}
\tablenotetext{c}{Values are quoted to the hundredths of magnitude, 
typical of MACHO observational uncertainties.} 
\tablenotetext{d}{This is the difference of the two baselines as defined above, not of the two medians as is the $\vr$ shown in the figures. This column is not directly available in the online table but can be deduced by subtracting col. (12) from col. (10).}
\end{deluxetable}
\subsection{Ellipsoidal variables in the samples}
\label{subsec:ell}
Ellipsoidal variability occurs in a close binary system when one (or both)
component(s) is (are) tidally distorted by the companion.
If the binary system is detached, as most systems in our samples are, 
the distorted stars assume the asymmetric, egg like shape of the Roche 
equipotential surface whereas in the case of contact system the shape of 
the common equipotential surface is more reminiscent of a dumbbell.
The light curve of an ellipsoidal variable system reveals a 
continuously varying profile, with two maxima and two minima 
per period, with the minima often having different depth, whereas
the maxima are usually equal.
The main reason for this variability is that, as the stars rotate, their 
projected areas on the sky vary, reaching a maximum at the two quadratures 
and a minimum at the two conjunctions; the measured flux thus varies in the
same way during a period.
More information on ellipsoidal variables can be found in 
\citep{hilditch01}
A large sample of ellipsoidal variables in the LMC has been 
released by the OGLE collaboration \citep{soszynski04}; an analysis
of ellipsoidal variables found in the MACHO database has been published
by \citet{derekas06}.
A binary system can present both eclipses and ellipsoidal variability but 
in many cases it may not be possible to clearly recognize an eclipse from 
visual inspection; this poses a problem for the compilation of 
EB catalogues since the light curves of EBs and non eclipsing 
ellipsoidal variables can be easily confused. 
\par
\begin{figure}
\footnotesize
\begin{center}
\plottwo{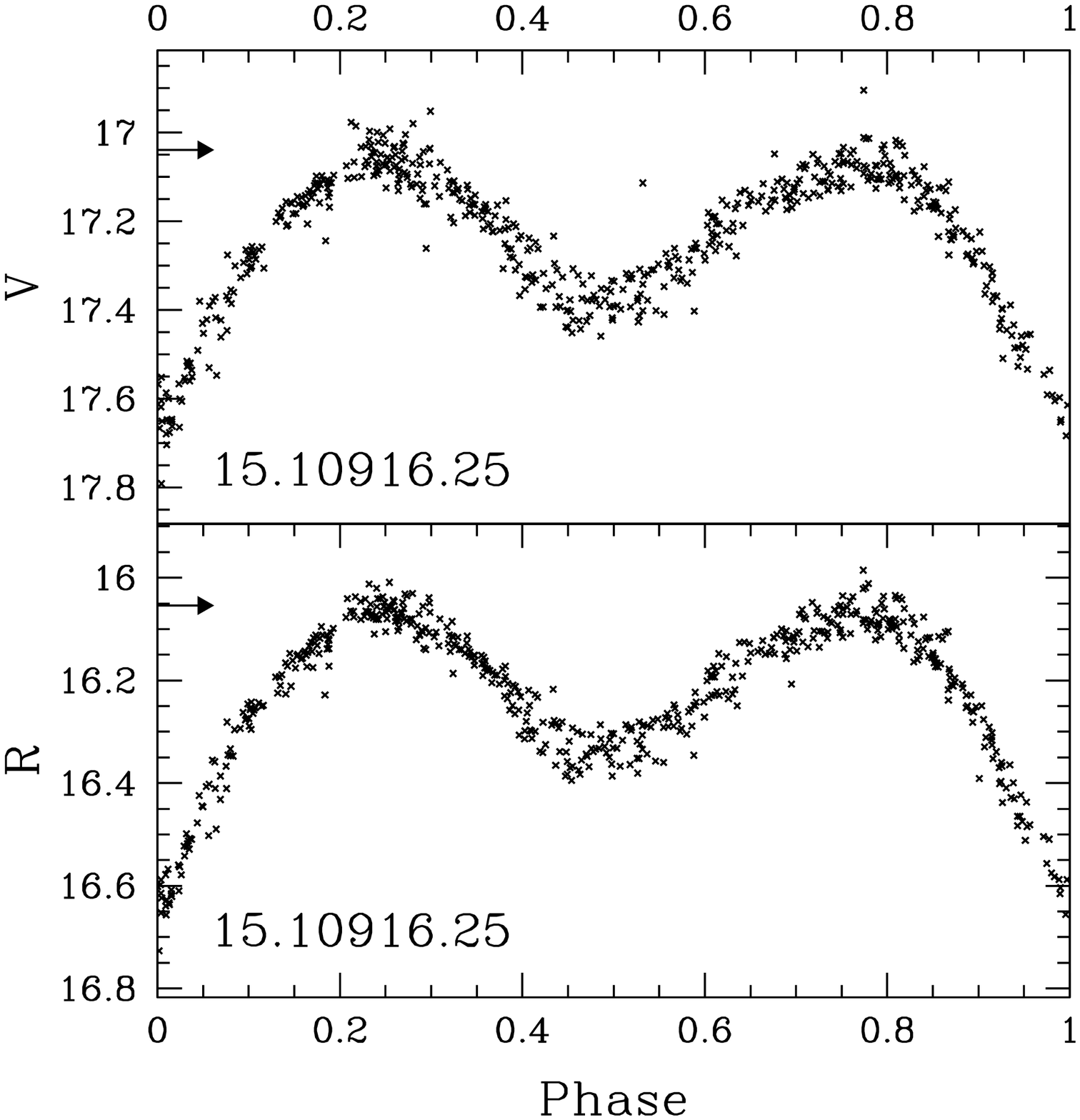}{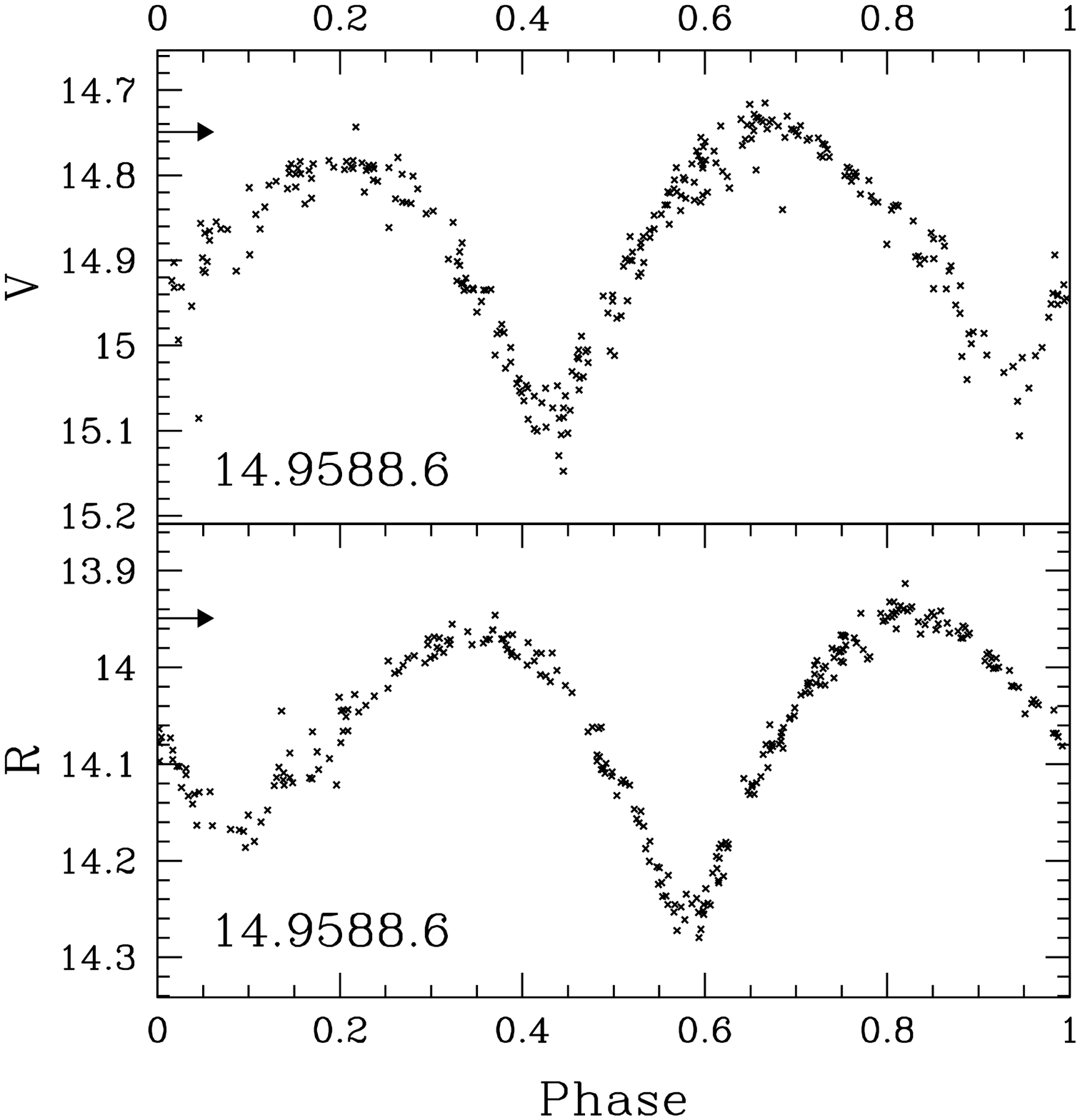}
\plottwo{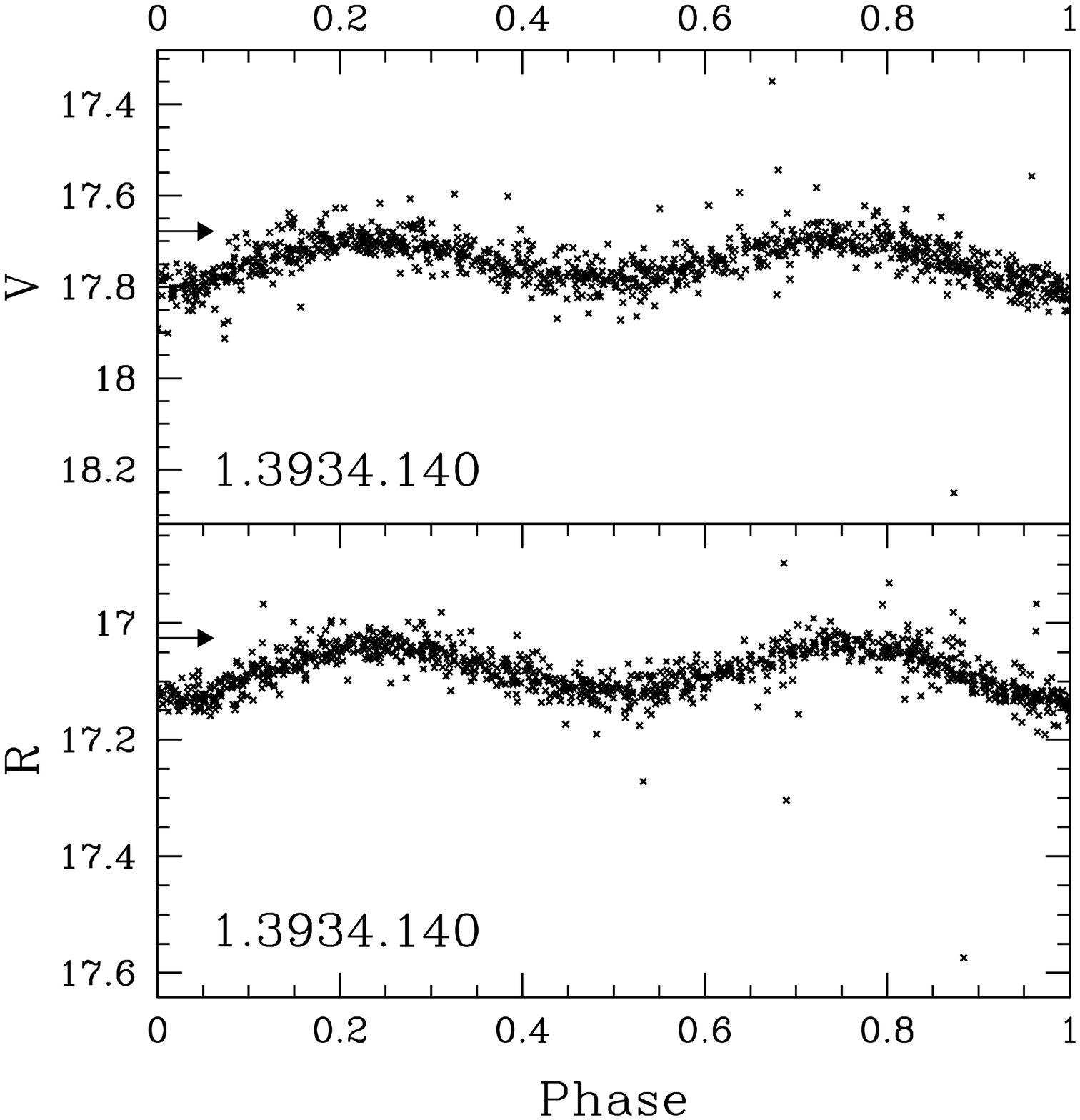}{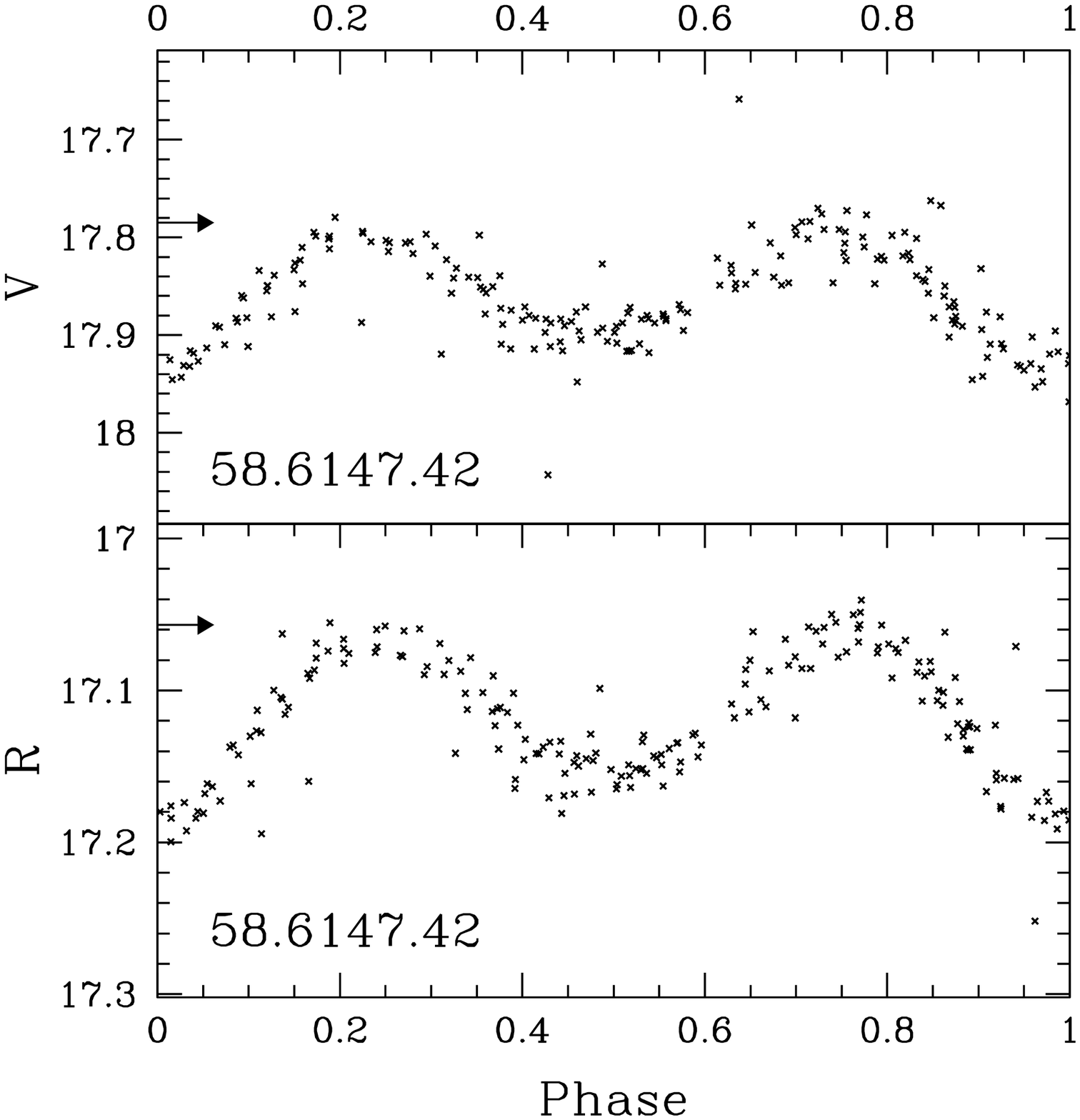}
\end{center}
\caption{Above: light curves for two long period EBs with strong 
ellipsoidal variability in the LMC sample.
Below: light curves of two non eclipsing ellipsoidal variables.}
\label{fig:ell}
\normalsize
\end{figure}

We looked for possible contamination by ellipsoidal variables in our sample
and found $\sim 120$ systems exhibiting ellipsoidal variability which 
we then visually checked more carefully than other stars which were clearly EBs.
We also attempted a less subjective approach by fitting these systems 
with the EBOP program \citep{etzel81,popper81,nelson72} following the prescriptions of \citet{lacy97}; however since EBOP is not designed for analyzing such distorted systems the final decision about whether or not to include a star exhibiting ellipsoidal variability in the sample was taken upon visual inspection.
Figure \ref{fig:ell} shows two examples of EB systems with pronounced 
ellipsoidal variability, basic data on these systems are given in Table 
\ref{tab:ell}.
\tabletypesize{\footnotesize}
\begin{deluxetable}{cccccccl}
\tablecolumns{8}
\tablewidth{0pc}
\tablecaption{Basic data for the LMC ellipsoidal variables shown in Figure 
\ref{fig:ell}.
The variables are arranged by ascending period.
\label{tab:ell}}
\tablehead{
\colhead{MACHO ID} & 
\colhead{RA(J2000)} & 
\colhead{DEC(J2000)} & 
\colhead{Period\tablenotemark{a}($\mathrm{d}$)} & 
\colhead{$V$ baseline\tablenotemark{b~d}} & 
\colhead{$R$ baseline\tablenotemark{b~d}} & 
\colhead{$\vr$\tablenotemark{c~d}} &
\colhead{Eclipsing}
}
\startdata
1.3934.140 & 05:04:23.977 & -68:49:21.65 & 85.88 & 17.68 & 17.03 & 0.65 & 
No\\
58.6147.42 & 05:18:03.677 & -66:27:20.17 & 105.08 & 17.78 & 17.06 & 0.72 & 
No\\
15.10916.25 & 05:47:21.831 & -71:10:46.49 & 355.28 &  17.04 &  16.05 & 0.99
& Yes\\
14.9588.6 & 05:39:28.498 & -71:00:46.01 &  411.04 & 14.75 & 13.95 & 0.80 &
Yes\\
\enddata
\tablenotetext{a}{Supersmoother provides different periods for the $V$
and $R$ unfolded light curves, but their difference is usually smaller than
the precision to which we report their values in this table.
On line summary tables provide both periods to $5$ significant digits.}
\tablenotetext{b}{See Table \ref{tab:lcs} for an explanation of the 
baseline calculation.}
\tablenotetext{c}{This is the difference of the two baselines, not of the 
two medians as is the $\vr$ shown in figures.}
\tablenotetext{d}{Values are quoted to the hundredths of magnitude, typical
of MACHO observational uncertainties.} 
\end{deluxetable}
\subsection{The Small Magellanic Cloud sample}
The SMC sample comprises $1509$ EBs selected via the same techniques 
as the LMC EBs and confirmed by visual inspection; the general 
considerations of the preceding subsection regarding search for 
variability apply here as well.
The sky coverage in the SMC corresponds to MACHO fields 206, 207, 208, 211,
212 and 213; field center coordinates for these fields are given in Table 
\ref{tab:smcfields}.
\tabletypesize{\footnotesize}
\begin{deluxetable}{cccc}
\tablecolumns{4}
\tablewidth{0pc}
\tablecaption{MACHO field coordinates for the SMC.
\label{tab:smcfields}}
\tablehead{
\colhead{Field ID} & \colhead{RA(J2000)} & \colhead{DEC(J2000)} & 
\colhead{Date}}
\startdata
206&1:05:21.70&-72:26:58.3&(J2000.0) \\
207&0:57:16.58&-72:34:57.0&(J2000.0) \\
208&0:48:03.19&-72:34:20.9&(J2000.0) \\
211&0:58:27.40&-73:04:55.3&(J2000.0) \\
212&0:49:10.27&-73:13:32.9&(J2000.0) \\
213&0:40:18.91&-73:08:49.5&(J2000.0) \\ 
\enddata
\end{deluxetable}
Magnitudes quoted for the SMC have been obtained by using 
transformations which differ slightly in the zero point from the LMC ones 
due to the larger exposure times in the SMC \citep{alcock99}; they are 
reported in Eq. \ref{eq:smcmachocal}.
\begin{eqnarray}\label{eq:smcmachocal}
V&=&V_{\mathrm{MACHO}}+24.97~\mathrm{mag}-0.18(V_{\mathrm{MACHO}}-R_{\mathrm{MACHO}})
\nonumber \\
R&=&R_{\mathrm{MACHO}}+24.73~\mathrm{mag}+0.18(V_{\mathrm{MACHO}}-R_{\mathrm{MACHO}}).
\end{eqnarray}
The SMC search gave $194$ duplicates and again we chose the field which had 
the highest total number of observations and the object is identified by that
corresponding FTS only.
\par
\begin{figure}
\footnotesize
\begin{center}
\plotone{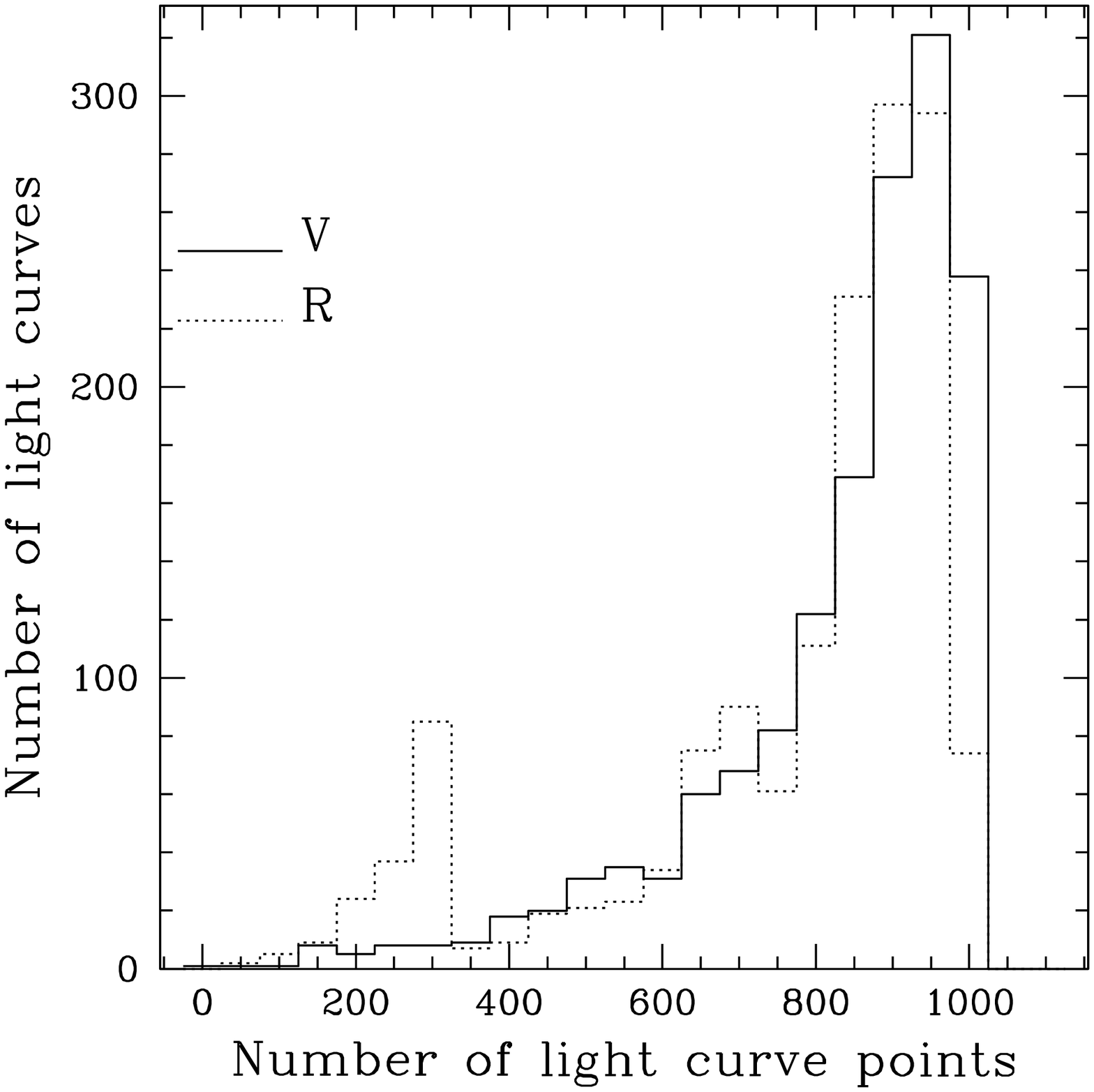}
\end{center}
\caption{Histogram of the number of light curve points for both bands for 
the SMC sample.}
\label{fig:smcpointhist}
\end{figure}
\begin{figure}
\footnotesize
\begin{center}
\plotone{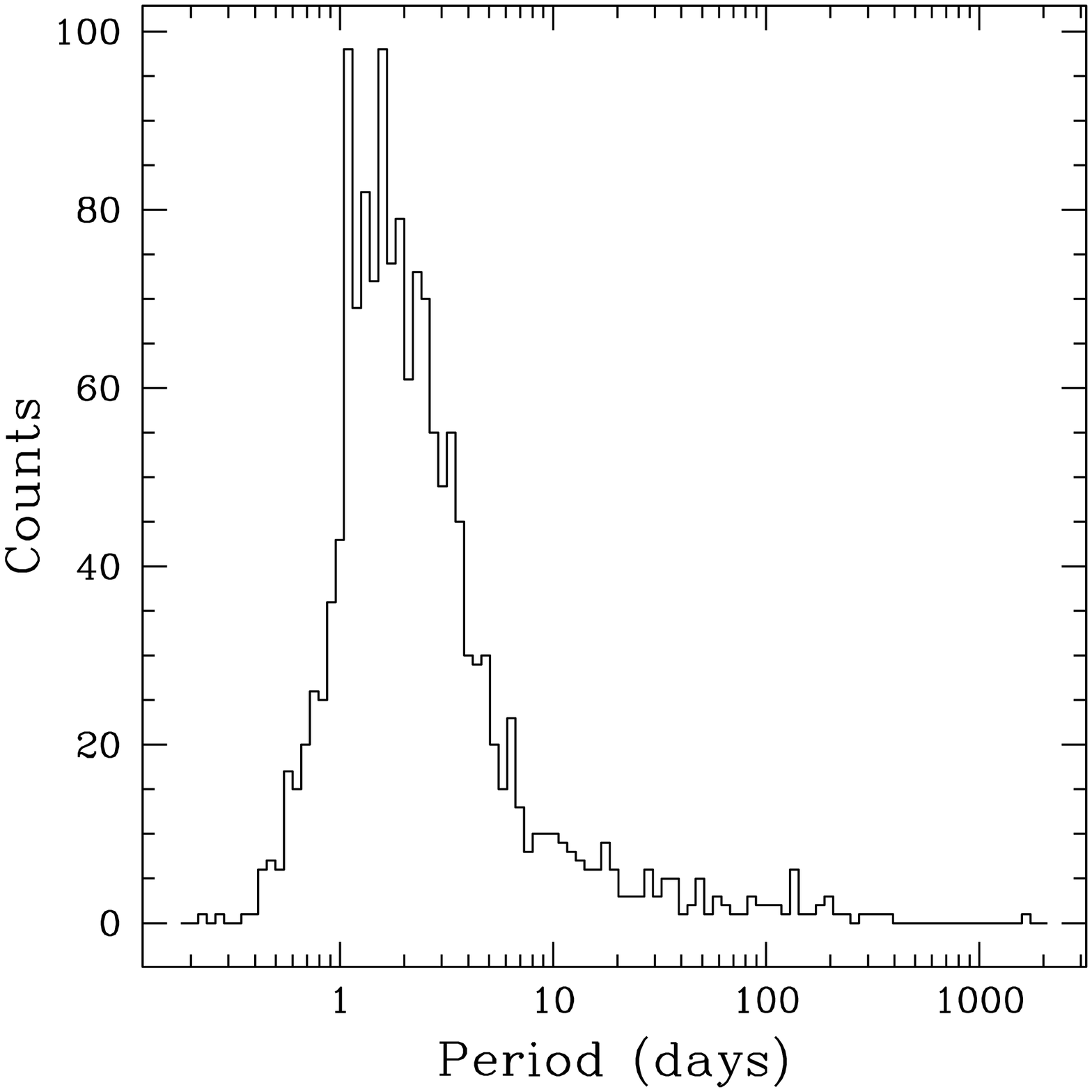}
\end{center}
\caption{Period histogram for $1509$ EBs in the SMC sample.
The size of the bins is $\sim 1/100$ of the span of the logarithms of the
periods.}
\label{fig:smcperhist}
\end{figure}
\begin{figure}\footnotesize
\begin{center}
\plotone{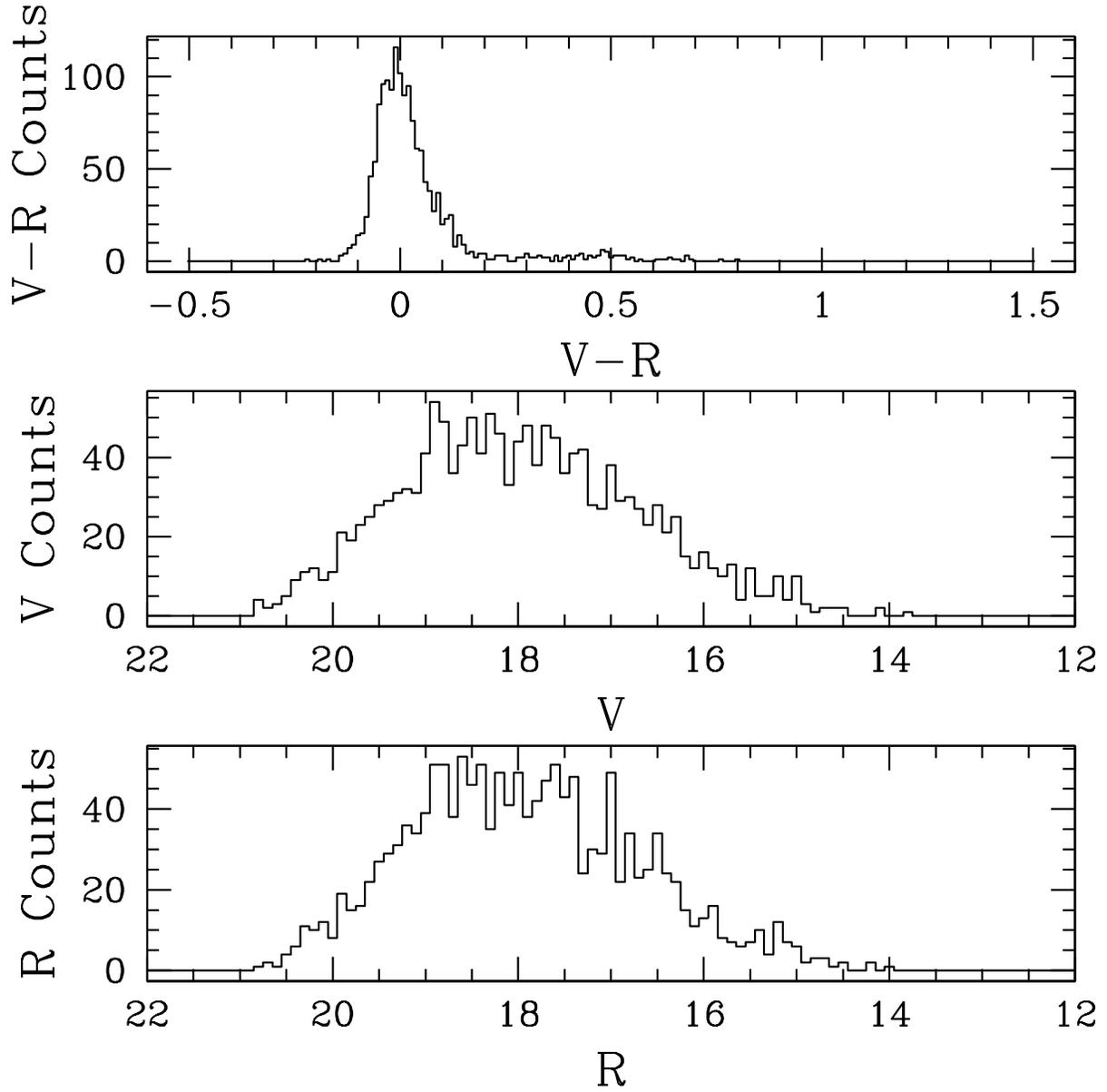}
\end{center}
\caption{$V$ and $R$ magnitudes and $\vr$ histograms for $1508$ EBs in the 
SMC sample.
The bin size is $0.1~\mathrm{mag}$ for the $V$ and $R$ histograms and 
$0.01~\mathrm{mag}$ for the $\vr$ one.
}
\label{fig:smcmaghist}
\end{figure}
\begin{figure}
\footnotesize
\begin{center}
\plotone{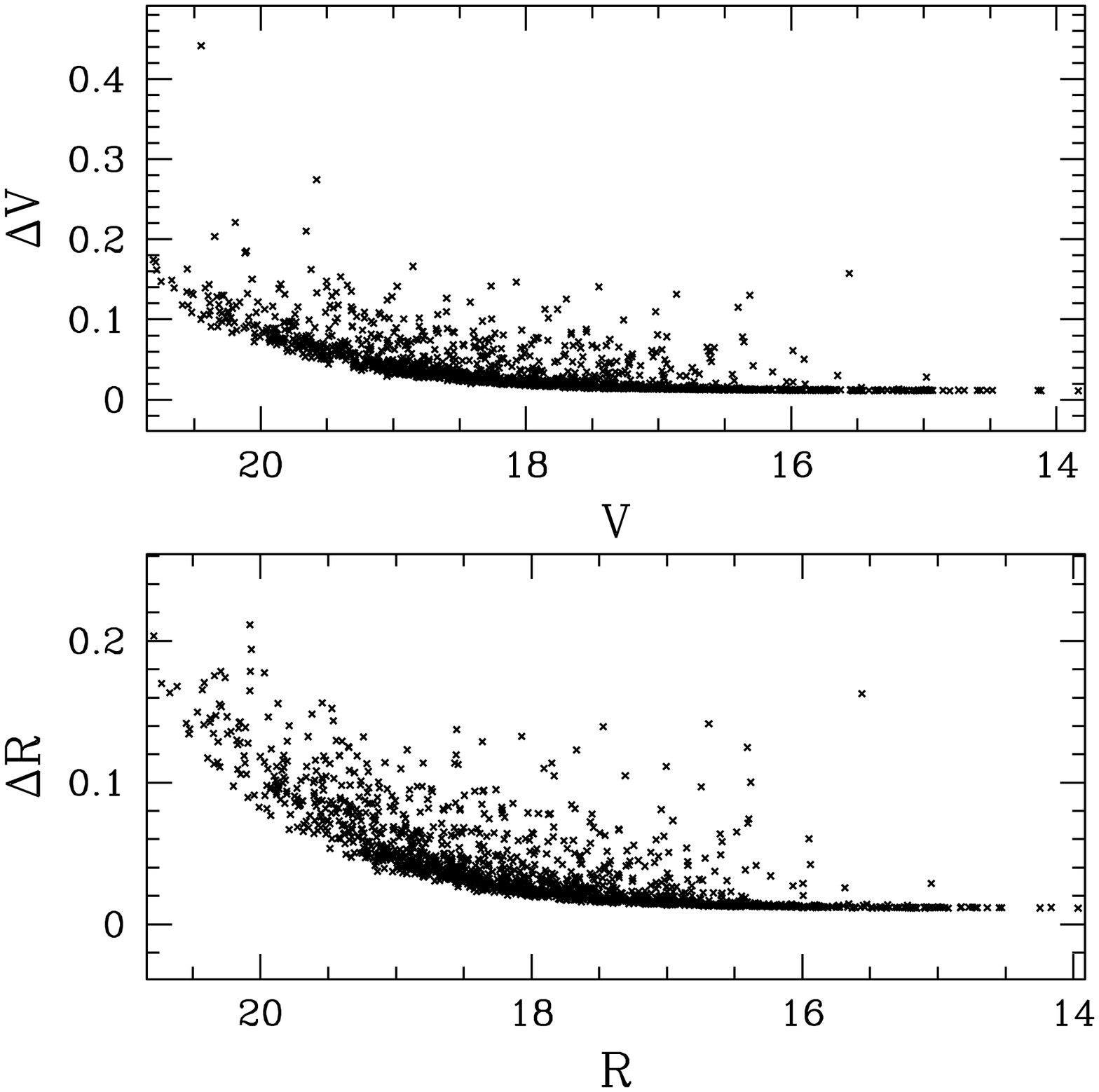}
\end{center}
\caption{Errors as a function of median magnitude for the SMC sample.
} 
\label{fig:smcmagerr}
\normalsize
\end{figure}
\begin{figure}
\begin{center}
\footnotesize
\plottwo{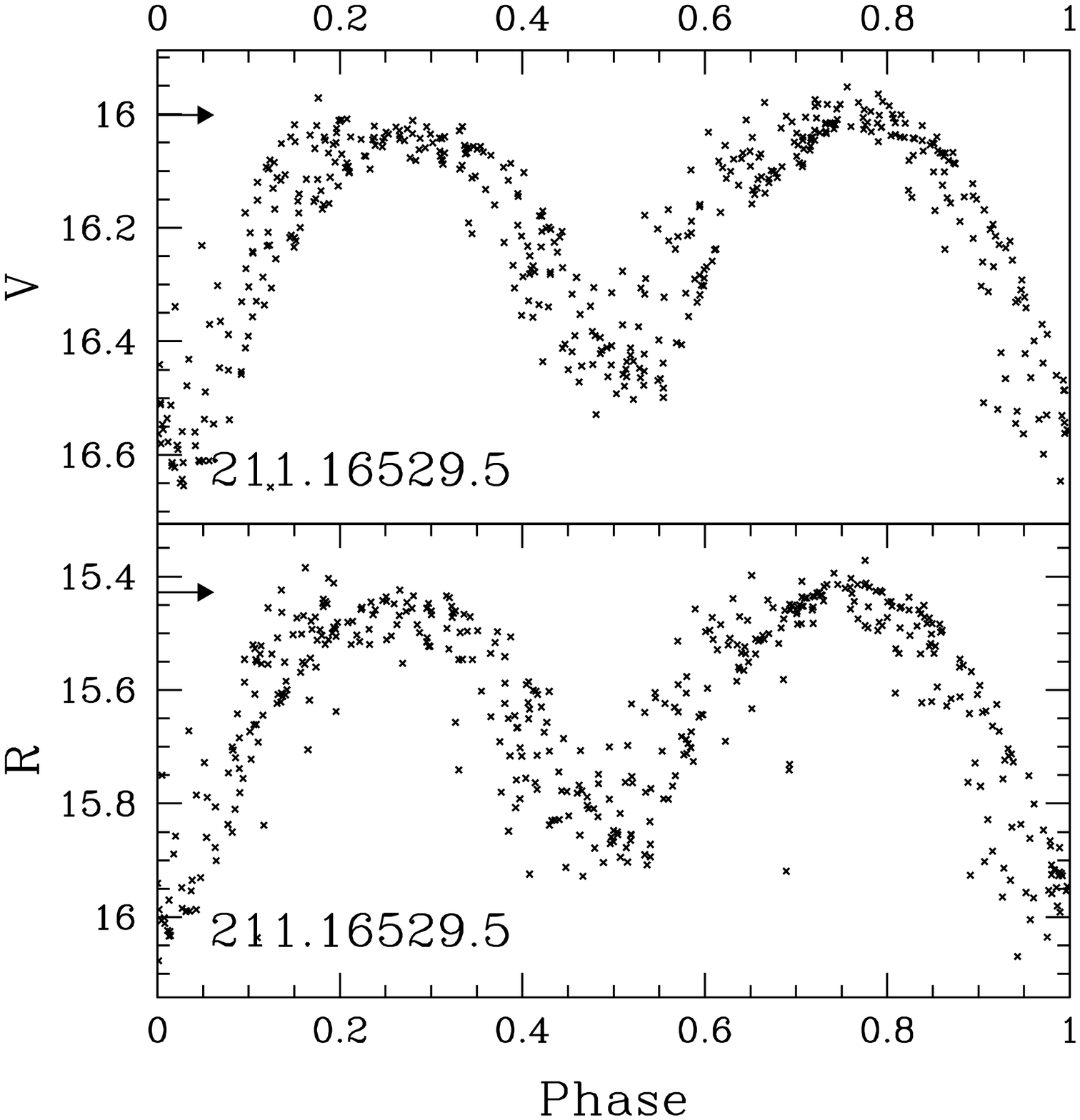}{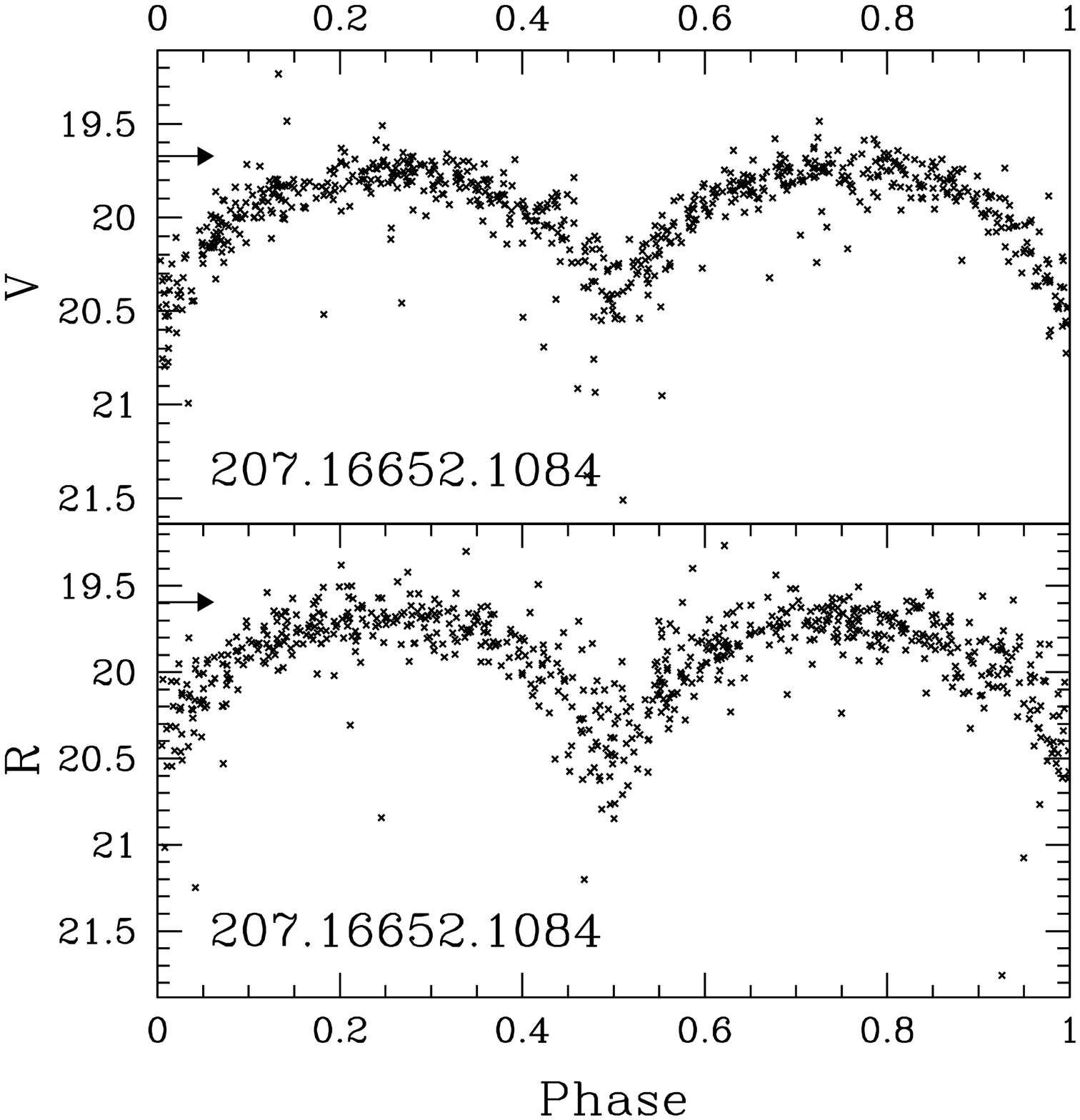}
\plottwo{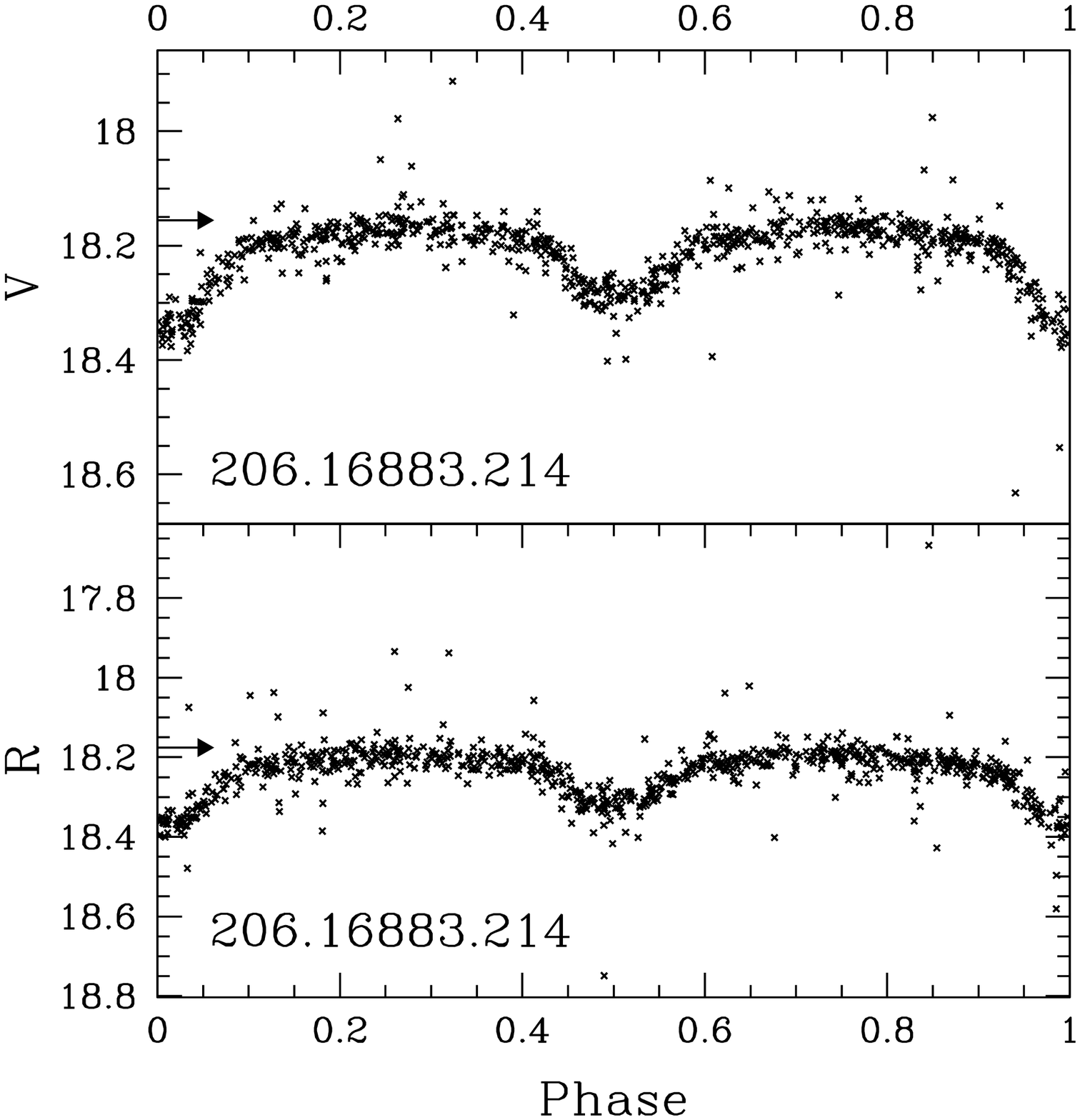}{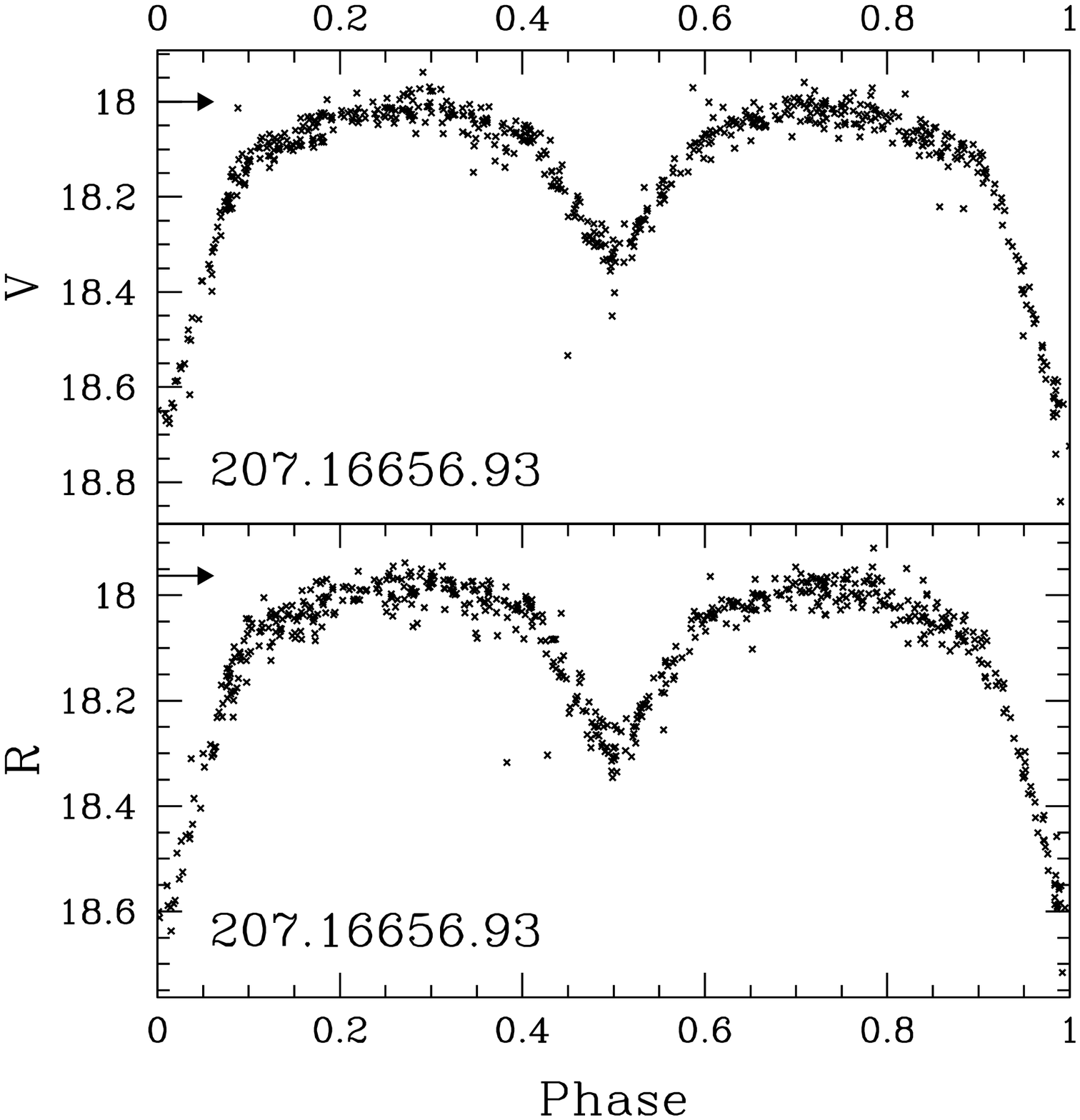}
\end{center}
\caption{Examples of light curves of EBs in the SMC sample; for basic data 
see Table \ref{tab:smclcs}.
The arrows show the baseline as defined in Table \ref{tab:smclcs}.
The EBs are arranged by ascending period.} 
\label{fig:smclcs1}
\normalsize
\end{figure}
\begin{figure}
\begin{center}
\footnotesize
\plottwo{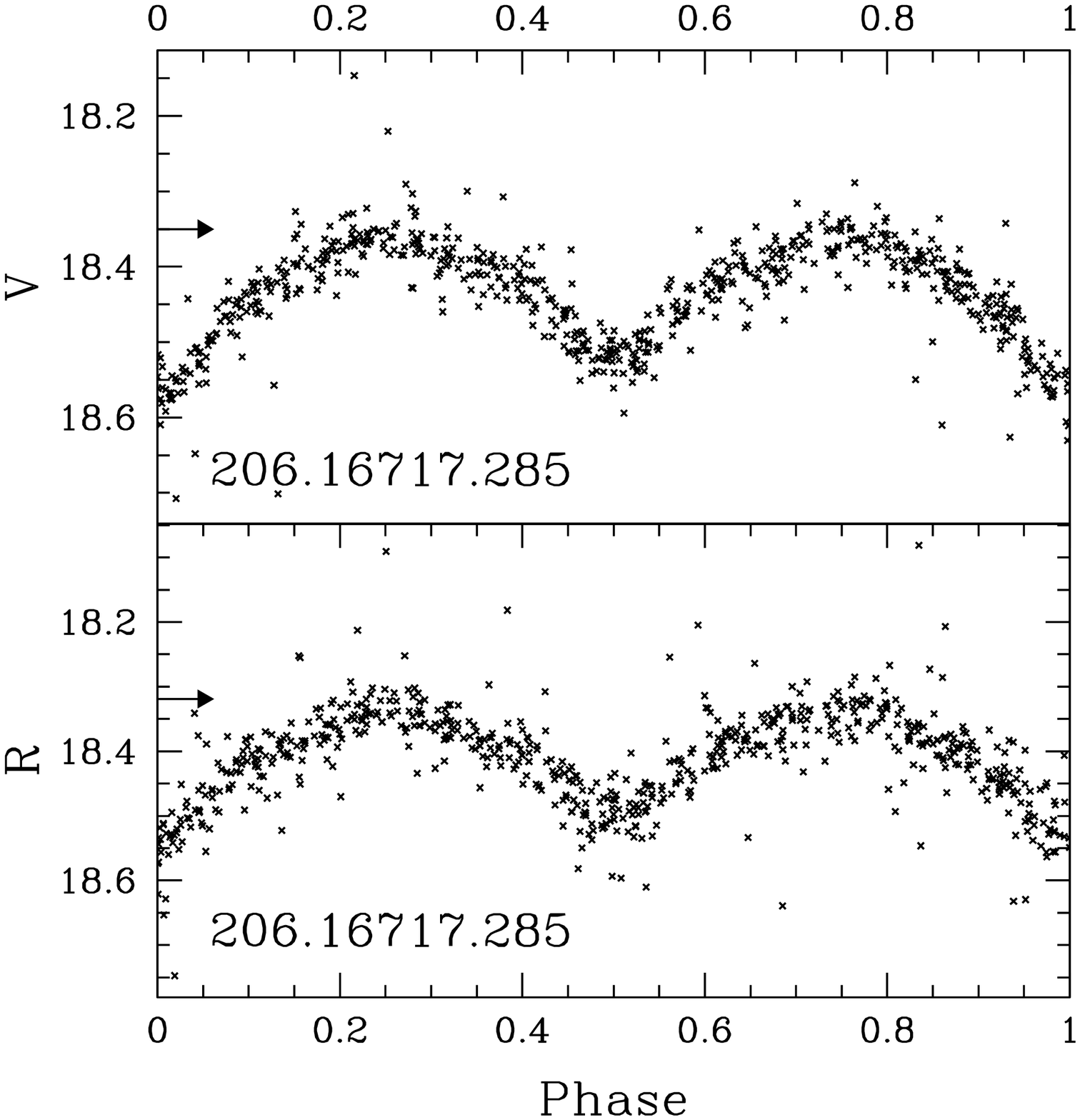}{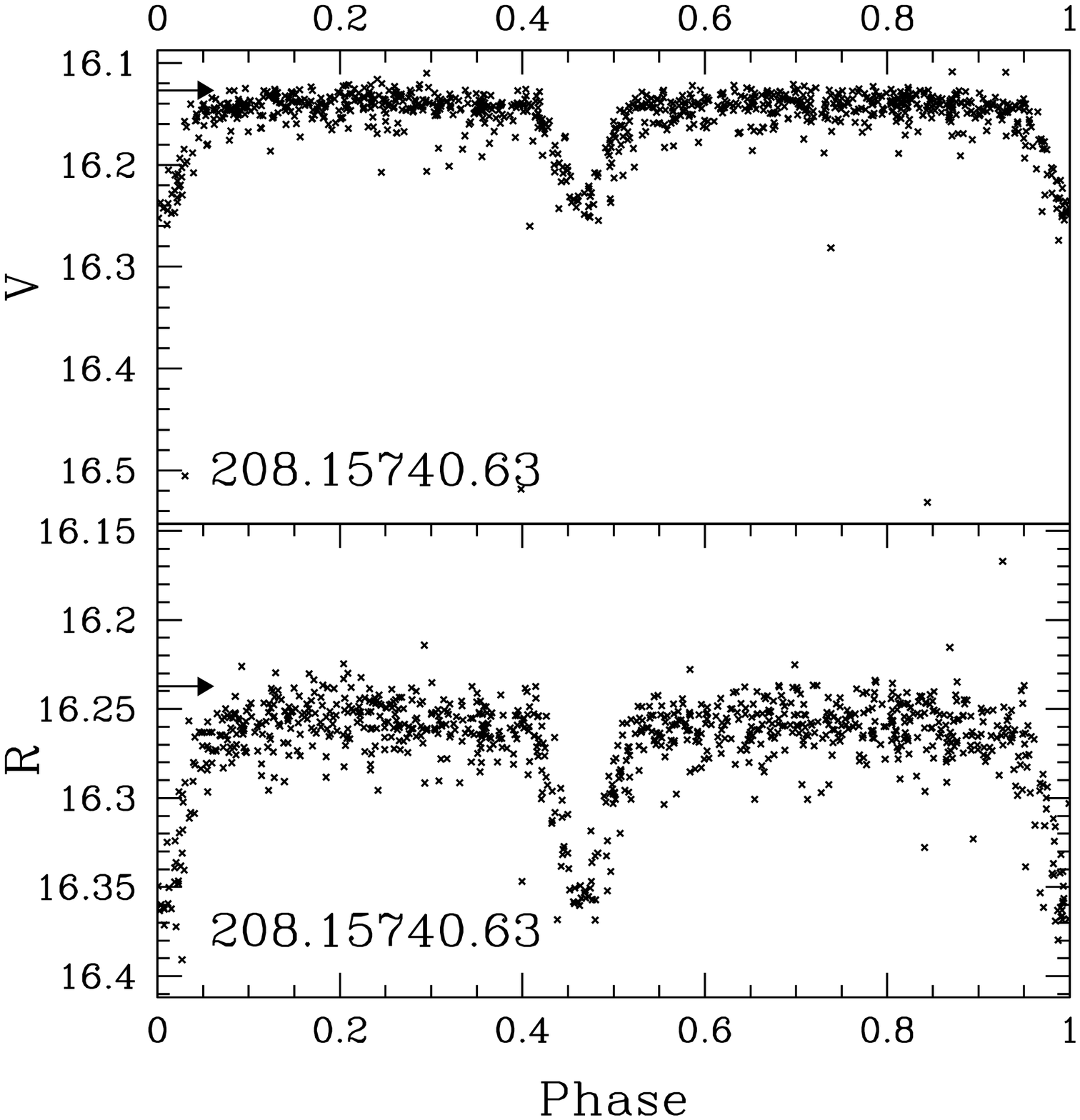}
\plottwo{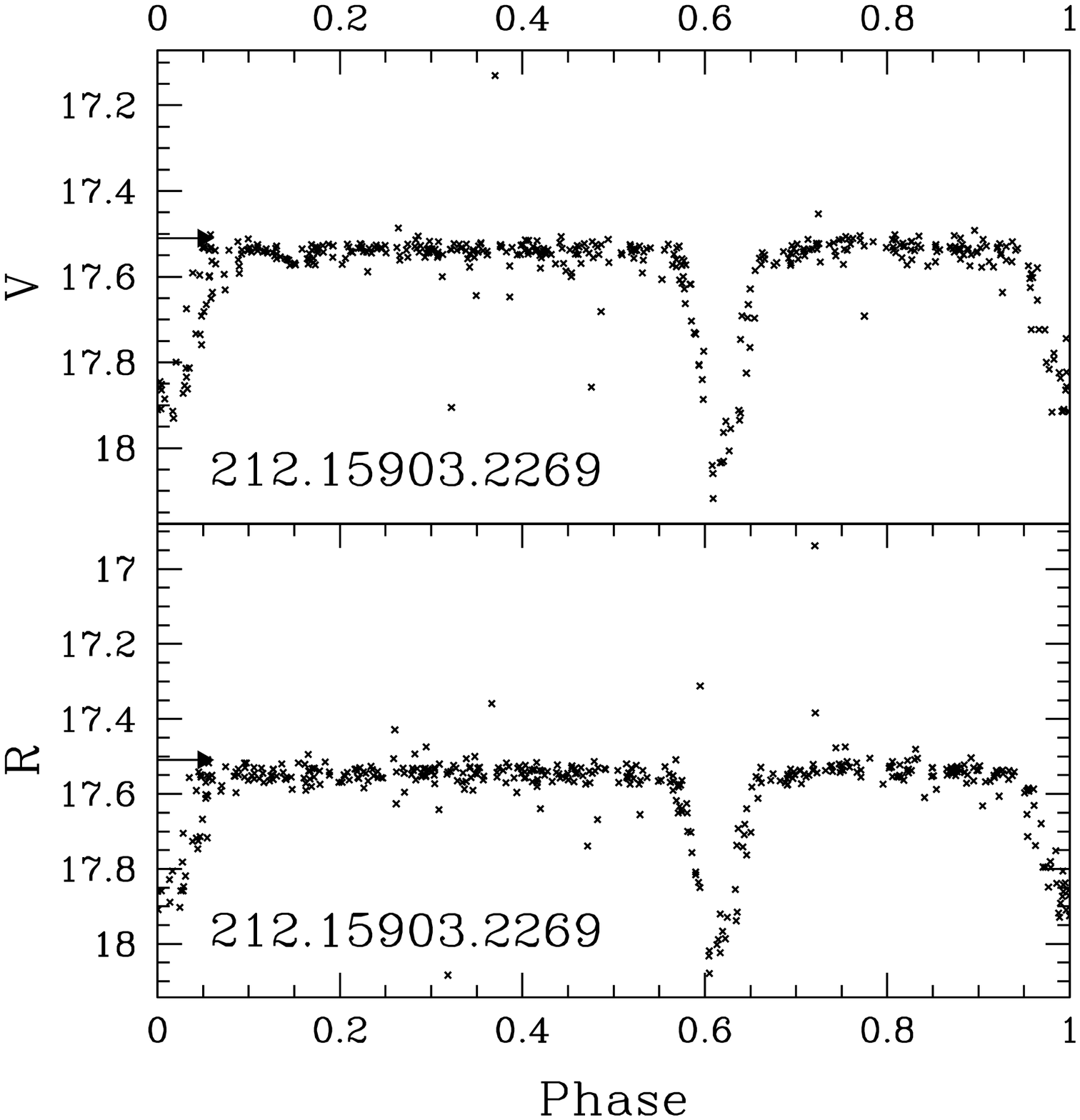}{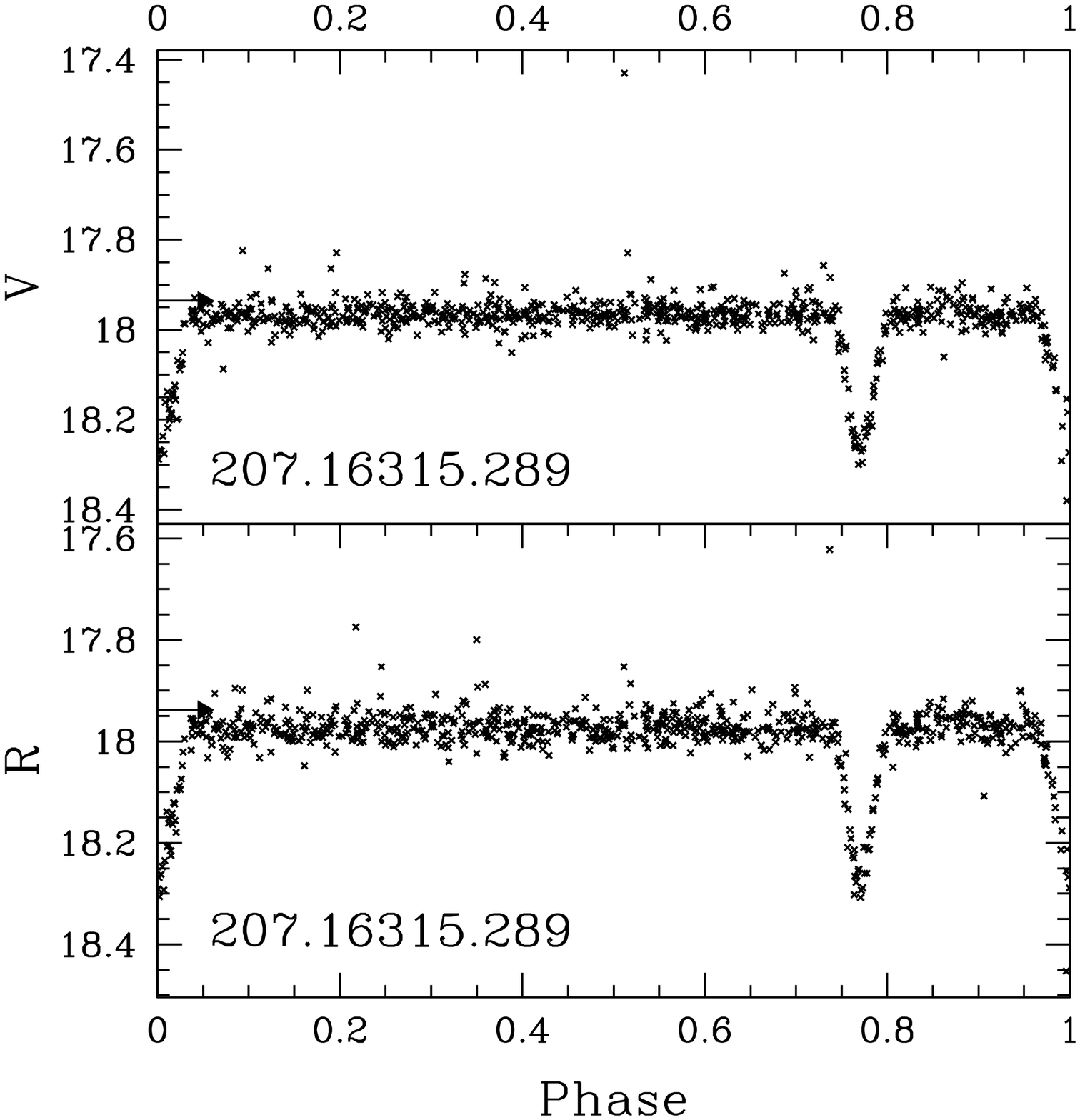}
\end{center}
\caption{Examples of light curves of EBs in the SMC sample; for basic data 
see Table \ref{tab:smclcs}.
The arrows show the baseline as defined in Table \ref{tab:smclcs}.
The EBs are arranged by ascending period.} 
\label{fig:smclcs2}
\normalsize
\end{figure}
\begin{figure}
\begin{center}
\footnotesize
\plottwo{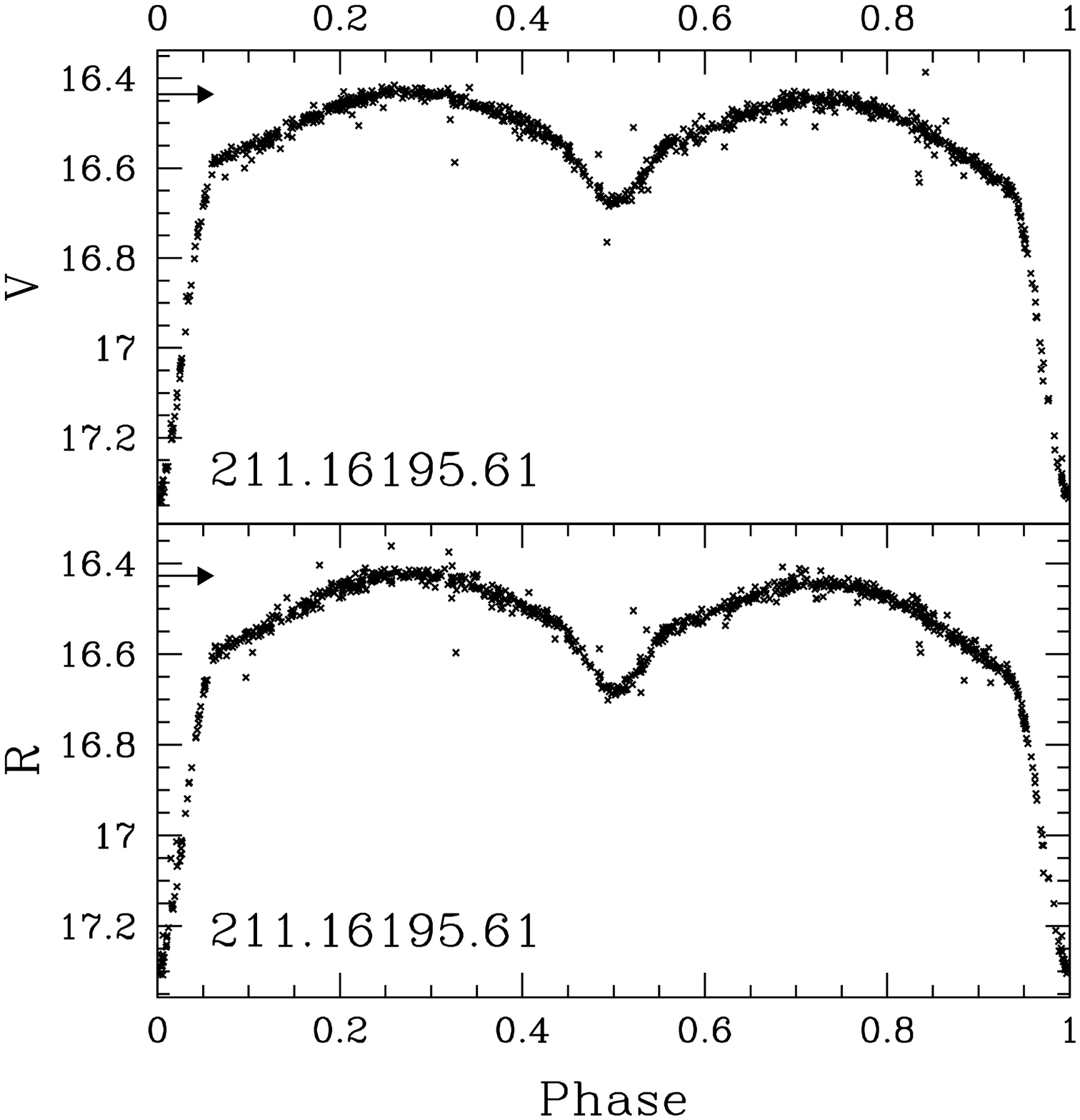}{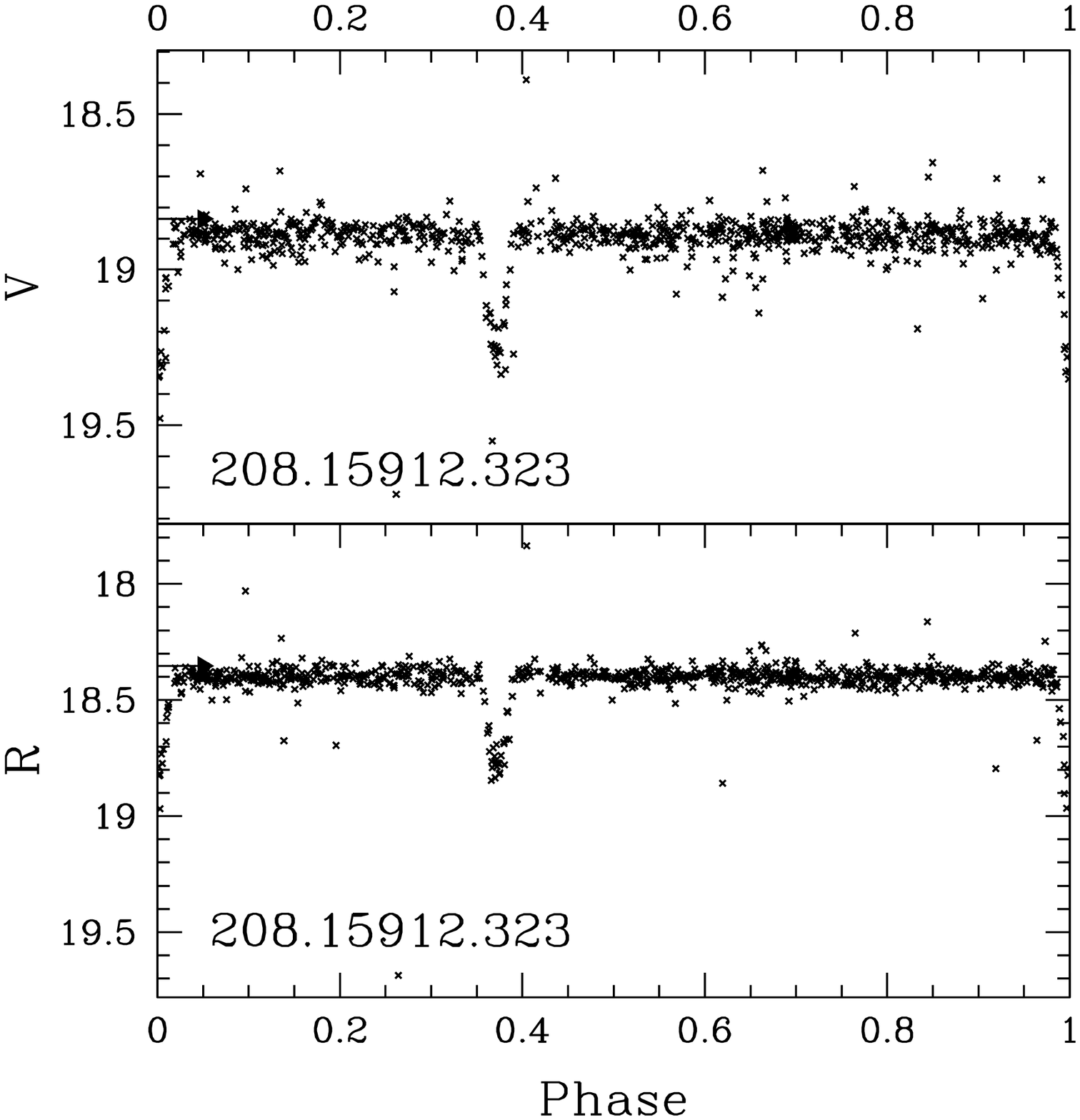}
\plottwo{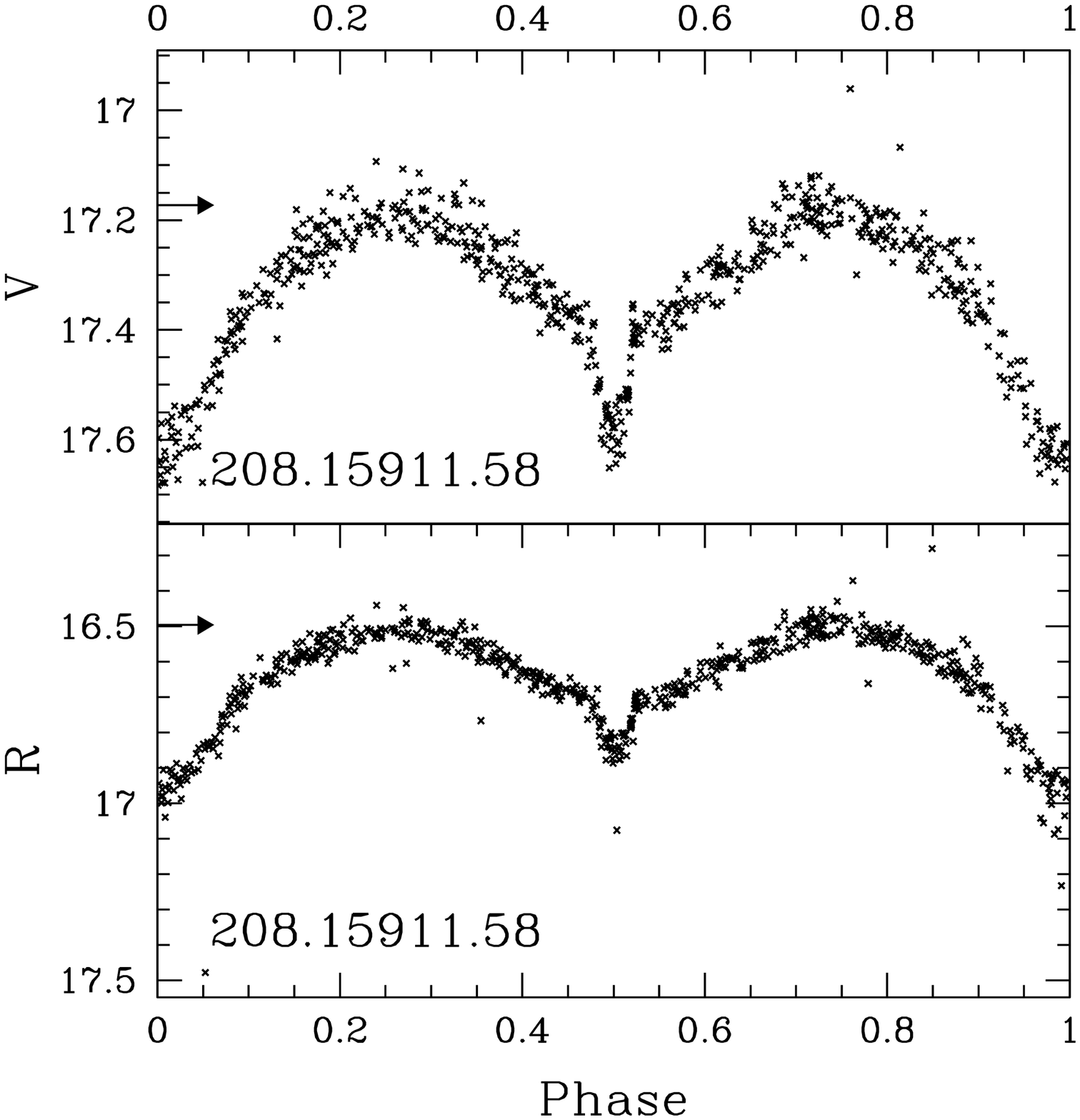}{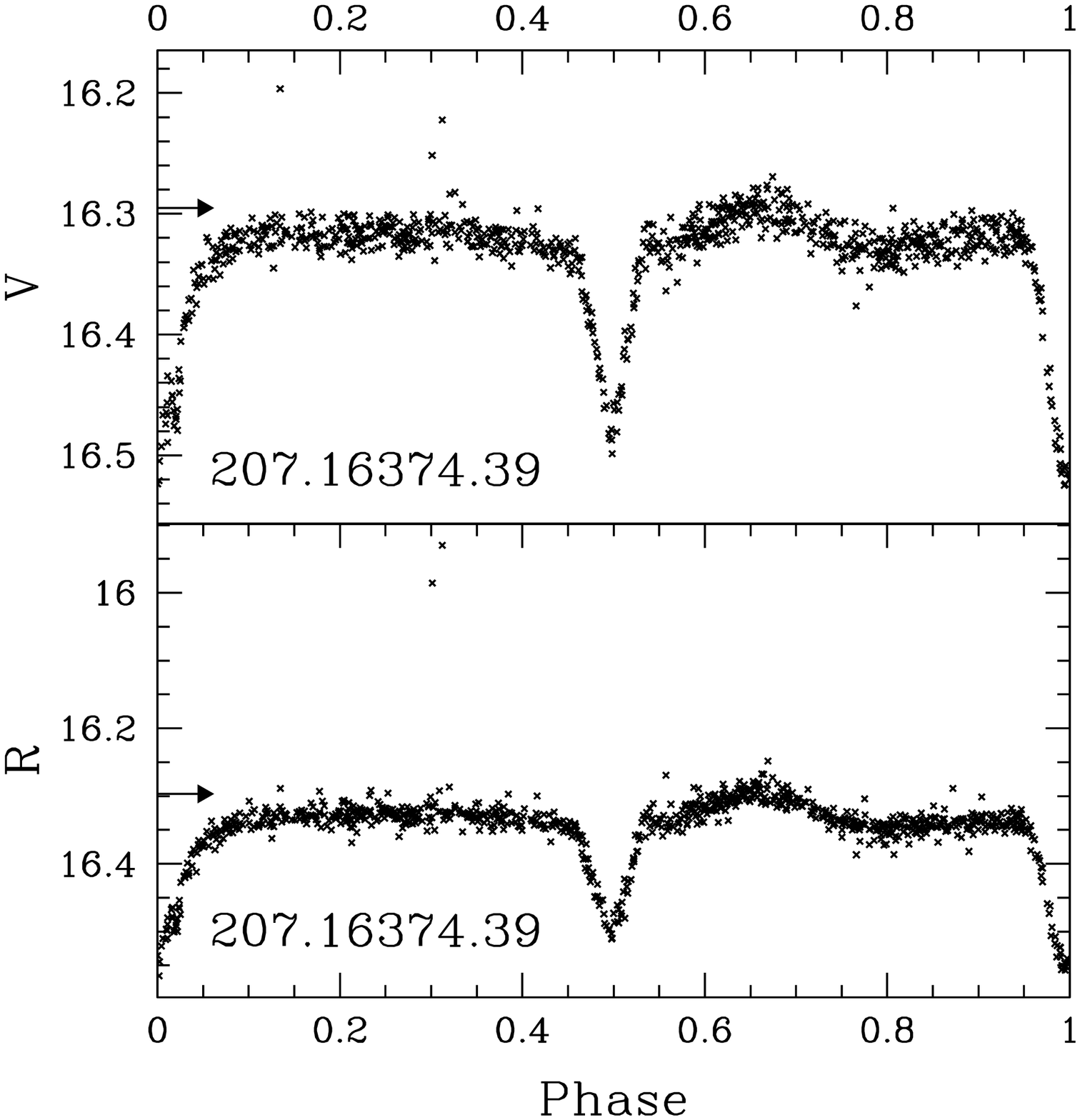}
\end{center}
\caption{Examples of light curves of EBs in the SMC sample; for basic data 
see Table \ref{tab:smclcs}.
The arrows show the baseline as defined in Table \ref{tab:smclcs}.
The EBs are arranged by ascending period.} 
\label{fig:smclcs3}
\normalsize
\end{figure}
\begin{figure}
\begin{center}
\footnotesize
\plottwo{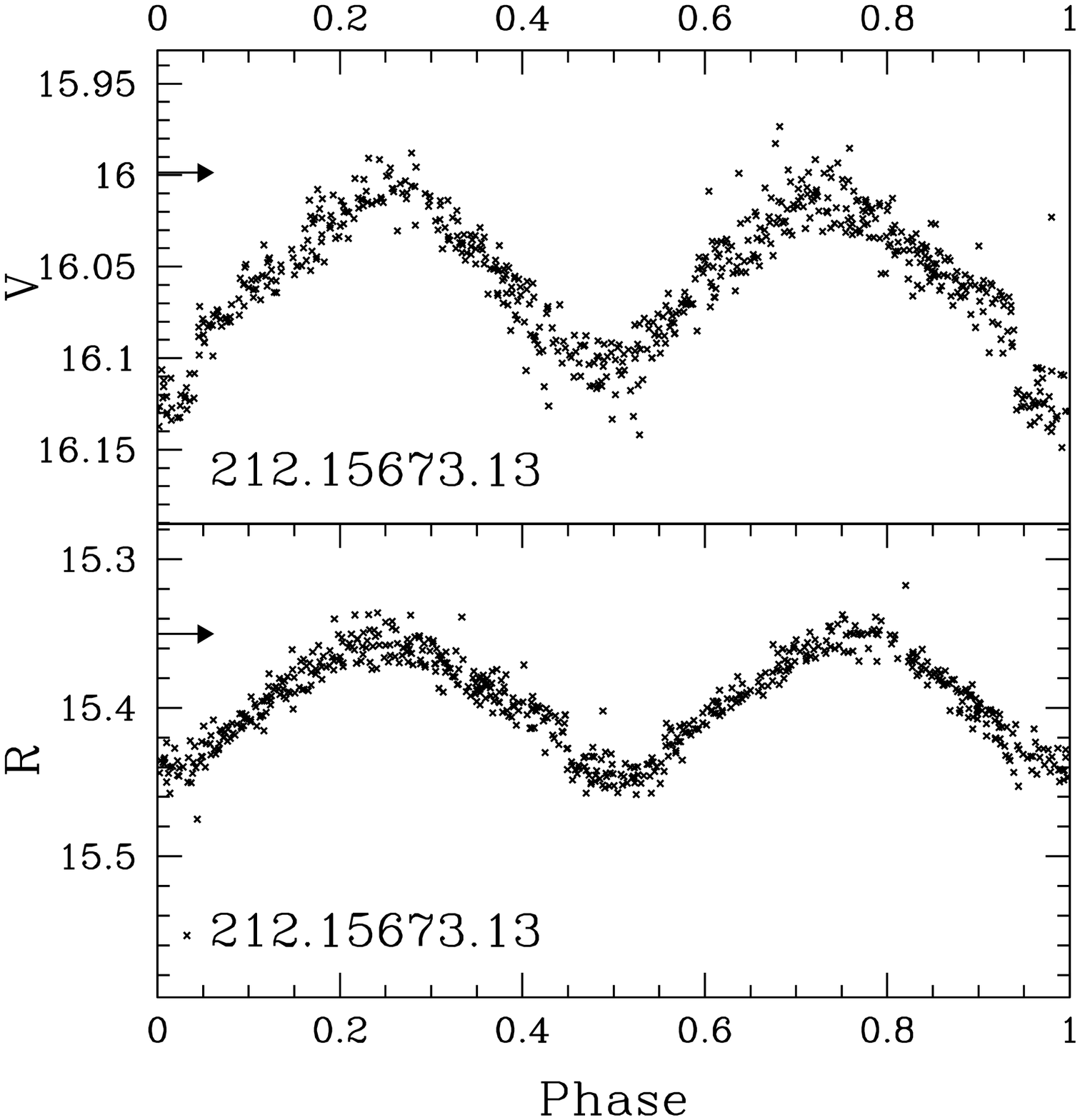}{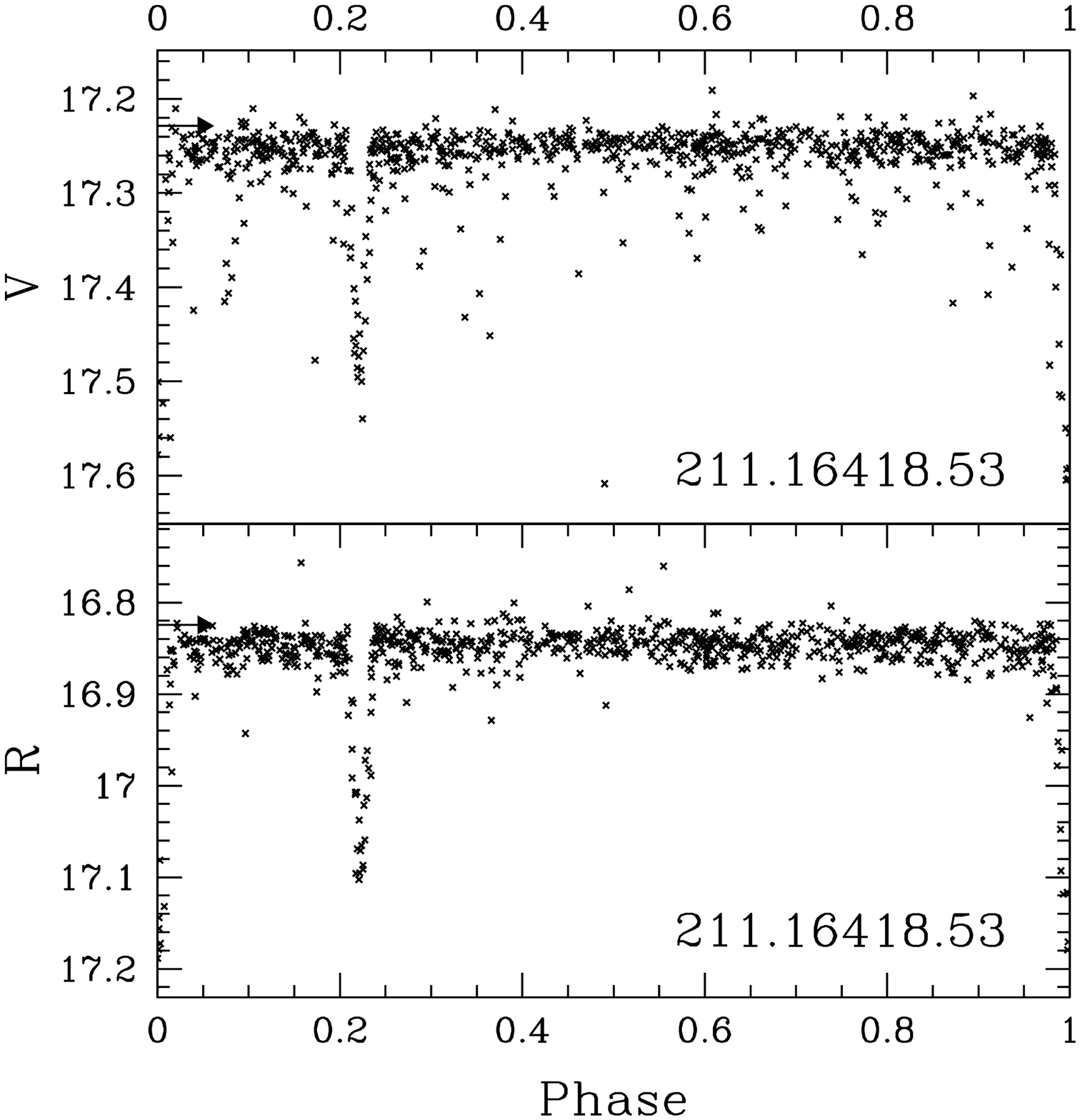}
\plottwo{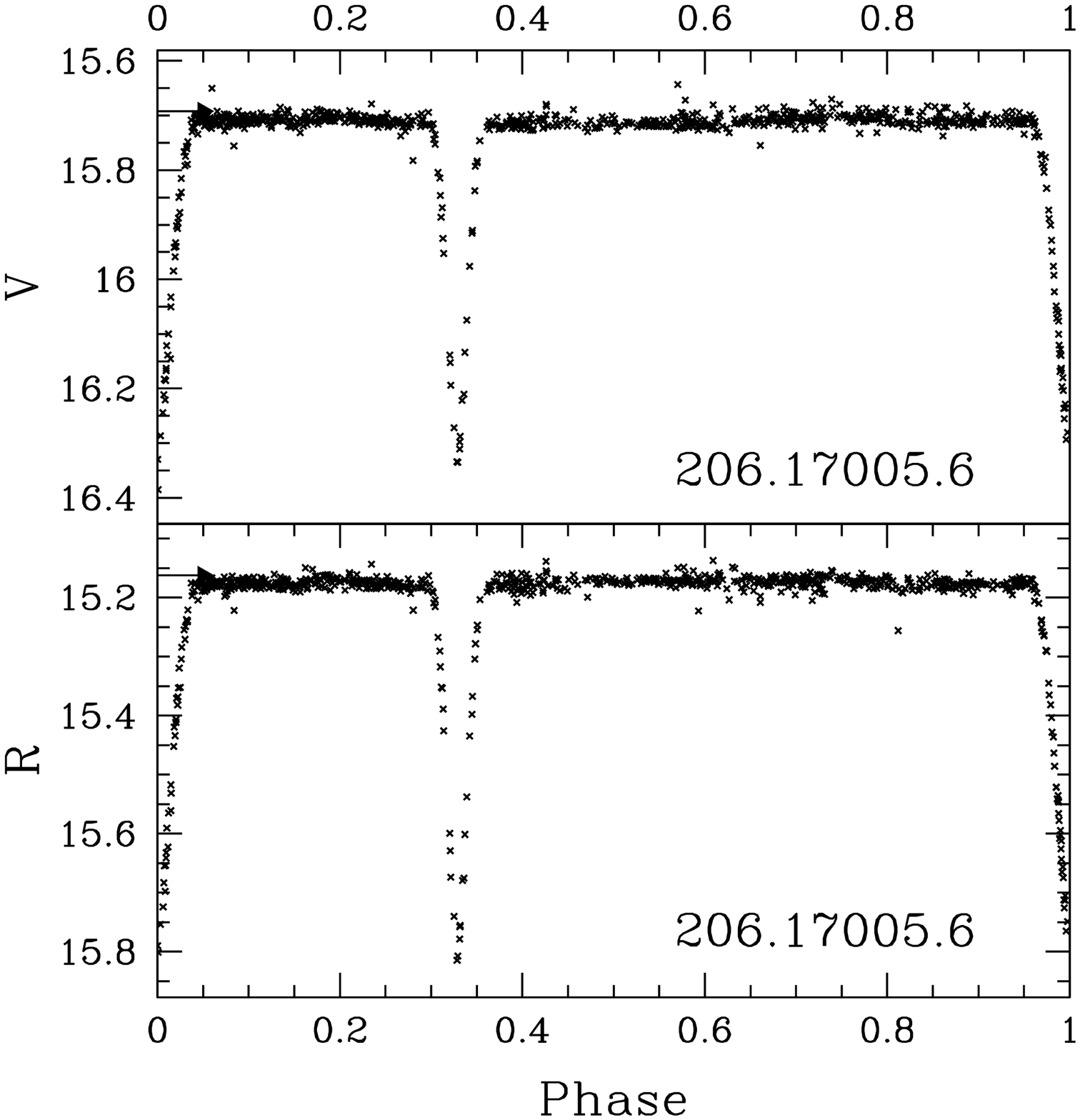}{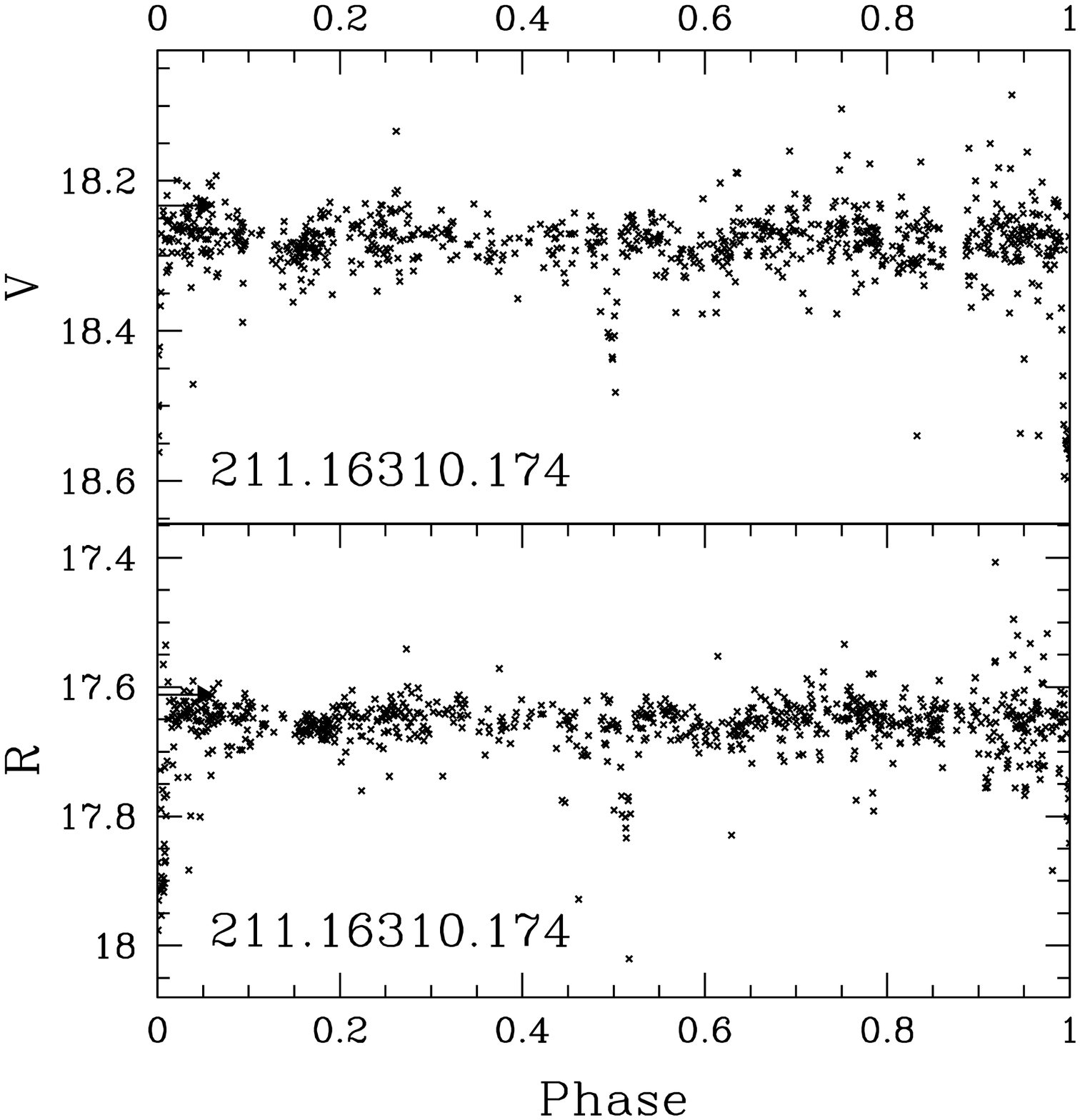}
\end{center}
\caption{Examples of light curves of EBs in the SMC sample; for basic data 
see Table \ref{tab:smclcs}.
The arrows show the baseline as defined in Table \ref{tab:smclcs}.
The EBs are arranged by ascending period.} 
\label{fig:smclcs4}
\normalsize
\end{figure}

\normalsize
Figure \ref{fig:smcpointhist} shows the histogram of the number of 
observations in both bands for the EBs in the sample.
Figure. \ref{fig:smcperhist} shows a logarithmic histogram of the period 
distribution.
\normalsize
Figure \ref{fig:smcmaghist}  shows the histograms of the magnitudes for both bands 
as well as for color.
Magnitudes range in values from $\sim 19~\mathrm{mag}$ to $\sim 14~\mathrm{mag}$ both in $V$ and $R$ bands, with a peak around $17~\mathrm{mag}$.
\normalsize
The average photometric error for the SMC is again $\sim 0.05~\mathrm{mag}$ 
in both instrumental bands; the error as a function of standard magnitude is 
shown in Figure \ref{fig:smcmagerr}.
Figures \ref{fig:smclcs1}, \ref{fig:smclcs2}, \ref{fig:smclcs3}, and 
\ref{fig:smclcs4} show some examples of light curves; their properties are 
summarized in Table \ref{tab:smclcs}.
The ``bump'' shown by the star labelled 207.16374.39 is probably due to star spots: we were able to roughly reproduce it by appropriately choosing spots on the components and fitting the light curve using the PHOEBE\footnote{\url{http://phoebe.fiz.uni-lj.si/}}\citep{prsa05} software package.
\tabletypesize{\footnotesize}
\begin{deluxetable}{cccccccl}
\tablecolumns{8}
\tablewidth{0pc}
\tablecaption{Basic data for the SMC EBs shown in Figs. \ref{fig:smclcs1},
\ref{fig:smclcs2}, \ref{fig:smclcs3} and \ref{fig:smclcs4}. 
The EBs are arranged by ascending period.
\label{tab:smclcs}}
\tablehead{
\colhead{MACHO ID} & 
\colhead{RA(J2000)} & 
\colhead{DEC(J2000)} & 
\colhead{Period\tablenotemark{a}($\mathrm{d}$)} & 
\colhead{$V$ baseline\tablenotemark{b~d}} & 
\colhead{$R$ baseline\tablenotemark{b~d}} & 
\colhead{$\vr$\tablenotemark{c~d}} &
\colhead{Comment}
}
\startdata
211.16529.5 & 00:59:31.368 & -73:26:56.04 & 
0.28 & 16.00 & 15.43 & 0.57 & Shortest period in sample\\
207.16652.1084 & 01:00:43.656 & -72:51:51.48 
& 0.36 & 19.67 & 19.59 & 0.08 & Very short period\\
206.16883.214 & 01:04:25.272 & -72:37:48.36 
& 1.14 & 18.16 & 18.18 & -0.02 & Fairly typical EB\\
207.16656.93 & 01:00:56.345 & -72:36:44.40
& 1.23 & 18.00 & 17.96 & 0.04 & Highly eccentric orbit\\
206.16717.285 & 01:01:41.400 & -72:19:25.68 
& 1.65 & 18.35 & 18.32 & 0.03 & Fairly typical EB\\
208.15740.63 & 00:46:13.949 & -72:52:37.03
& 1.74 & 16.13 & 16.24 & -0.11 & Bluest in sample\\
212.15903.2269 & 00:49:18.192 & -73:21:55.44 
& 2.42 & 17.51 & 17.51 & 0.00 & Highly eccentric orbit\\
207.16315.289 & 00:55:48.168 & -72:29:33.36 
& 3.34 & 17.93 & 17.94 & -0.01 & Highly eccentric orbit\\
211.16195.61 & 00:53:59.729 & -72:56:56.13
& 4.73 & 16.44 & 16.43 & 0.01 & Fairly typical EB\\
208.15912.323 & 00:49:10.440 & -72:46:37.56 
& 120.51 & 18.84 & 18.35 & 0.49 & Highly eccentric orbit\\
208.16.58 & 00:49:28.392 & -72:49:40.80 
& 137.89 & 17.17 & 16.50 & 0.67 & Reddest in sample\\
207.16374.39\tablenotemark{d} & 00:56:25.872 & -72:22:15.96 
& 186.34 & 16.30 & 16.30 & 0.00 & Long Period  \\
212.15673.13 & 00:45:46.824 & -73:31:32.52 
& 200.27 & 16.00 & 15.35 & 0.65 & Long period\\
211.16418.53 & 00:57:5.304 & -73:15:10.44 
& 234.64 & 17.23 & 16.82 & 0.41 & Highly eccentric orbit\\
206.17005.6 & 01:06:10.224 & -72:06:24.48 
& 371.89 & 15.69 & 15.16 & 0.53 & Highly eccentric orbit\\
211.16310.174 & 00:55:30.024 & -72:52:49.44 
& 1559.81 & 18.23 & 17.61 & 0.62 & Longest Period in sample\\
\enddata
The information in Table \ref{tab:smclcs} is also available in its entirety via the link to the machine-readable version above. The EBs are arranged by ascending period. Units of right ascension are hours, minutes, and seconds, and units of declination are degrees, arcminutes, and arcseconds. 
\tablenotetext{a}{Supersmoother provides different periods for $V$
and $R$ unfolded light curves, but their difference is usually smaller than
the precision to which we report their values in this table.
On line summary tables provide both periods to $5$ significant digits.}
\tablenotetext{b}{The baseline is calculated in the following way.
First outlying points have been eliminated by dividing the light curve in 
boxes of $\sim 50$ data points and eliminating the points which were more 
than 2 standard deviations away from the mean in each box.
Then the median of the $10\%$ most luminous points was taken.
This value is not the median of the whole light curve that is shown in the 
figures.}
\tablenotetext{c}{Values are quoted to the hundredths of magnitude, typical 
of MACHO observational uncertainties.}
\tablenotetext{d}{
This is the difference of the two baselines as defined above, not of the two medians as is the $\vr$ shown in the figures. This column is not directly available in the online table but can be deduced by subtracting col. (12) from col. (10)}
\tablenotetext{e}{There is a curious ``bump'' in the light curve of this 
long period EB (Figure \ref{fig:smclcs3}) that suggests further investigation.}
\end{deluxetable}
\section{Color Magnitude Diagram and Color Period Diagram }
\label{sec:diagrams}
\subsection{The Large Magellanic Cloud sample}
\begin{figure}
\footnotesize
\plotone{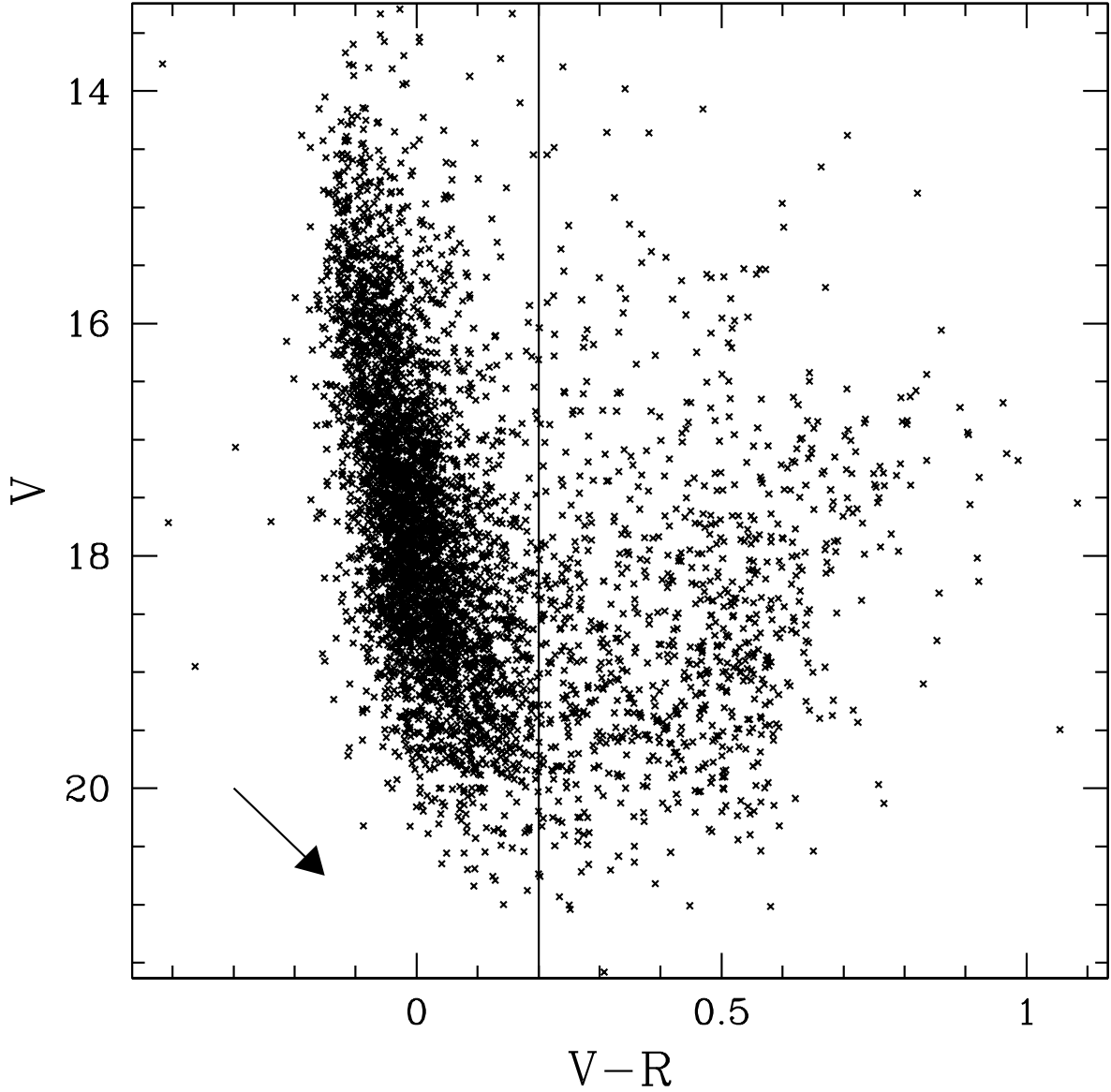}
\caption{CMD for $4634$ EBs in the LMC sample.
The sample is made up mostly by young luminous stars (with at least one 
component of spectral type B or A).
A fairly high number of EBs ($\sim 19\%$) are evolved; there are no real 
``gaps'' between the young star region and the evolved star region.
The vertical line shows the cut used to separate the young star region
from the evolved star one.
The reddening vector is $\frac{A_V}{E(\vr)}=5$; the adopted values of 
$\langle A_V\rangle$ and $\langle E(\vr)\rangle$ are $0.64~\mathrm{mag}$ 
and $0.128~\mathrm{mag}$.}
\label{fig:cmd}
\normalsize
\end{figure}

Figure \ref{fig:cmd} shows the CMD for the $4634$ EBs in the LMC sample; 
the lower magnitude limit is $V\sim 21~\mathrm{mag}$.
We estimated the reddening by using the LMC extinction map described in 
the LMC photometric survey of \citet{zaritsky04}.
The extinction catalog produced by the survey is available for query on 
line\footnote{\url{http://ngala.as.arizona.edu/dennis/lmcext.html}} and we 
retrieved the values of $A_V$ specified in Table \ref{tab:lmcextinction}, 
based on the \emph{hot} stars only found by the survey ($T>12000\mathrm{K}$)
in a radius of $12\arcmin$ (the maximum allowed) around the positions specified in Table \ref{tab:lmcextinction}, which sample the EB position distribution.
\tabletypesize{\footnotesize}
\begin{deluxetable}{ccc}
\tablecolumns{3}
\tablewidth{0pc}
\tablecaption{Extinction values in the LMC.
\label{tab:lmcextinction}}
\tablehead{
\colhead{RA(J2000)} & 
\colhead{DEC(J2000)} & 
\colhead{$A_V$} 
}
\startdata
06:06:00 & -69:05:00 & 0.48\\
06:06:00 & -72:43:00 & 0.97\\
05:40:00 & -65:30:00 & 0.75\\
05:40:00 & -69:05:00 & 0.80\\
05:40:00 & -72:30:00 & 0.78\\
05:20:00 & -65:30:00 & 0.53\\
05:20:00 & -69:05:00 & 0.50\\
05:20:00 & -72:30:00 & 0.56\\
05:00:00 & -65:30:00 & 0.50\\
05:00:00 & -69:05:00 & 0.46\\
05:00:00 & -72:43:00 & 0.45\\
04:40:00 & -69:05:00 & 0.62\\
04:40:00 & -72:30:00 & 0.97\\

\enddata
\end{deluxetable}
From  the values of Table \ref{tab:lmcextinction} we derive a mean
value for $A_V$ of $0.64~\mathrm{mag}$ which we use to characterize the average LMC 
$V$ extinction.
We use the reddening vector $\frac{A_V}{E(\vr)}=5$ from 
\citet[][and references therein]{machovar4} and find 
$\langle E(\vr)\rangle=0.128~\mathrm{mag}$, more than a factor of two and a half 
larger than the value $0.049~\mathrm{mag}$ found by \citet{machovar4}.
This is likely due to the fact that the reddening for the EBs are likely 
to be along lines of sight toward young, hot stars in the 
\citet{zaritsky04} catalog which are derived to have higher $A_V$, while 
the \citet{machovar4} $A_V$ was derived from observations of RR Lyrae stars.
As the CMD shows the sample is made up mostly of bright
early type stars: from the range in magnitudes and assuming an LMC distance
modulus of $18.5~\mathrm{mag}$ \citep{marel02} we see that in most cases at least one 
component is of spectral type B or A \citep{alcock00b,cox00}.
We employ the term \emph{young star region} to describe the main feature on
the left part of the CMD, rather than Main Sequence because our sample 
contains some bright and short lived stars that may not be burning 
hydrogen in their core while still being on the blue part of the CMD; 
likewise we employ the term \emph{evolved star region} to indicate the 
feature on the red part of the CMD.
We define the young star region as $\vr<0.2~\mathrm{mag}$ and the evolved 
star region as $\vr>0.2~\mathrm{mag}$: with these definitions we find 
$3760$ EBs in the young star region  and $874$ EBs in the evolved star region.
We used a simple cut to separate the young star region from the evolved 
star region rather than a more precise one because our cut can be easily 
seen both in the CMD and in the Color Period Diagrams.
An interesting feature of the CMD is the lack of a clear gap between the 
young star region and the evolved star region; there is instead a fairly 
continuous transition, with a higher number of systems filling the 
Hertzsprung Gap that would be expected from CMDs of single stars.
This may indicate that these systems are composed of a more massive and 
hence more evolved and redder star and a less massive, less evolved, bluer 
one.
\begin{figure}
\footnotesize
\begin{center}
\plottwo{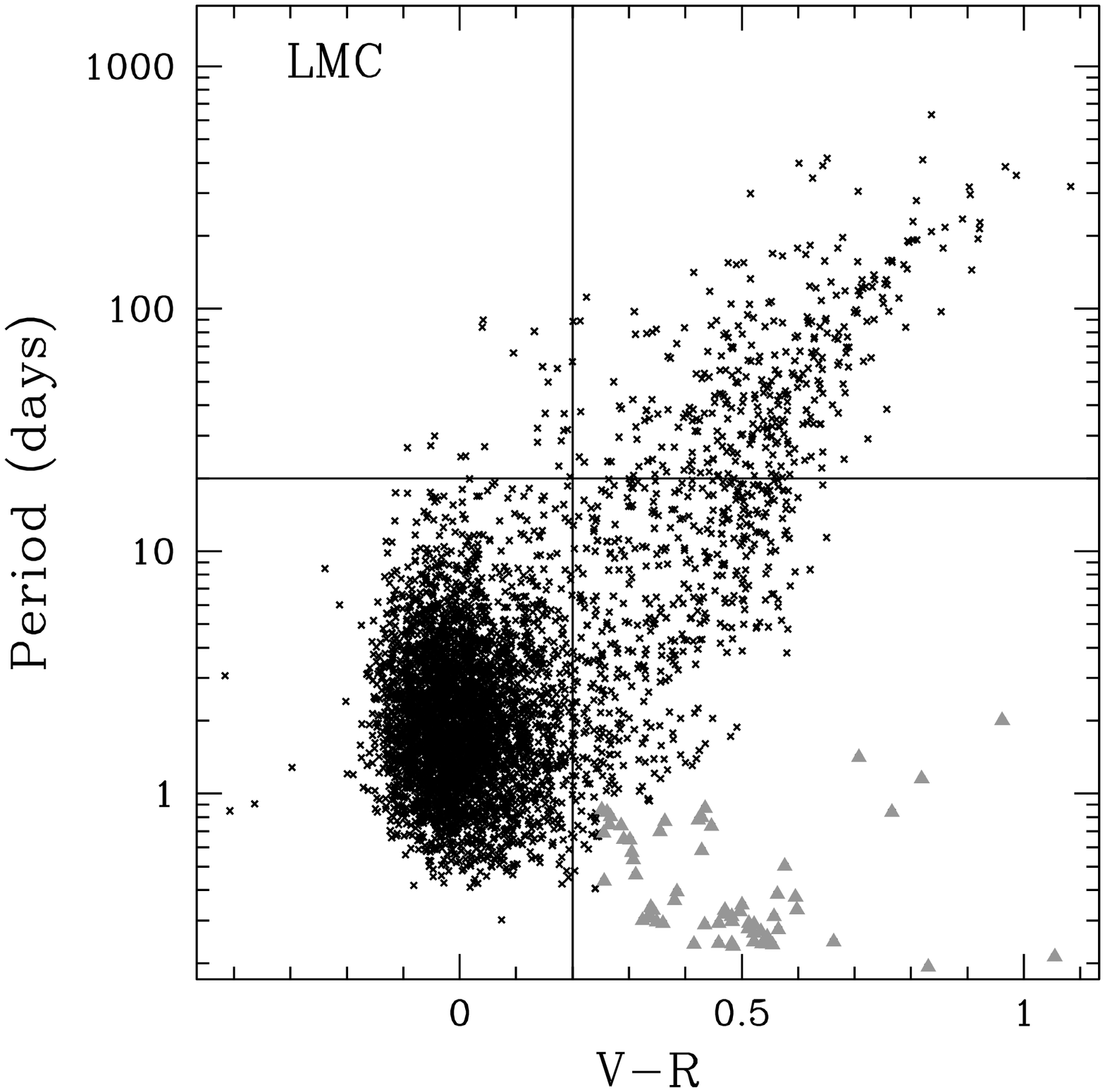}{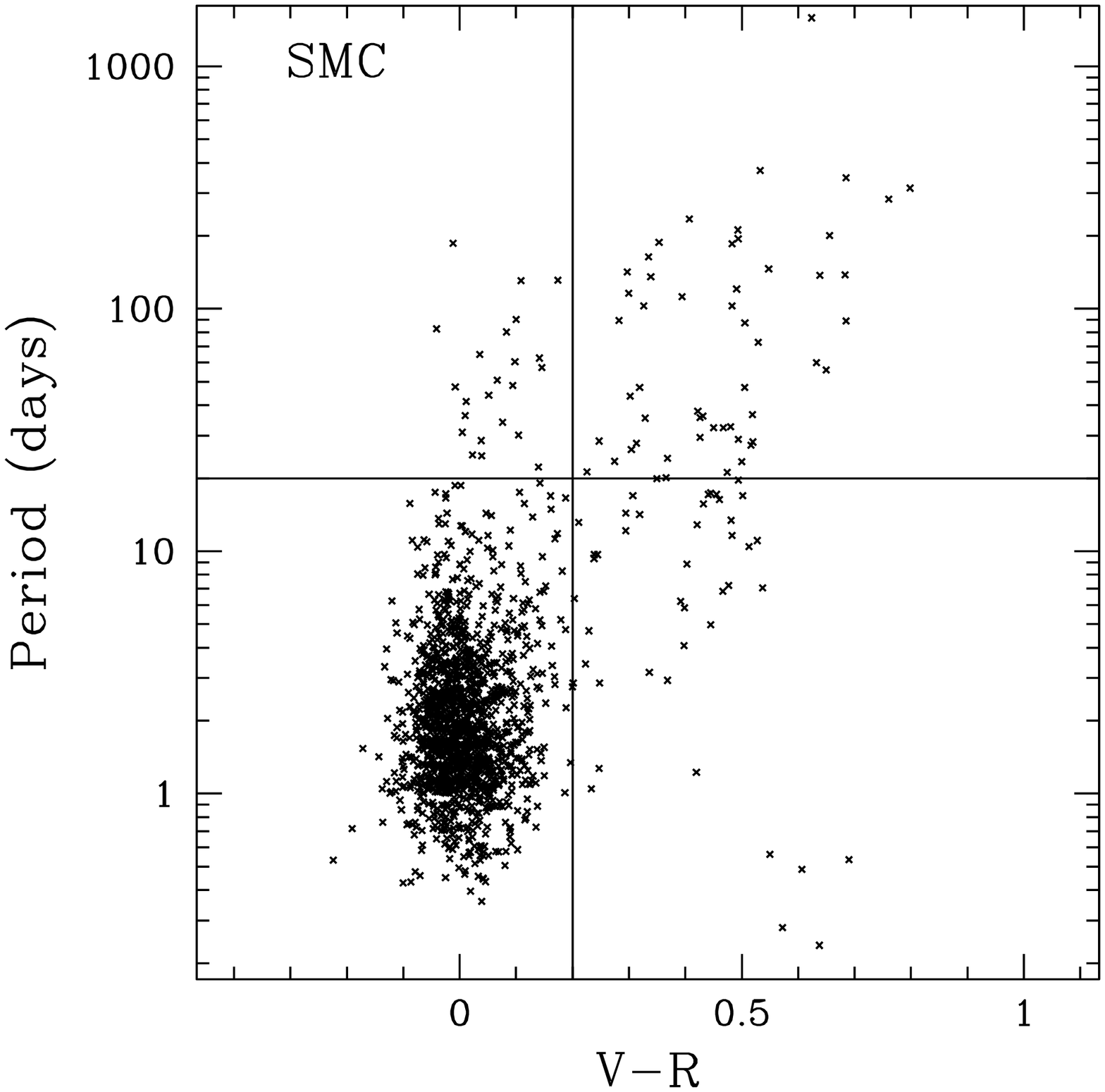}
\end{center}
\caption{
Left Panel: Color Period Diagram for $4634$ EBs in the LMC.
The gray filled triangles represent the foreground population.
Right Panel: Color Period Diagram for $1508$ EBs in the SMC.
The figure shows the higher fraction of long period EBs that belong to the 
young star region in the SMC than in the LMC.}
\label{fig:smcvslmc}
\normalsize
\end{figure}
\par
The Color Period Diagram is shown in the left panel of Figure \ref{fig:smcvslmc};
this diagram clearly shows the young EBs with periods 
$P\lesssim 20\mathrm{d}$ and a second population 
($333$ objects) of long period, evolved EBs with periods 
$P>20\mathrm{d}$ and $\vr>0.2~\mathrm{mag}$.
Several interesting features emerge in Figure \ref{fig:smcvslmc}: there is a 
paucity of long period objects on the young star region of the CMD, and a 
corresponding lack of short period objects with very red colors.
Furthermore, the red ($\vr>0.2~\mathrm{mag}$) population shows a positive 
correlation between period and color.
There are virtually no long period, blue objects or short period, red objects 
in this group.  
In contrast, the young star region shows no such correlation.
This structure in the Color Period Diagram is a consequence of (i) Kepler's 
Third Law, (ii) the probability that an EB is favorably oriented in space 
to allow eclipses to be detected ($\mathrm{Prob}=(R_1+R_2)/a$) where 
$R_1$ and $R_2$ are the radii of the primary and secondary stars, 
respectively and $a$ is the semi major axis, and (iii) that objects usually 
evolve in this diagram at constant period, from blue to red.
Long period binary stars have large semi-major axes. 
When both stars are on the young star region, their relatively small radii 
yield a relatively low probability that they will eclipse when seen from 
our vantage point.
When one of the pair evolves away from the young star region, one of these 
radii (typically $R_1$) will increase.
The consequence of this is an increase in the probability that eclipses 
will be detected. 
This accounts for the presence of red, long period stars and the absence of 
corresponding young progenitors.
The situation is different for short period systems. 
These are relatively likely to be detected because of their small semi-major
axes, and are prominent on the blue side.
As one of these stars evolves and expands rapidly, it may engulf the 
companion and enter a stage of \emph{common envelope evolution} in which 
the expanding star overflows the \emph{second} Lagrangian point $L_2$
\citep{paczynski76}; leading to the disappearance of eclipses.
Common envelope system differ from contact binaries 
\citep{kal99,hilditch01,shore94}\footnote{Called ``over contact'' in \citep{kal99}} 
in which two young stars overflow their first Lagrangian point ($L_1$) and their Roche equipotential surface assumes a dumbbell shape.
Contact systems usually show ellipsoidal variation and, if the orbital inclination 
is large enough, also eclipses; EBs of the W UMa type belong to this category.
\citet{shore94} and \citet{iben93} provide more information on common 
envelope binaries.
The correlation between period and color among red objects reflects the 
general correlation between radius and color for the evolved partner.
\subsection{Foreground objects}
\begin{figure}
\footnotesize
\plotone{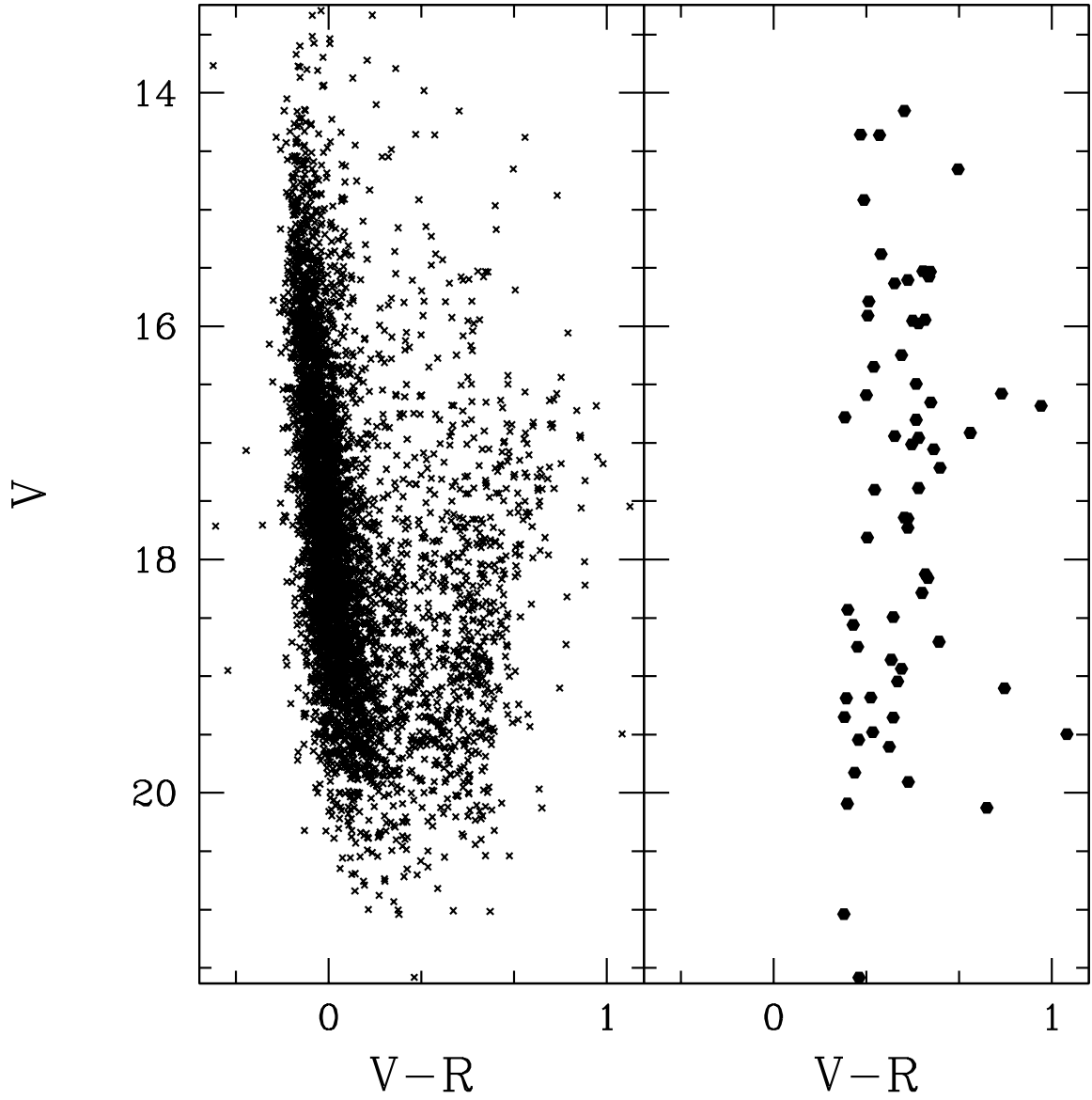}
\caption{CMD of the LMC sample (left panel) and of the foreground
population (right panel,filled hexagons).}
\label{fig:cmdforeground}
\normalsize
\end{figure}

The left panel of Figure \ref{fig:smcvslmc}
reveals a population of $\sim 63$ EBs with low 
periods ($P\lesssim 2\mathrm{d}$) and high color ($\vr>0.3~\mathrm{mag}$).
These objects are probably foreground galactic EBs composed of late type 
stars.
This interpretation is suggested by several factors.
First, due to the large angular extent of the LMC, there is foreground 
contamination in the LMC MACHO fields \citep{alcock00b}; in particular the 
feature marked ``H'' in the CMD of their Figure $1$ indicates foreground 
galactic disk stars and is centered at $\vr\sim 0.5~\mathrm{mag}$ as is our 
presumptive foreground population.
Second, both the short period of these EBs, and the shape of their light 
curves which are either detached or mildly distorted, strongly suggests 
that the stars making up this population are small, late type stars; this 
is further borne out by their color, again typical of a solar like star;
Finally, the CMD of this population, shown in Figure \ref{fig:cmdforeground} 
clearly shows what appears to be a turnoff feature at 
$\vr\sim 0.4~\mathrm{mag}$, with few evolved objects ($\vr>0.7~\mathrm{mag}$).
It is interesting to note that the overall shape of this population in the 
Color Period Diagram shows, on a smaller scale, the same features of the 
LMC diagram; a Main Sequence\footnote{We employ this term since these foreground
stars are probably not very massive and therefore are in their core hydrogen 
burning phase.} is clearly visible in Figure \ref{fig:cmdforeground} and the 
evolved EBs show the same Color Period correlation of their LMC counterparts
in Figure \ref{fig:smcvslmc}.
\subsection{The Small Magellanic Cloud sample}
\begin{figure}
\footnotesize
\begin{center}
\plotone{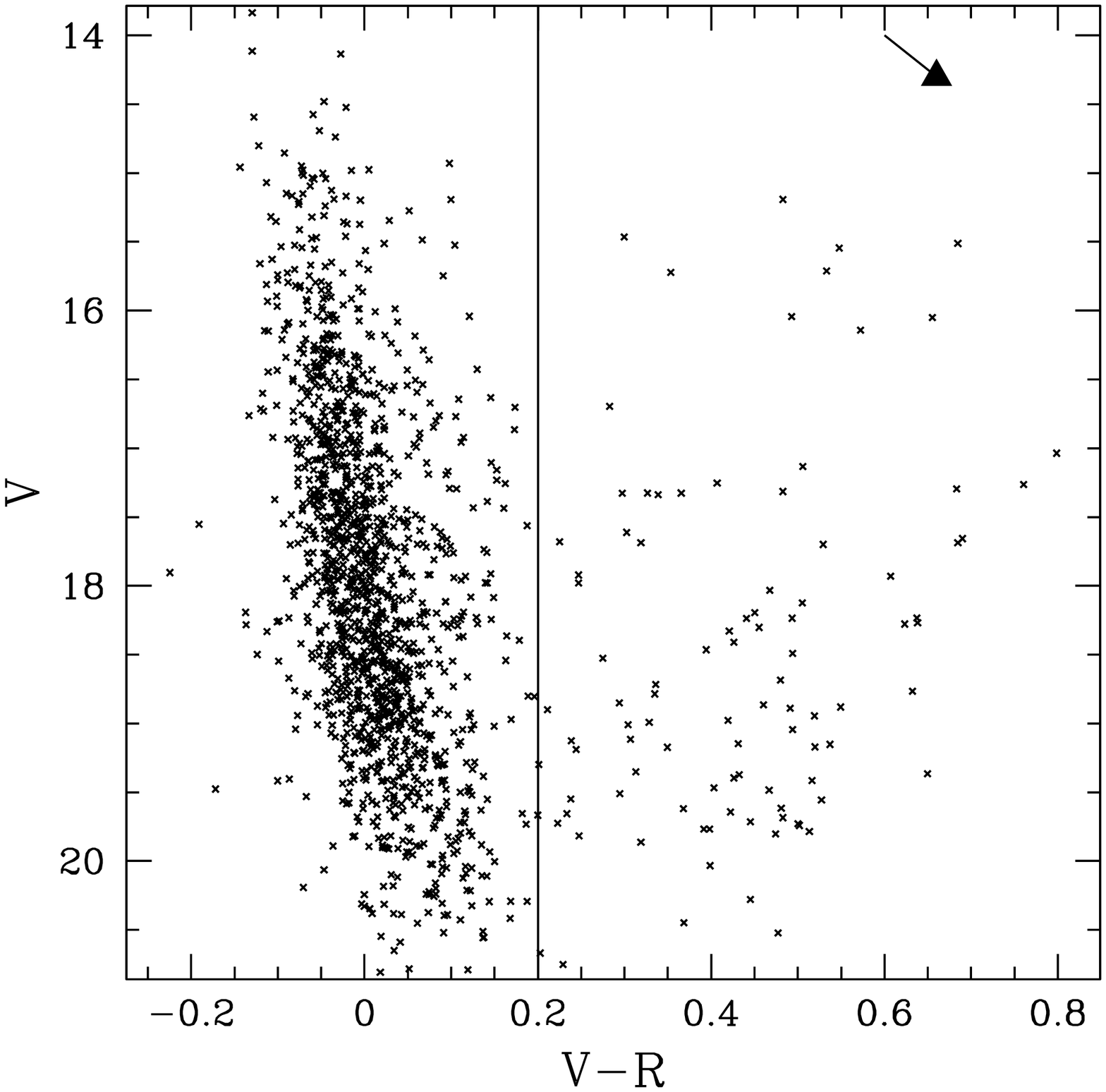}
\end{center}
\caption{CMD for $1508$ EBs in the SMC sample.
The reddening vector is $\frac{A_V}{E(\vr)}=5$; the adopted values of 
$\langle A_V\rangle$ and $\langle E(\vr)\rangle$ are $0.3~\mathrm{mag}$ 
and $0.06~\mathrm{mag}$.}
\label{fig:smccmd}
\normalsize
\end{figure}

Figure \ref{fig:smccmd} shows the CMD for $1508$ EBs in 
the SMC sample out of the $1509$ in the sample; one EB which has valid data
only in the $R_{\mathrm{MACHO}}$ band is not shown because it was not possible to
determine the standard magnitudes via Eq. \ref{eq:smcmachocal}.
The general remarks made for the LMC CMD apply here as 
well and we used a reddening vector with the same inclination as the LMC.
The figure clearly shows the young star region which is  composed of $1412$ 
EBs ($94\%$) whereas there are just $96$ evolved EBs.
We estimate the SMC reddening from \citet{zaritsky02}; from their Figure $19$
we infer a mean $A_V\sim 0.3~\mathrm{mag}$ for their hotter SMC population, 
relevant to our sample which is composed mostly of early type hot stars on the 
young star region.
We use the same reddening vector as the LMC, $\frac{A_V}{E(\vr)}=5$, and 
find a mean $E(\vr)=0.06~\mathrm{mag}$.
The Color Period Diagram is shown in the right panel of Figure \ref{fig:smcvslmc};
the general considerations made for the LMC Color Period Diagram apply here as well.
The diagram shows $5$ EBs with low period ($P<1\mathrm{d}$) and red color 
($\vr\sim 0.6~\mathrm{mag}$) which are probably foreground objects.
The much smaller number of foreground objects in the SMC is probably due both to its smaller angular size and to its higher galactic latitude
($l\sim -44^\circ$ compared to $l\sim -33^\circ$ for the LMC).
We can further test the hypothesis that these two short period, red populations in the LMC 
and SMC are due to foreground objects by comparing the ratio of their numbers to the
ratio of the areas of the LMC and SMC, which we can estimate from Figures \ref{fig:pos} and \ref{fig:smcpos}; if these objects are indeed foreground these two ratios should be roughly equal.
From the figures we can estimate the sky area of the LMC as $\sim 110\Box^\circ$ and the sky area of the SMC as $\sim 10\Box^\circ$; their ratio is thus similar to the ratio of the numbers of EBs in the two populations as expected.
\section{Comparison between the Large Magellanic Cloud and the Small 
Magellanic Cloud samples}
\label{sec:comparison}
\begin{figure}
\footnotesize
\begin{center}
\plottwo{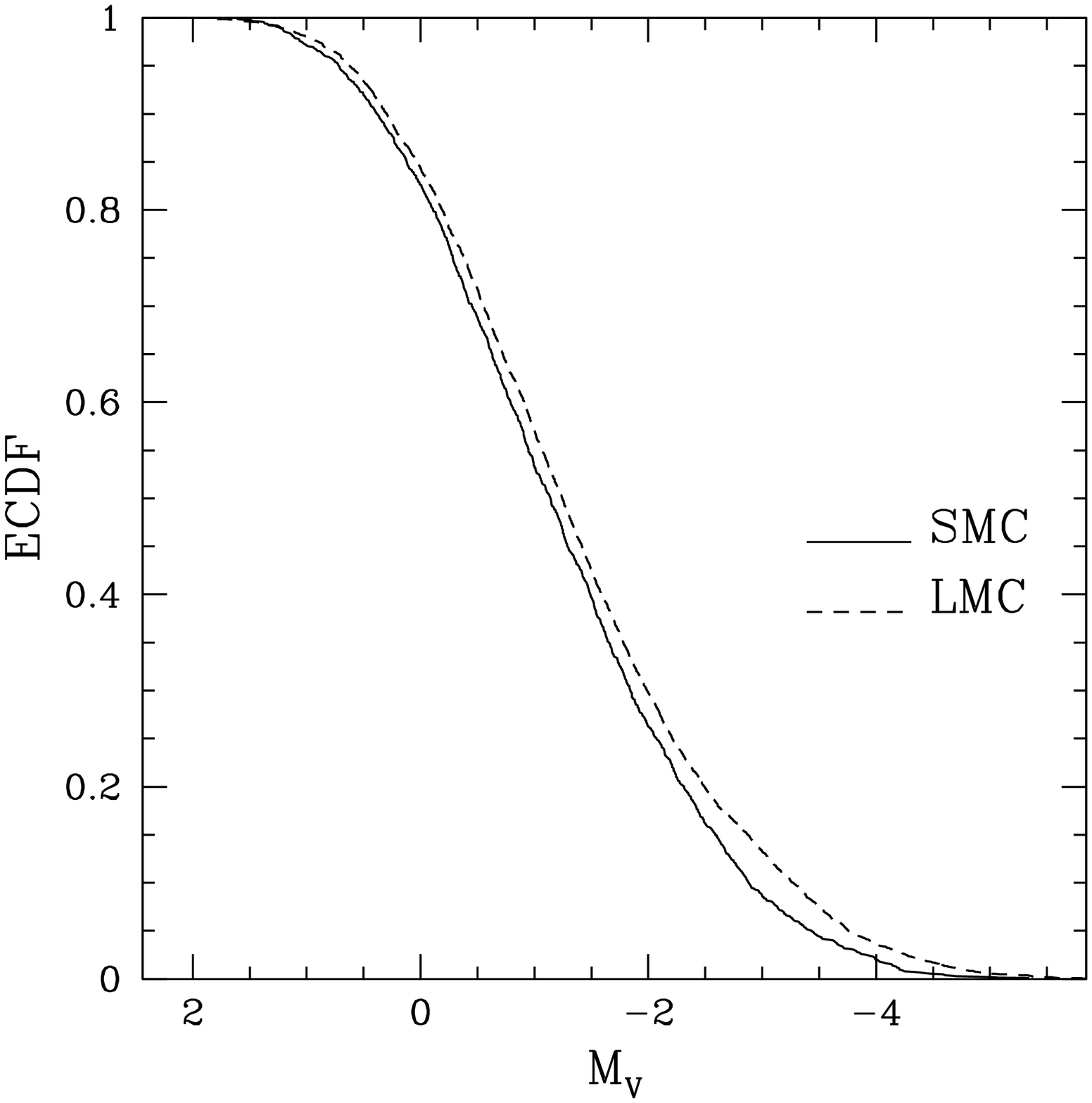}{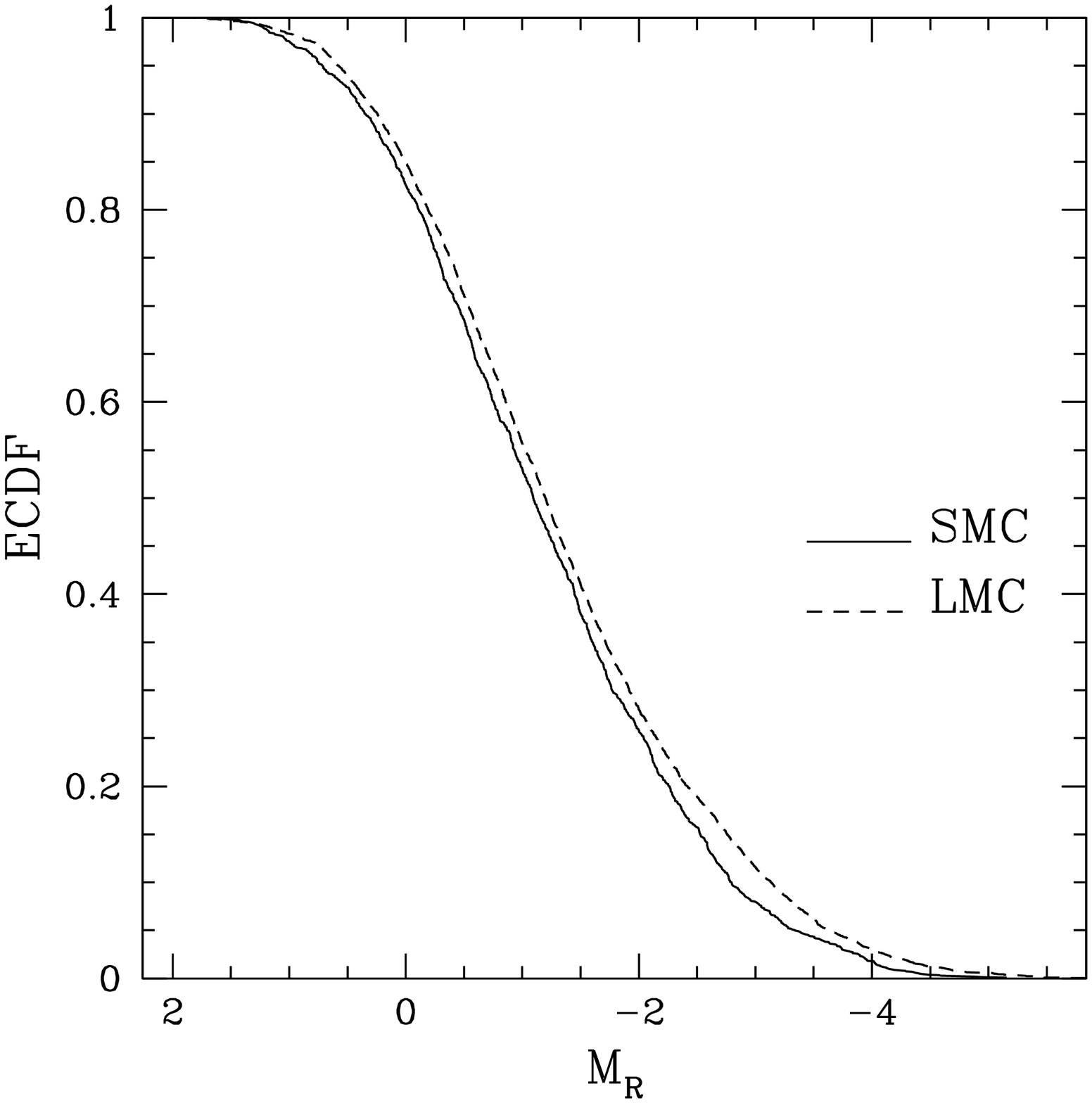}
\plottwo{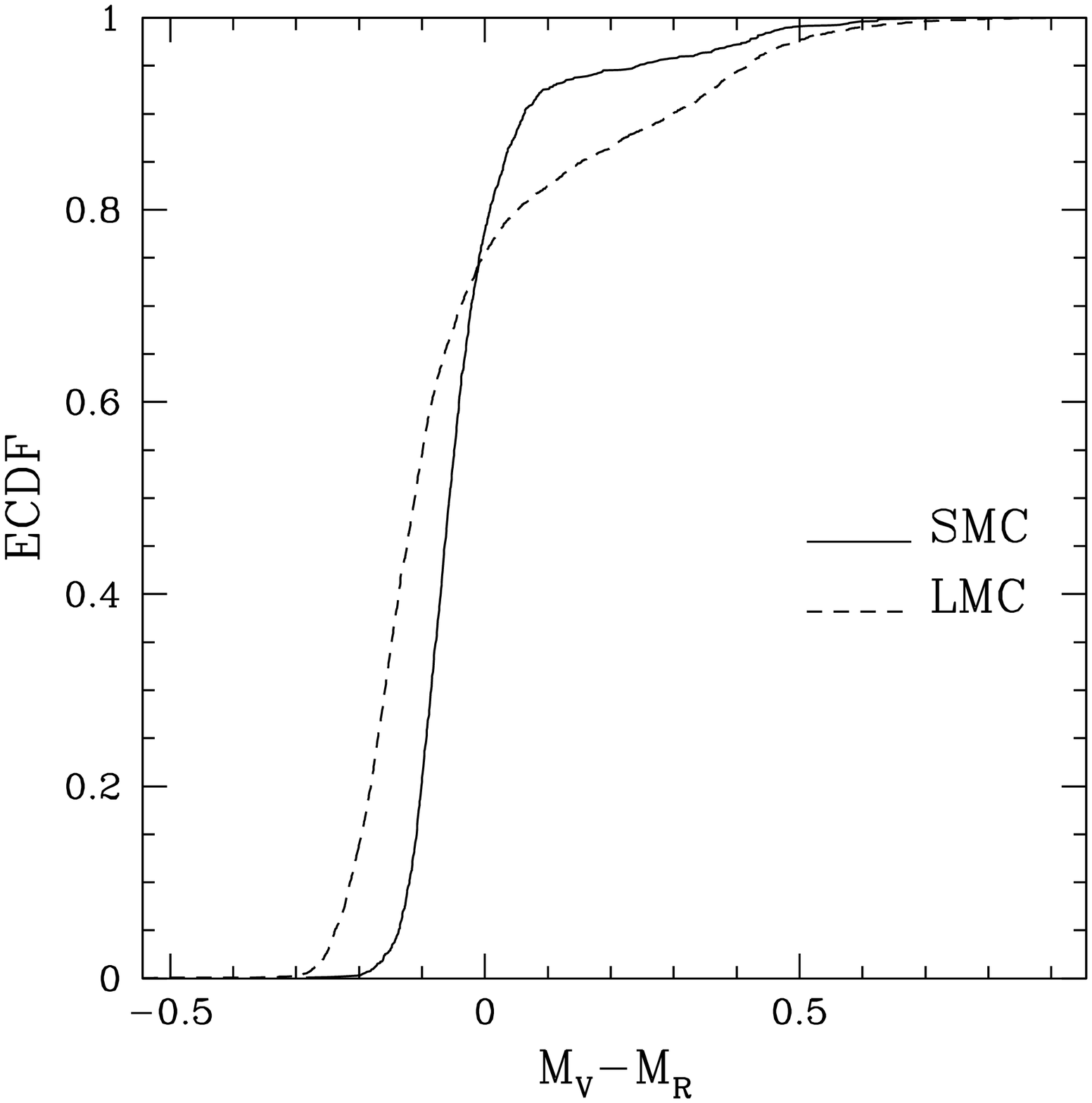}{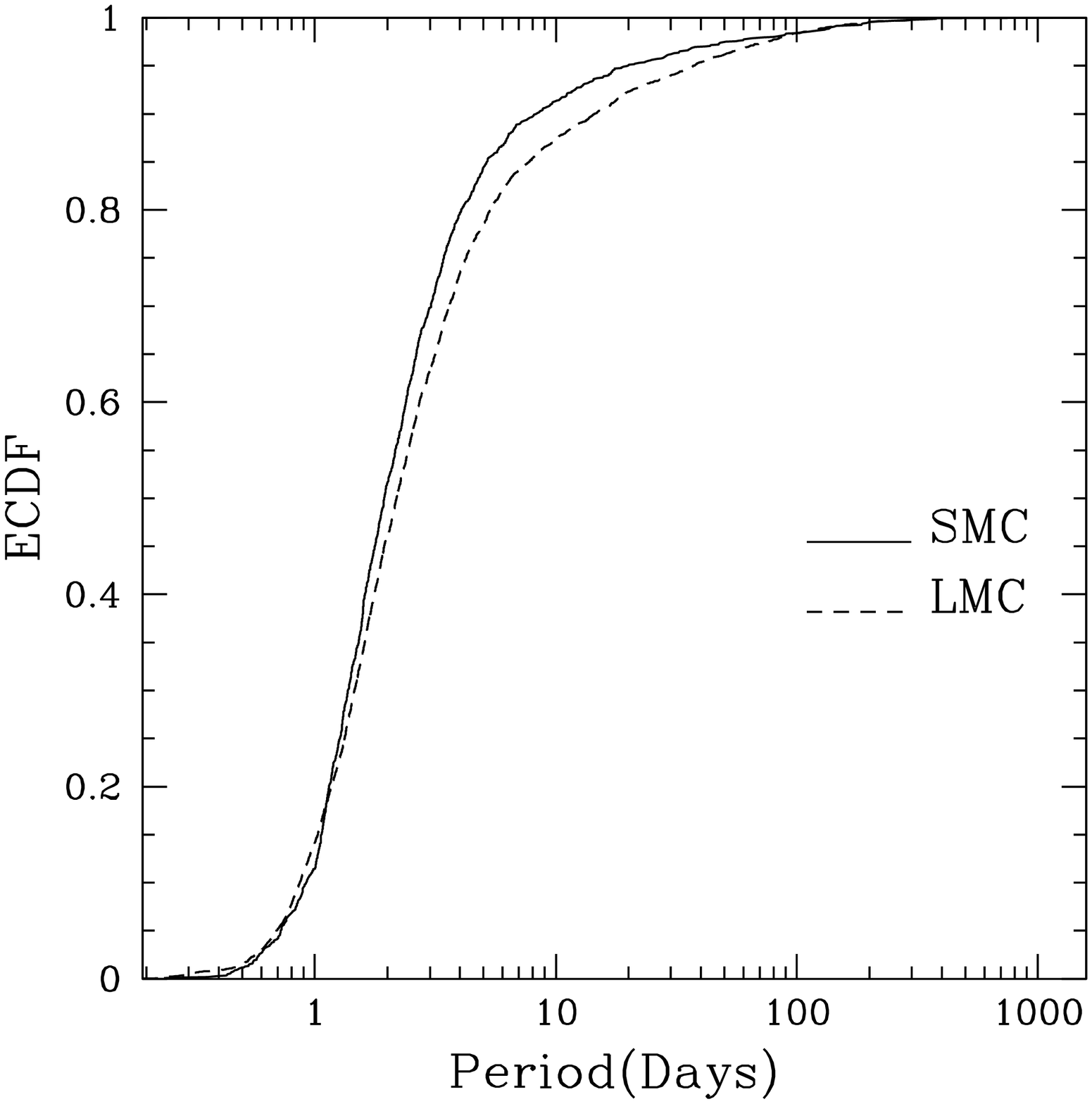}
\end{center}
\caption{ECDFs for the distribution of $M_V$ (above left), $M_R$ (above 
right), $M_V-M_R$ (Below left) and $P$ (below right) for the SMC (continuous 
line) and the LMC (dashed line).}
\label{fig:ecdf}
\normalsize
\end{figure}

Although the basic features of the CMD and the Color 
Period Diagram are the same for LMC and SMC, there are some differences, 
shown by Figure \ref{fig:smcvslmc} and Table \ref{tab:smcvslmc}.
\tabletypesize{\footnotesize}
\begin{deluxetable}{ccccccc}
\tablecolumns{7}
\tablewidth{0pc}
\tablecaption{Summary of long period EBs in the SMC and LMC.
\label{tab:smcvslmc}}
\tablehead{
\colhead{Galaxy} &
\colhead{Total} &
\colhead{Young stars}\tablenotemark{a} &
\colhead{Evolved}\tablenotemark{b} &
\colhead{Long Period}\tablenotemark{c} &
\colhead{Long Period young stars}\tablenotemark{a~c} &
\colhead{Long Period evolved stars}\tablenotemark{b~c}
}
\startdata
LMC &
$4634$ &
$3760 (81\%)$ &
$874 (19\%)$ &
$356$ &
$23 (6\%)$ &
$333 (94\%)$ \\
SMC & 
$1508$ &
$1412 (94\%)$\tablenotemark{d} &
$96 (6\%)$\tablenotemark{d} &
$75$ &
$23 (31\%)$ &
$52 (69\%)$ \\
\enddata
\tablenotetext{a}{Defined as $\vr<0.2~\mathrm{mag}$.}
\tablenotetext{b}{Defined as $\vr>0.2~\mathrm{mag}$.}
\tablenotetext{c}{Defined as $P>20\mathrm{d}$.}
\tablenotetext{d}{One SMC EB has no valid $V_{\mathrm{MACHO}}$ data, hence the sum of the young star and evolved star numbers for the SMC is $1508$.}
\end{deluxetable}
The fraction of blue ($\vr<0.2~\mathrm{mag}$) EBs is higher in the 
SMC than in the LMC.
More striking, the fraction of blue long period 
($\vr<-0.2~\mathrm{mag}$, $P>20\mathrm{d}$) 
EBs is much higher in the SMC than in the LMC. 
To further investigate the differences between the two samples we carried 
out a Kolmogorov-Smirnov (KS) test on the distributions of the absolute 
magnitudes $M_{V}$ and $M_{R}$, their difference $M_V-M_R$, and the periods
$P$ of the two samples.
We adopted a distance modulus of of $18.88$ for the SMC \citep{dolphin01} and
$18.5$ for the LMC \citep{marel02}.
Magnitudes and colors were dereddened using $\langle A_V\rangle=0.64~\mathrm{mag}$, $\langle E(V-R)\rangle=0.128~\mathrm{mag}$ for the LMC and  
$\langle A_V\rangle=0.3~\mathrm{mag}$, $\langle E(V-R)\rangle=0.06~\mathrm{mag}$ for the SMC before subtracting the distance moduli.
The Empirical Cumulative Distribution Functions (ECDFs) of these quantities
are shown in Figure \ref{fig:ecdf}.
A KS test confirms that the distributions of $P$ and $M_V-M_R$ are different
at $>99.9\%$ confidence level; for the distributions of $M_V$ and $M_R$ the KS 
test gives a probability of $\sim 99.4\%$ and $\sim 98.5\%$ respectively for 
them being different.
The $M_V-M_R$ plot shows that EBs in the SMC tend to be bluer, not surprising 
given that a higher percentage of objects belong to the young star CMD 
region in the SMC than in the LMC.
The period distributions show that the SMC EBs have on average shorter 
periods, again not surprising given the much higher percentage of evolved 
systems in the LMC than in the SMC and the fact that evolved systems have 
on average higher periods than the young systems (as shown by the Color
Period Diagrams in Figure \ref{fig:smcvslmc}).
\section{Cross correlation with the OGLE-II samples}
\label{sec:crossogle}
The EBs in our samples have been cross correlated with the corresponding 
OGLE-II samples.
Stars in the samples were identified if their right ascension (RA) and 
declination (DEC) differed by less than $27.2\arcsec$ and if their periods 
differed by $<1\%$.
We used a very large search radius to be conservative. 
The astrometric precision for both surveys is typically $\sim 1\arcsec$, 
but a few stars which had much larger differences in RA and/or 
DEC turned out to be matches upon inspection of their periods:
in particular we found in the LMC $32$ matches with a position difference bigger 
than $10\arcsec$ and $2$ matches with a position difference bigger than
$20\arcsec$.
However most of the matches were within narrower radii: for the LMC roughly half of 
the matches ($534$ out of $1236$) were found within of $\sim 2\arcsec$, compatible with the astrometric precision of both MACHO and OGLE surveys; almost all of them ($1019$) were within 
$\sim 4\arcsec$.
We tested the robustness of our method of finding matches by investigating the 
probability for two periodic objects with a period difference of $<1\%$
to be within $27.2\arcsec$ of each other.
To do this we selected a random sample of $5000$ objects out of the 
$\sim 66000$
periodic ones found by MACHO in the LMC and we counted the frequency of pairs of
objects with both periods from the red and the blue lightcurves (as found by Supersmoother)
differing by $<1\%$ and positions within $27.2\arcsec$.
Most of the matches we found were due to the same object being observed in different tiles and only in one case did we find a possibly genuine match; we thus conclude that the probability of two objects being erroneously classified as a match is 
$\sim 0.02\%$ and therefore our method of finding matches is robust. 
\subsection{The Large Magellanic Cloud sample}
\begin{figure}\footnotesize
\plotone{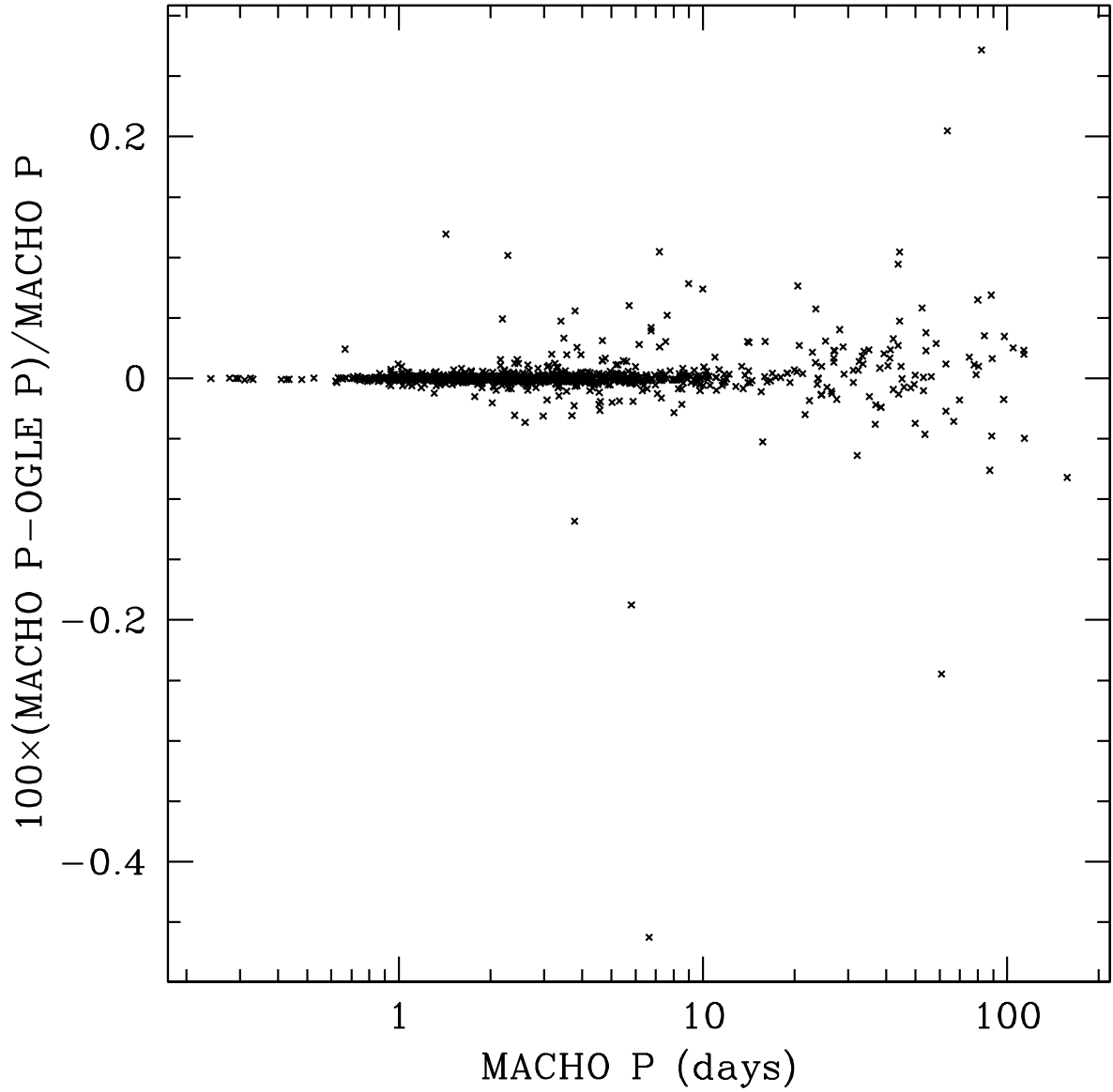}
\caption{Percentage difference for MACHO vs. OGLE-II period for the 1236 
OGLE-II matches in the LMC sample.
}
\label{fig:crossper}
\end{figure}

\begin{figure}
\plottwo{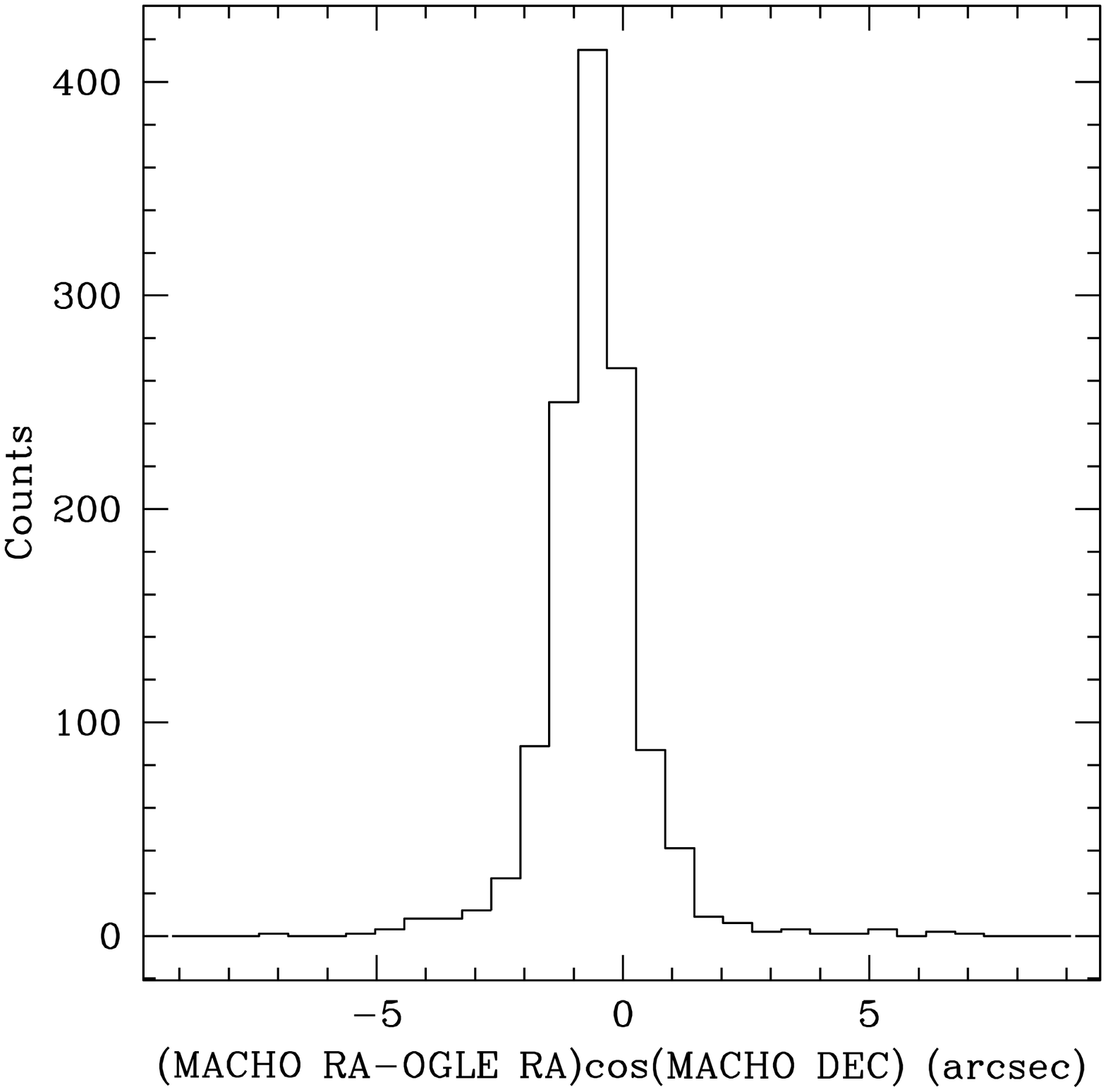}{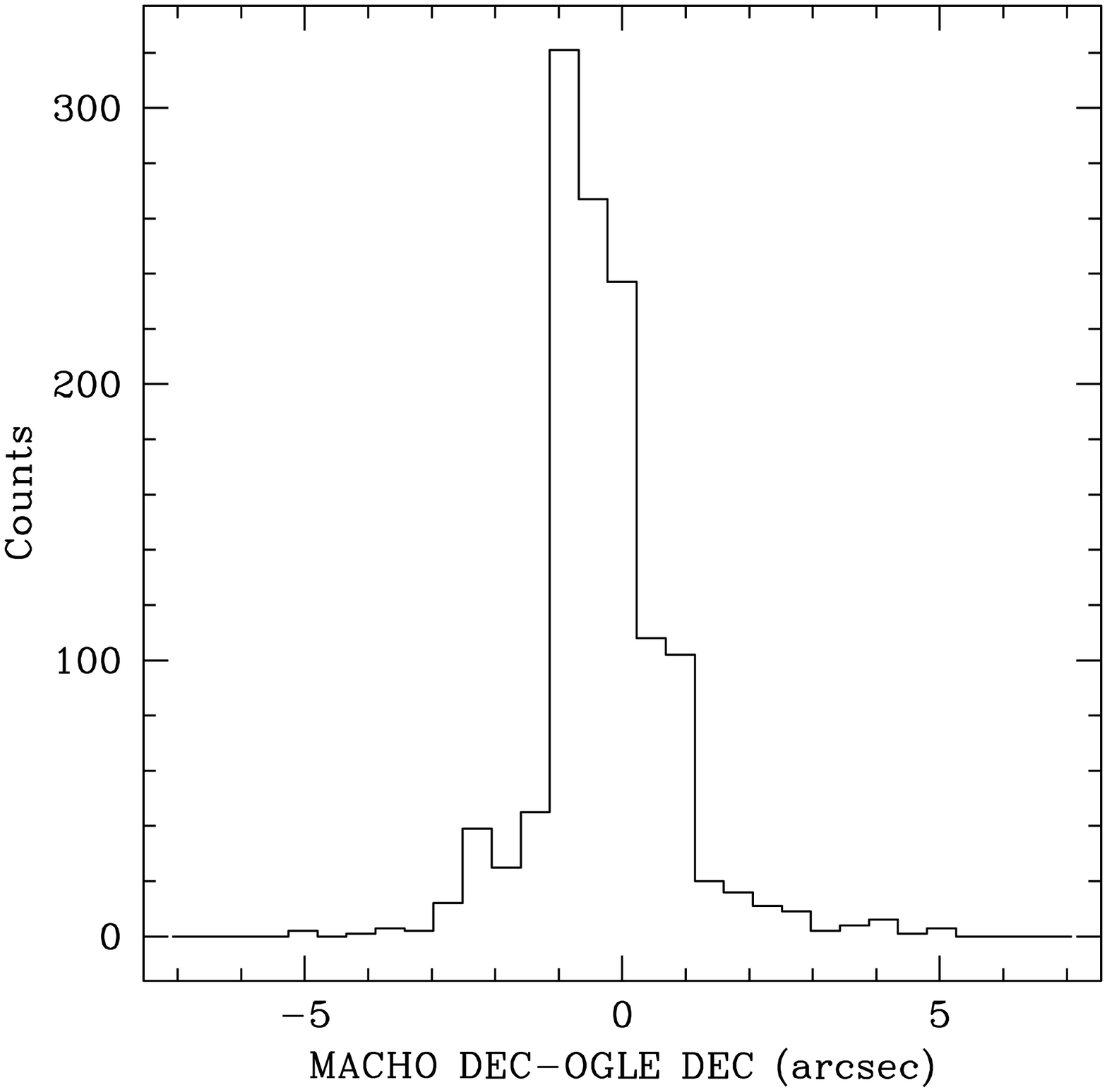}
\caption{Left Panel: histogram of the differences between Right Ascensions 
for $1236$ MACHO OGLE-II matches in the LMC.
Right Panel: histogram of the differences between declinations.
The bin size is equal to $1/30$ of the range of the differences in both
cases.
}
\label{fig:diffposhist}
\end{figure}

\begin{figure}
\footnotesize
\begin{center}
\plotone{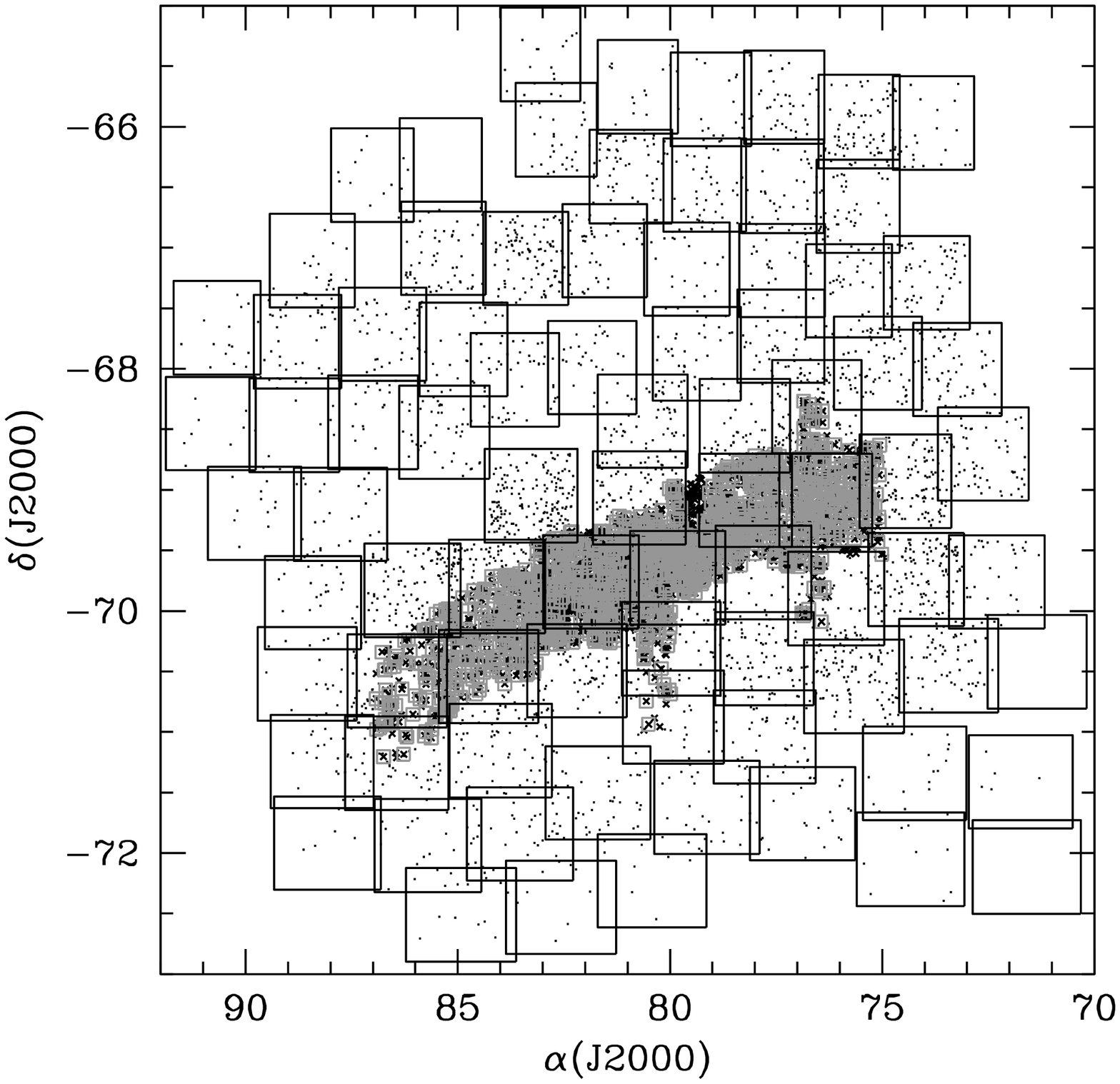}
\end{center}
\caption{Cross correlation between MACHO and OGLE-II LMC samples: points 
represent MACHO stars, crosses OGLE stars and gray empty boxes the matches.}
\label{fig:pos}
\normalsize
\end{figure}

\begin{figure}
\footnotesize
\begin{center}
\plotone{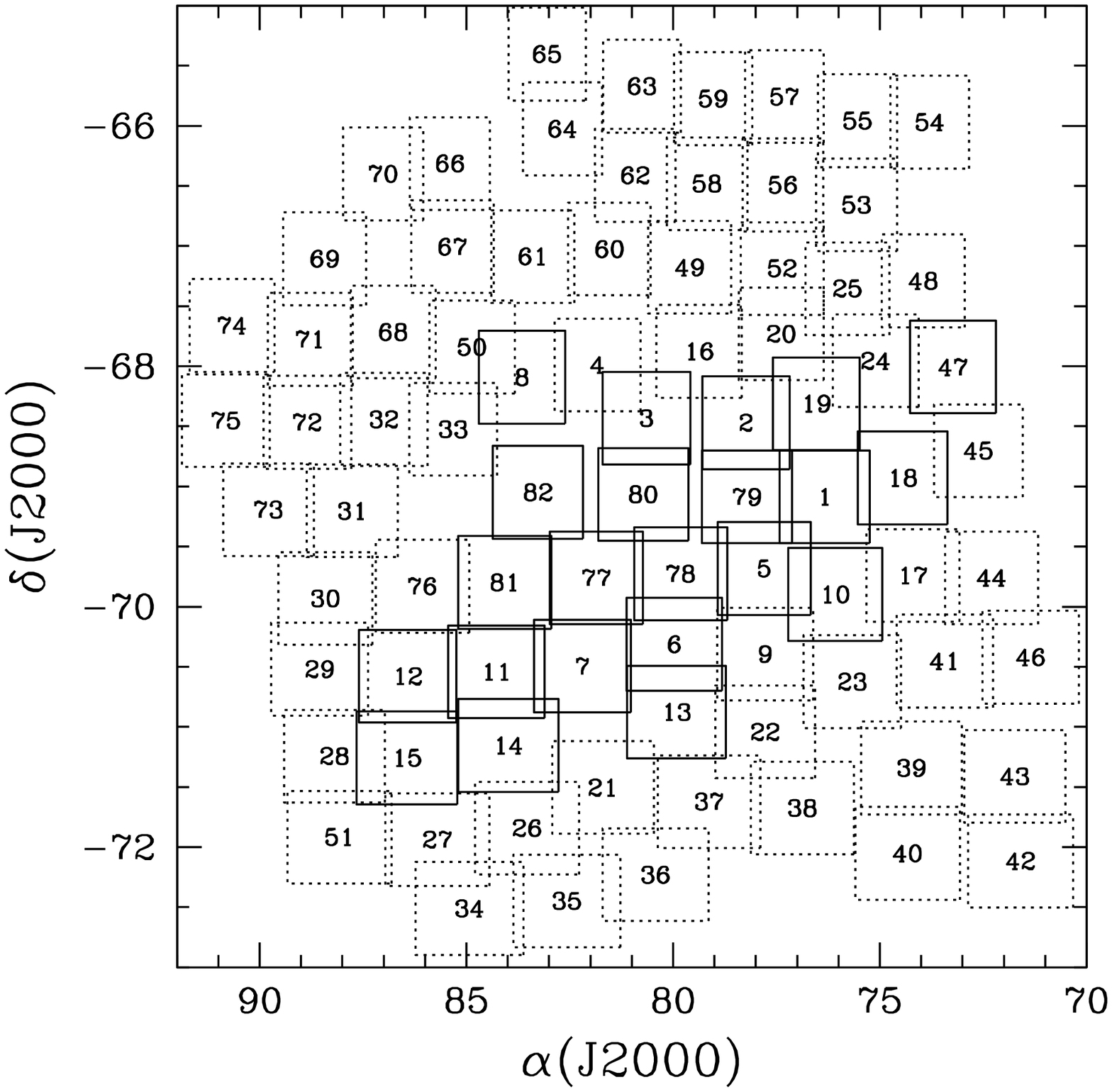}
\end{center}
\caption{MACHO LMC field numbers.}
\label{fig:machofields}
\normalsize
\end{figure}

Our search produced $1236$ matches in the LMC.
The MACHO and OGLE-II periods agree to high accuracy as shown in Figure 
\ref{fig:crossper}; usually much better than the $1\%$ cut we imposed.
Histograms of position differences are shown by Figure \ref{fig:diffposhist}.
Both panels show the entire span of the differences; the differences in Right Ascension
range from $\sim -20\arcsec$ to $\sim +20\arcsec$ but the left panel shows 
this range multiplied by the cosine of the declination ($\sim -70^\circ$) which 
gives a range from $\sim -7\arcsec$ to $\sim +6\arcsec$.
\normalsize
\normalsize
The sky coverage of the two surveys was different: OGLE-II, from which the 
sample was derived, covered about $4.5\Box^\circ$ in the central region 
of the LMC whereas the sky coverage of MACHO was larger.
Figure \ref{fig:pos} shows the positions of the EBs in both catalogues and 
Figure \ref{fig:machofields} shows the corresponding MACHO field number; the 
fields at the center of the LMC are drawn with continuous lines, the ones 
at the periphery with dashed lines.
\subsection{Comparison between the center and the periphery of the LMC}
\begin{figure}
\footnotesize
\begin{center}
\plottwo{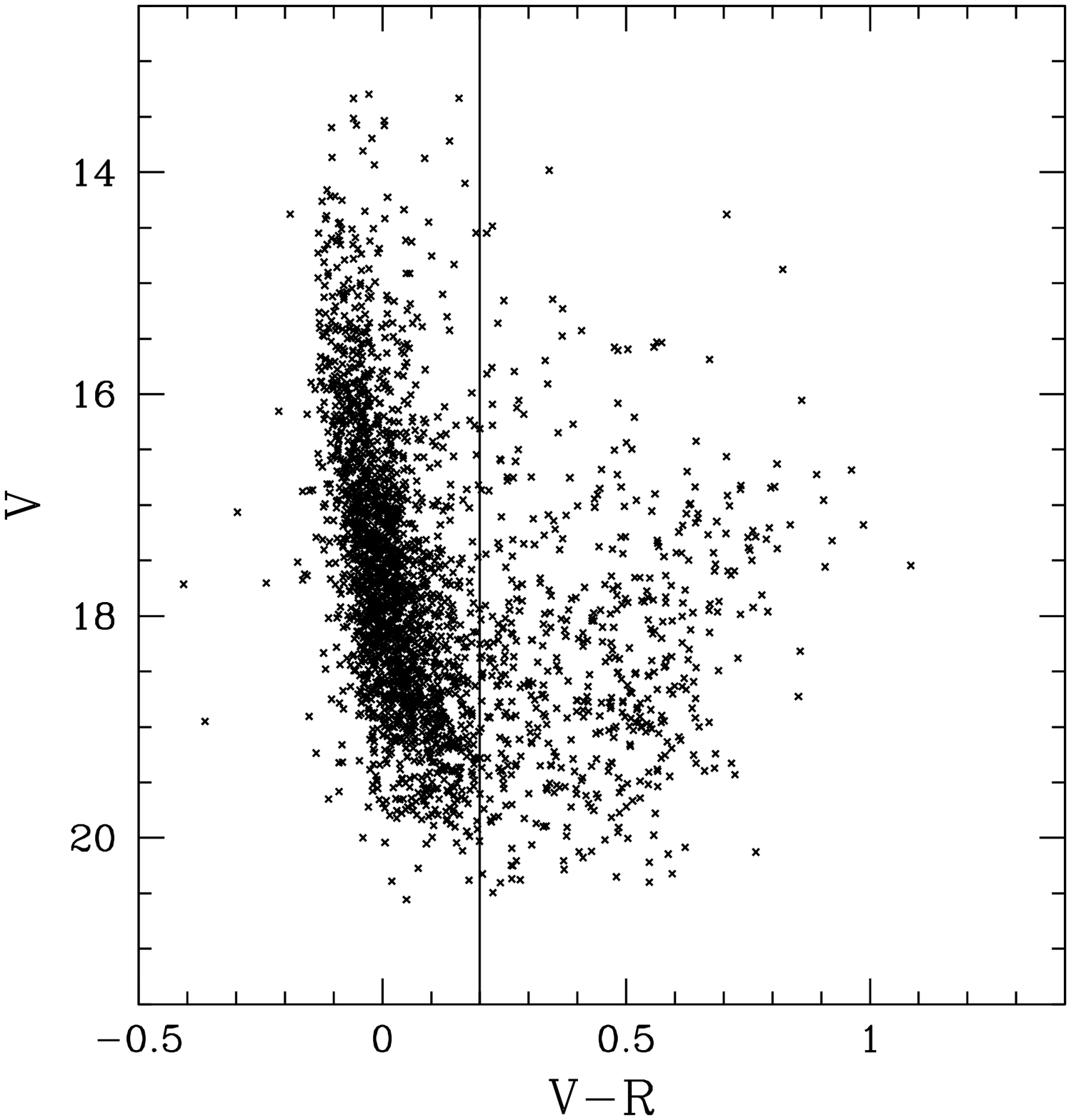}{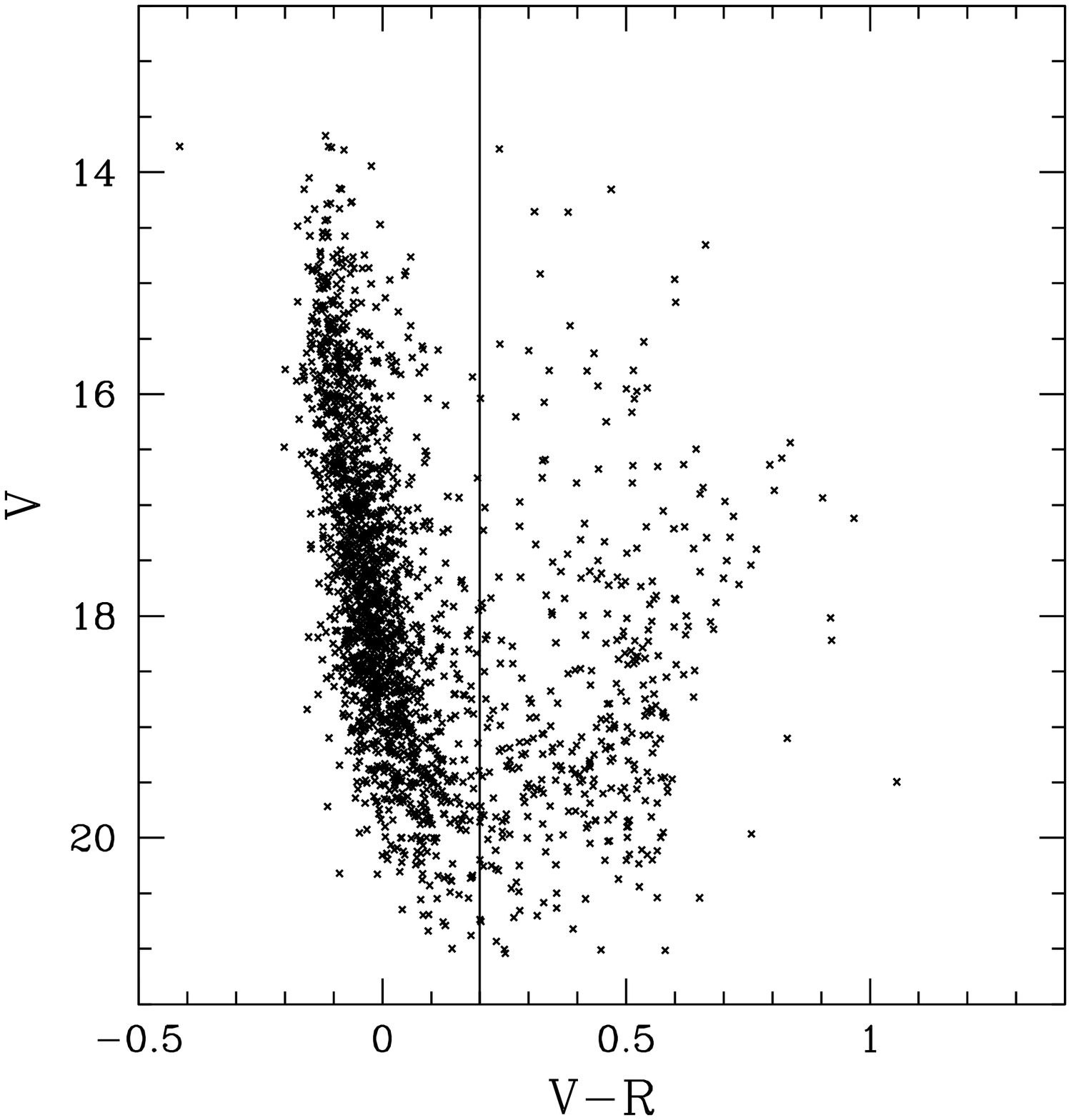}
\plottwo{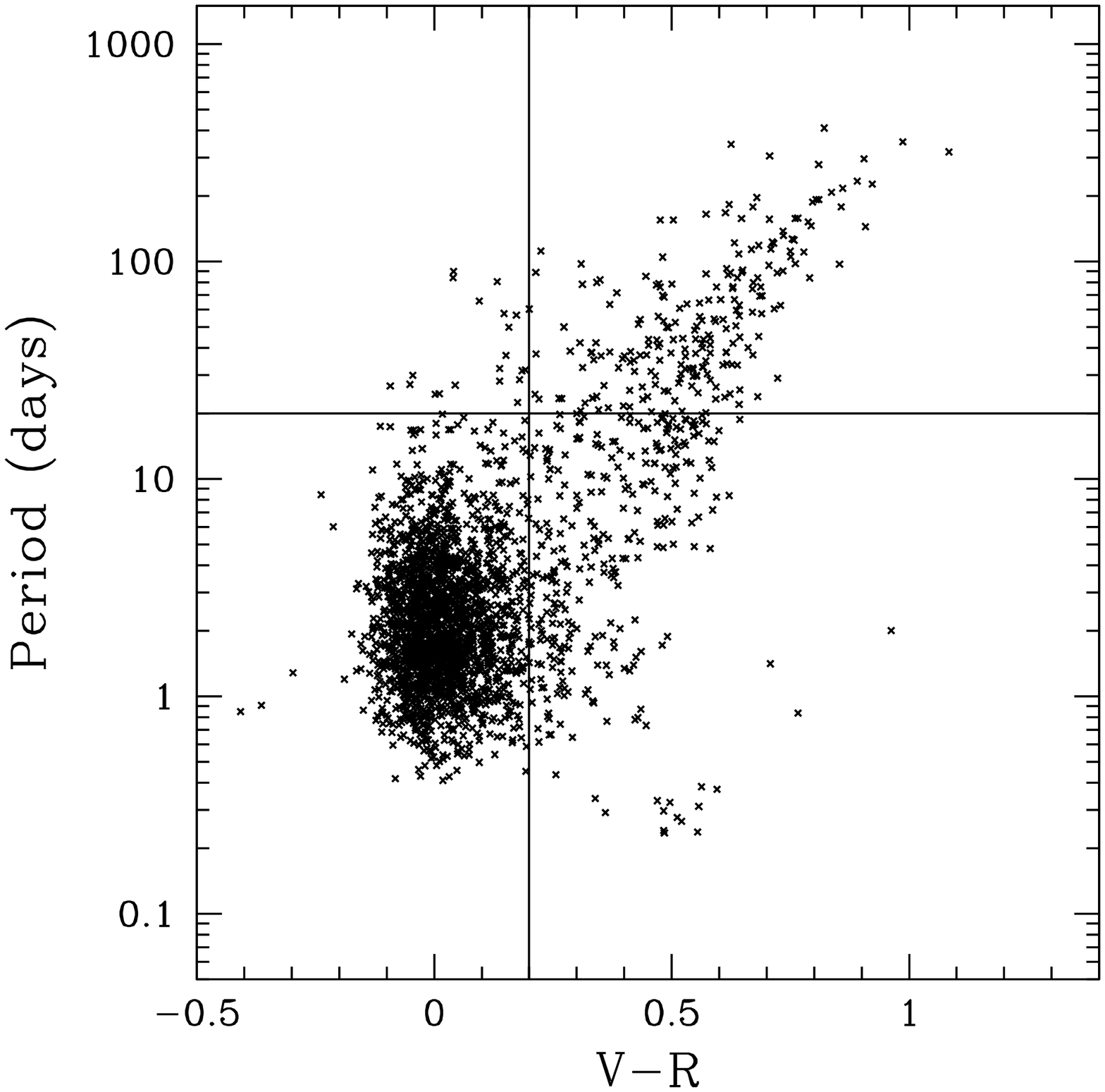}{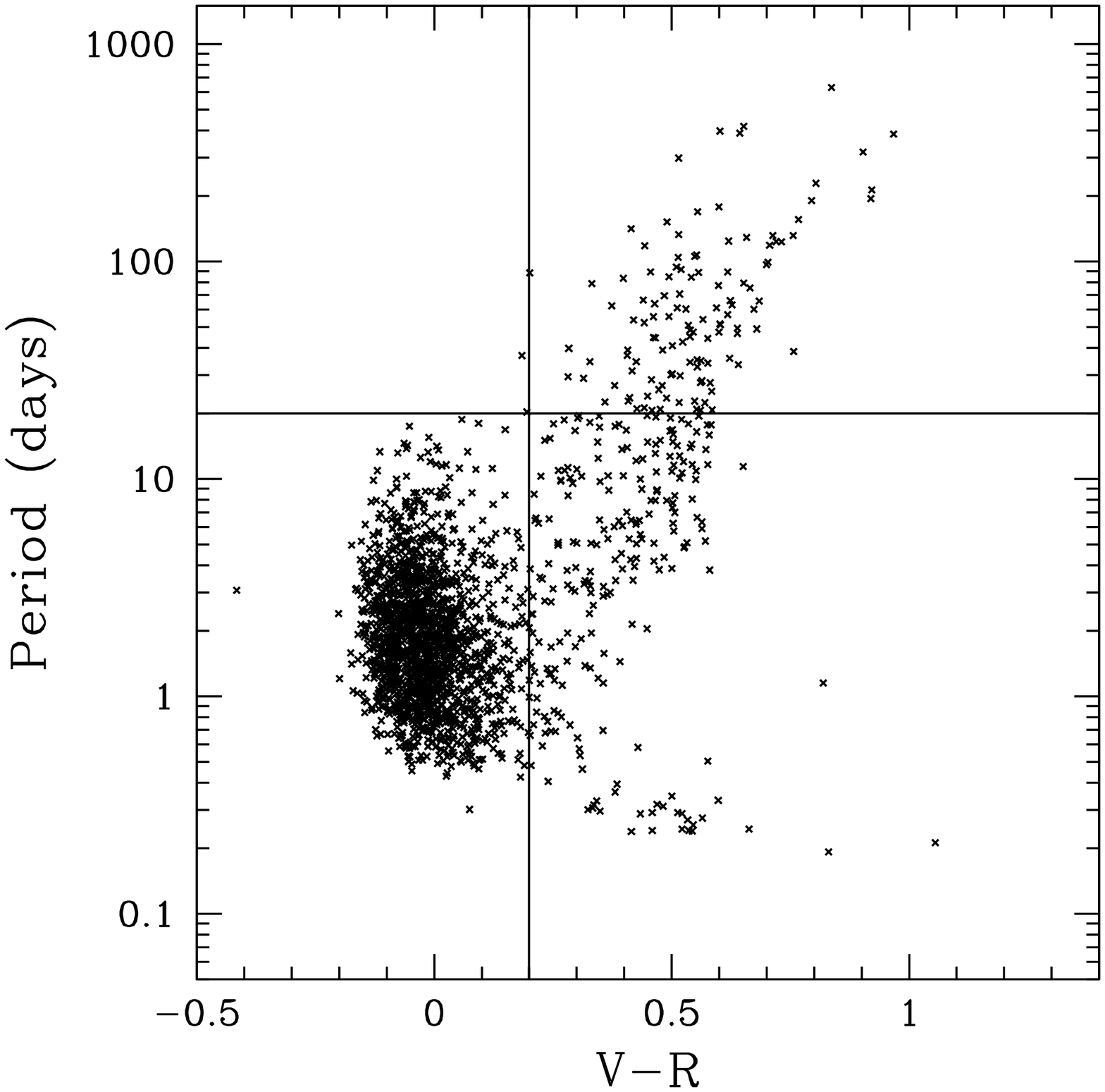}
\end{center}
\caption{Upper Left Panel: CMD for $2620$ EBs in the center of 
the LMC.
Upper Right panel: CMD for $2014$ EBs at the periphery of the 
LMC.
The CMDs suggest a more continuous transition from the young star region to 
the evolved star region in the center than in the periphery, especially for 
$V<19~\mathrm{mag}$.
\newline
Lower panels: Color Period Diagrams for the same populations.
Lower Left Panel: center.
Lower Right panel: periphery.
The Color Period Diagrams reveal the presence of a long period $(20-100\mathrm{d}$), 
relatively unevolved ($\vr\sim 0.2~\mathrm{mag}$) population in the center but not in the periphery.
}
\label{fig:cmdcenter}
\normalsize
\end{figure}

\begin{figure}
\footnotesize
\begin{center}
\plottwo{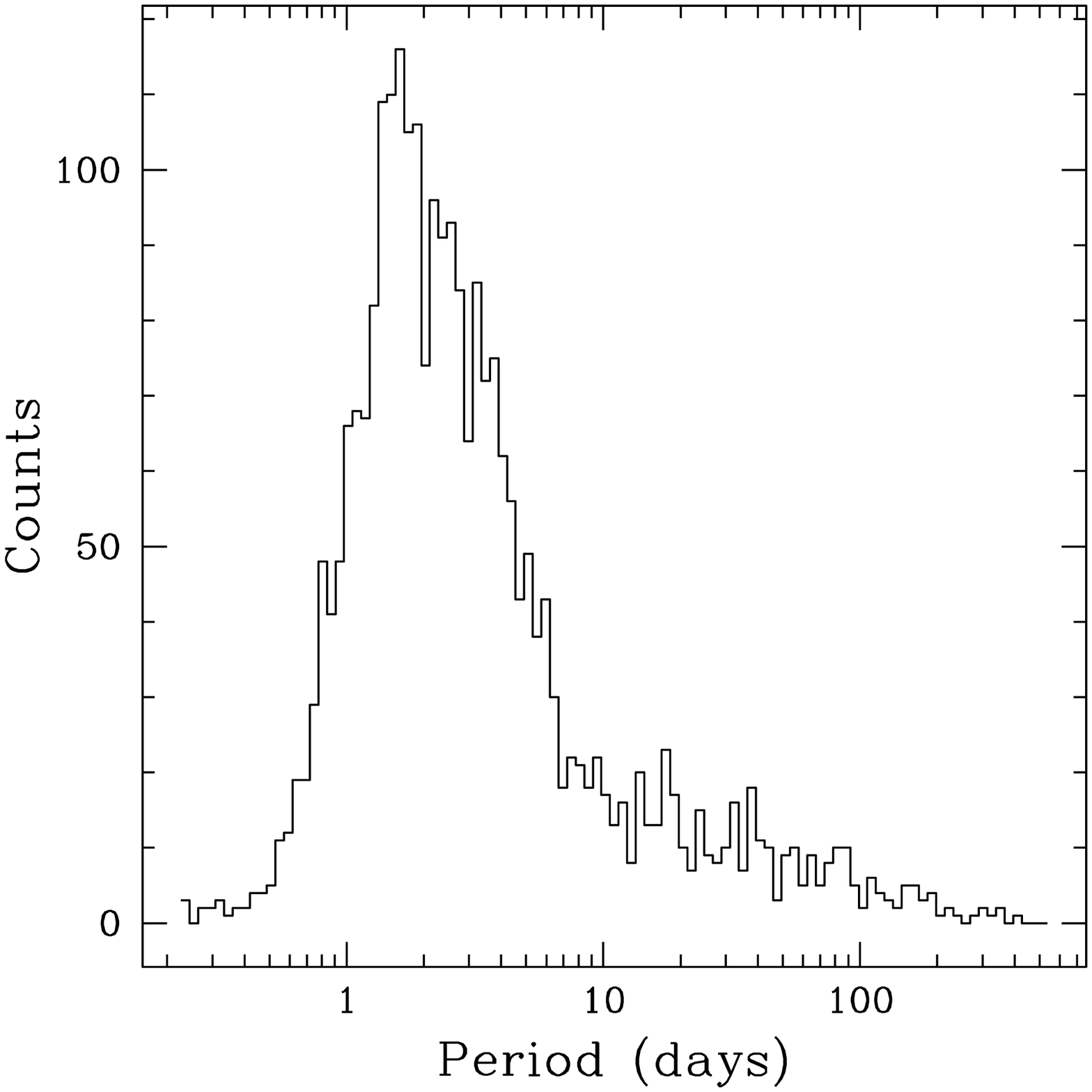}{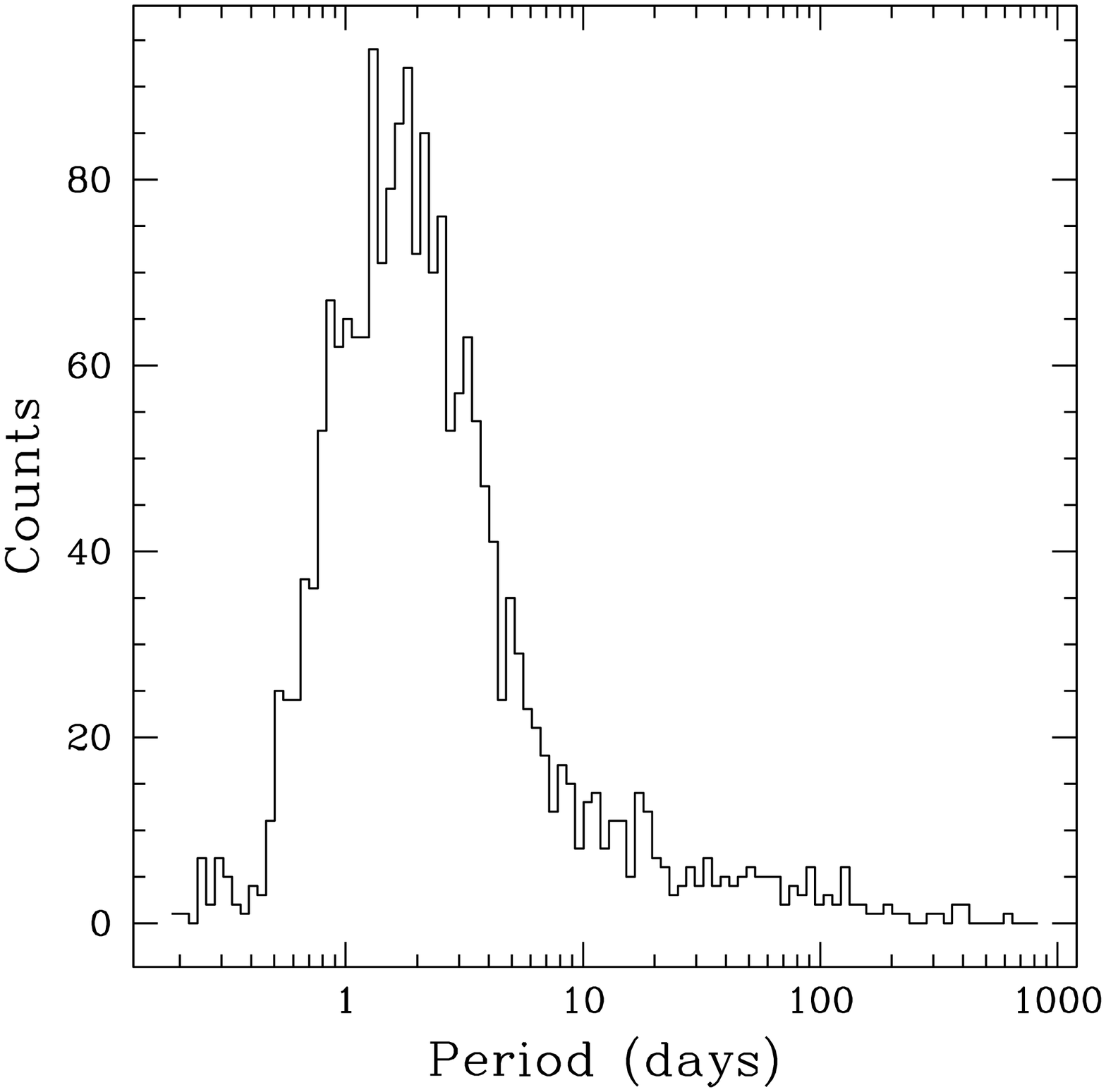}
\plottwo{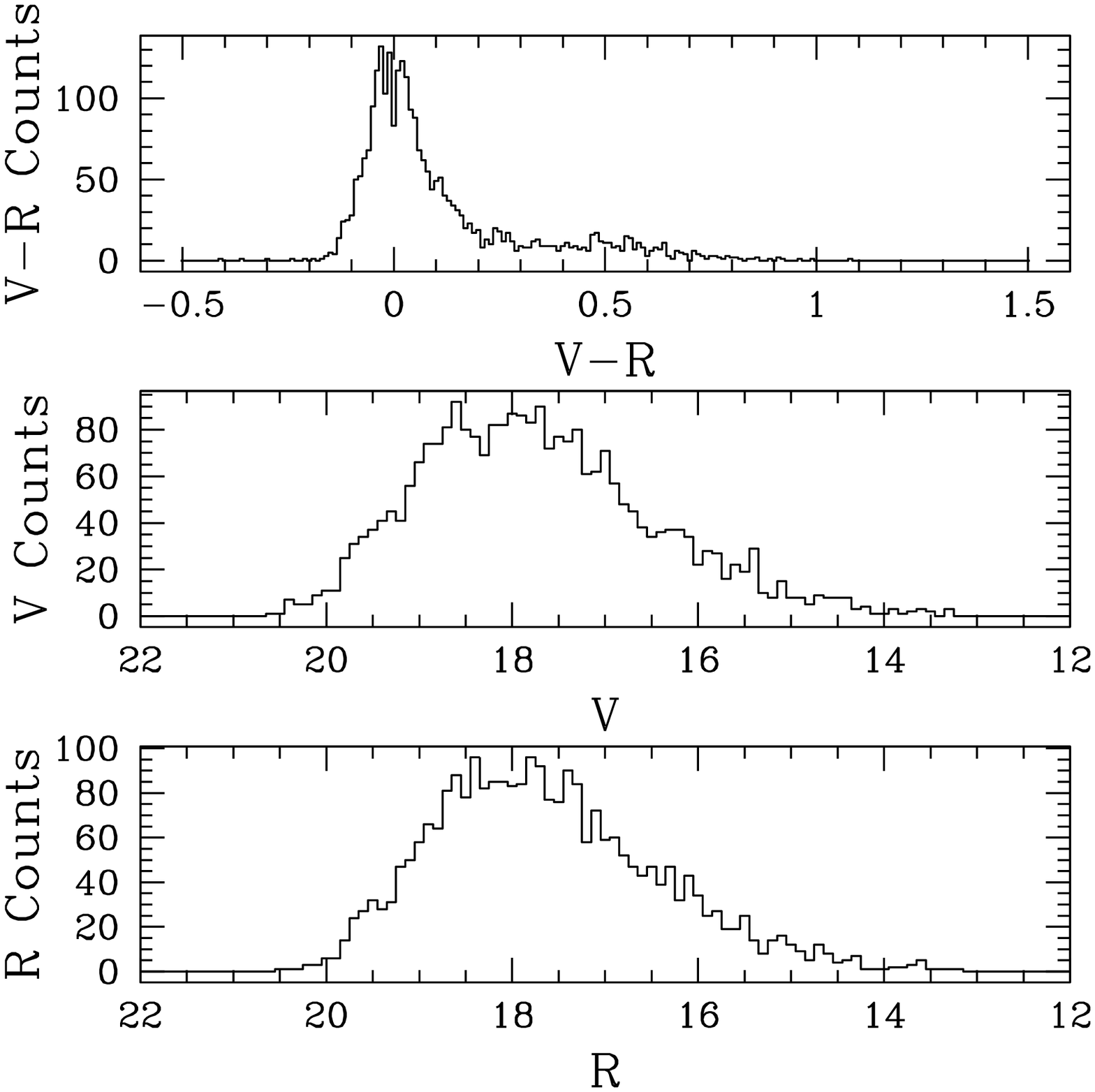}{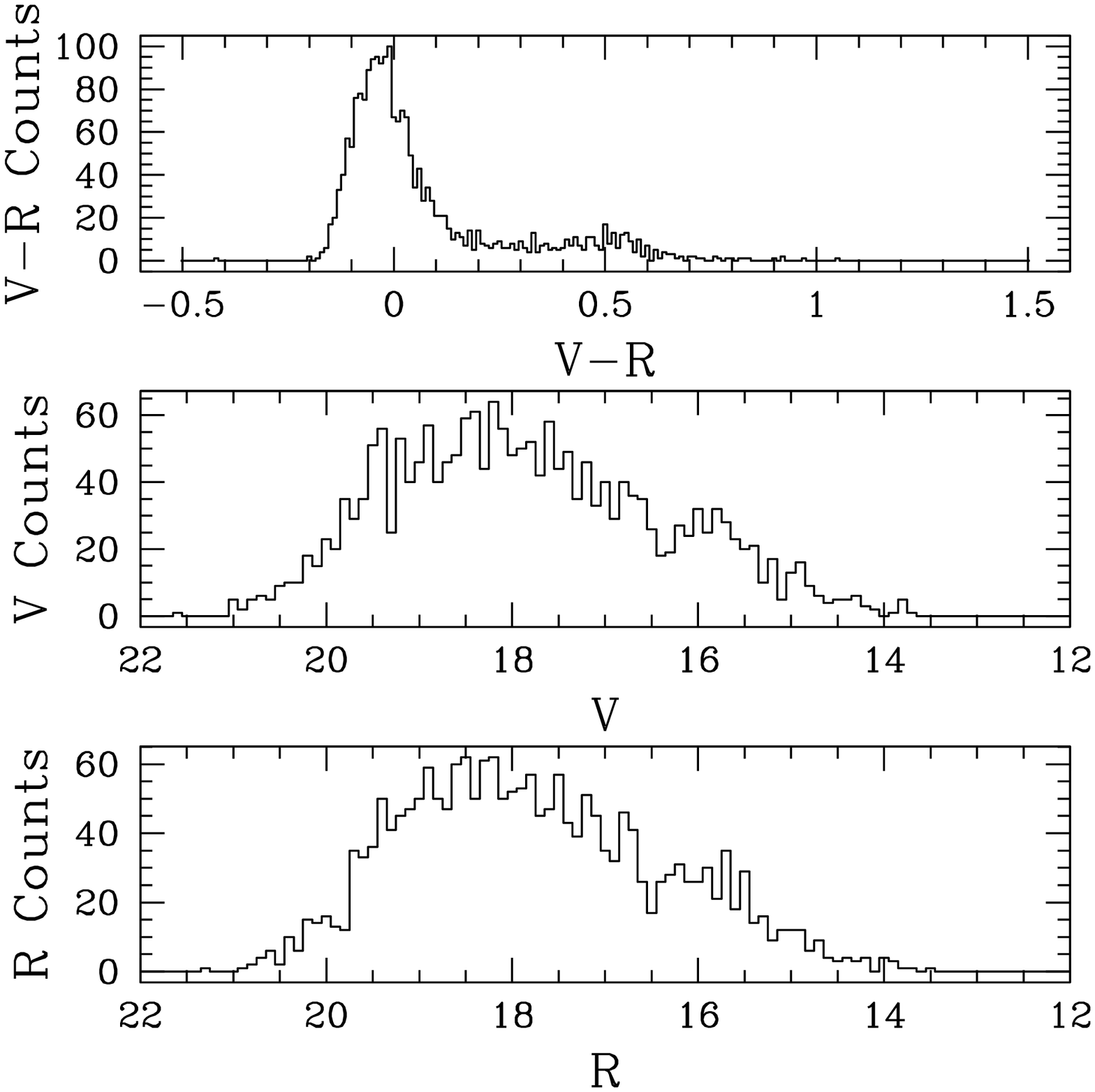}
\end{center}
\caption{Upper Left Panel: Period histogram for $2620$ EBs in the center of the LMC.
Upper Right panel: period histogram for $2014$ EBs at the periphery of the LMC.
The size of the bins is $\sim 1/100$ of the span of the logarithms of the
periods.
\newline
Lower panels: Magnitude and Color histograms.
Lower Left Panel: center. 
Lower Right panel: periphery.
The bin size is $0.1~\mathrm{mag}$ for the $V$ and $R$ histograms and 
$0.01~\mathrm{mag}$ for the $\vr$ one.}
\label{fig:perhistcenter}
\normalsize
\end{figure}

In view of the different sky coverage of the MACHO and OGLE-II surveys, 
it is interesting to analyze separately the stars in MACHO fields covering 
the center of the LMC (which roughly correspond to the OGLE-II sky coverage) 
and the stars in the MACHO fields at the periphery.
Figure. \ref{fig:machofields} shows that the fields in the center are 1, 2, 
3, 5, 6, 7, 8, 10, 11, 12, 13, 14, 15, 18, 19, 47, 77, 78, 79, 80, 81 and 82.
There are $2620$ EBs in the center and $2014$ in the periphery.
Figure \ref{fig:cmdcenter} 
shows the CMD and the Color Period Diagram for the center and the 
periphery of the LMC respectively; Figure \ref{fig:perhistcenter} 
shows the histograms for the magnitudes, the color and the period.
The figures reveal several differences between the two samples; to check 
these we performed a KS test for $V$, $\vr$ and $P$ and found that at high 
confidence level ($>99.9\%$)  the distributions are different. 
These differences are statistically significant but not large enough to be 
considered astrophysically important.
In particular the difference in period distribution could be attributed to 
the lower sampling in the outer fields, making it less likely to detect the
longer period  EBs.
The sampling might also be the cause of the periphery having a higher 
percentage of bright EBs, since these are easier to find with fewer epochs.
\subsection{Discussion of OGLE-II MACHO comparison}
\begin{figure}
\footnotesize
\plotone{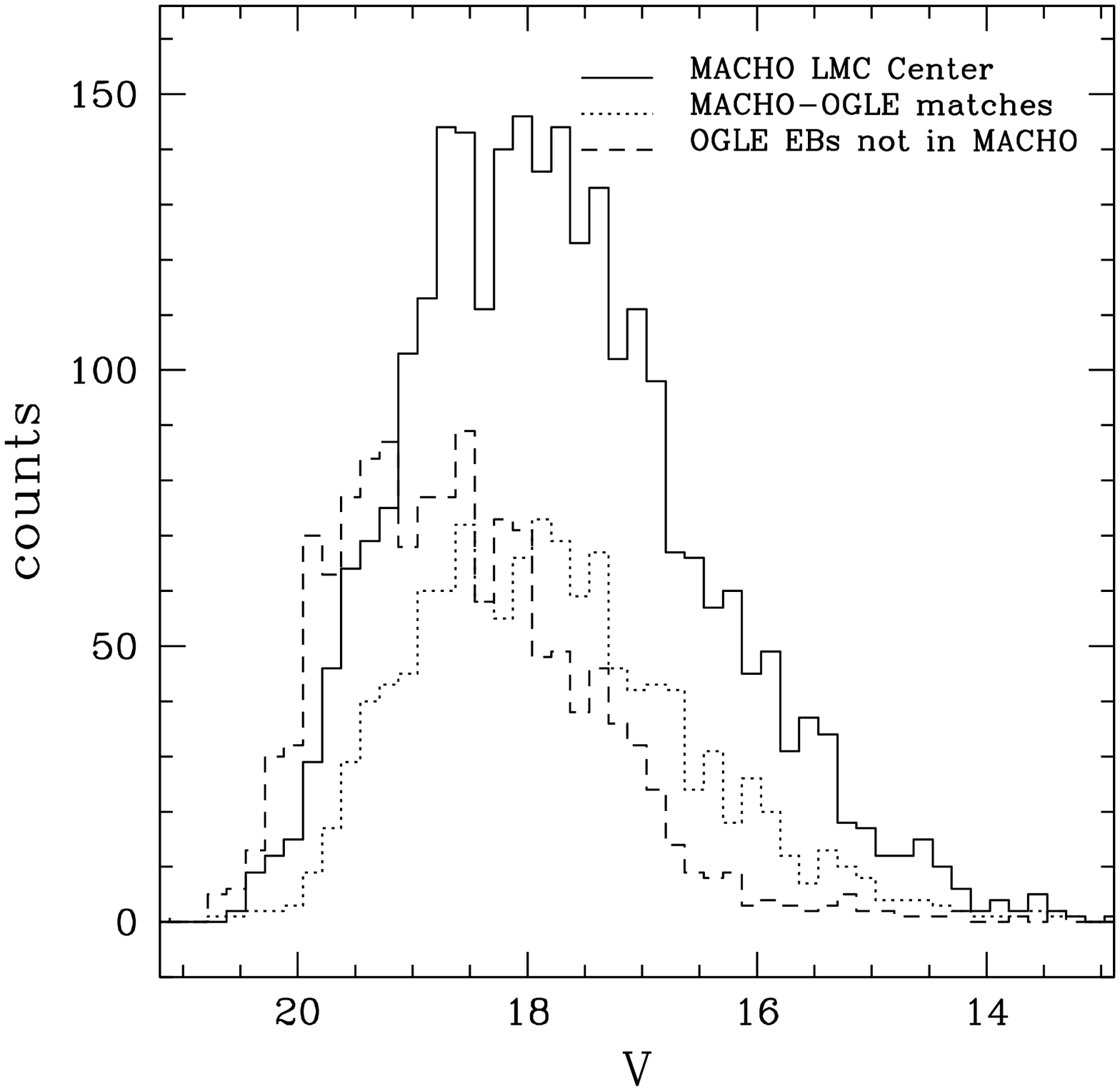}
\caption{$V$ distribution for MACHO $2620$ EBs in the central region of the 
LMC (continuous line), $1198$ OGLE-II MACHO matches (dashed line), and $1327$ 
OGLE-II EBs without MACHO counterparts (long dashed line).
The figure shows that OGLE-II EBs with MACHO counterparts are on average 
brighter than the ones without, and the shape of their $V$ distribution 
more closely resembles the MACHO $V$ distribution.
This is confirmed by a KS test.}
\label{fig:v_macho_and_ogle}
\normalsize
\end{figure}

We finally investigated why we did not find more matches with OGLE-II.
The most important reason, we think, is the fact that the techniques 
employed in assembling the samples are different: the OGLE team built their 
samples via neural networks \citep{wyr03,wyr04}.
The two surveys have roughly comparable limiting magnitudes, $V\sim 21.5~\mathrm{mag}$; nevertheless we checked how the performance of the two surveys varied with magnitude. 
We compared the distribution of the $V$ magnitudes for the $2620$ EBs in the central region of the LMC in our sample with those of $1198$ matches and the $1327$ EBs from OGLE-II without MACHO counterparts (these two numbers do not add up to $2580$, the size of the OGLE-II sample, because for some EBs the $V$ magnitude was not reported).
The histograms of these three distributions are shown in Figure 
\ref{fig:v_macho_and_ogle}.
The figure shows that OGLE-II EBs with MACHO counterparts, peaking at 
$V\sim 18~\mathrm{mag}$ like the MACHO sample, are on average 
brighter than the ones without MACHO counterparts, which peak at 
$V\sim 19.5~\mathrm{mag}$.
The shape of the $V$ distribution of OGLE-II EBs with MACHO counterparts 
much more closely resembles the MACHO $V$ distribution; a KS test gives a 
probability of the two distributions being the same of $\sim 46\%$.
The distributions of MACHO $V$ and OGLE-II $V$ without MACHO counterparts, 
as well as those of these two OGLE-II populations are, on the other hand, 
shown to be different at $>99.9\%$ confidence level.
Both distributions vanish at $V\sim 20.5~\mathrm{mag}$, showing that their 
limiting magnitudes are comparable.
\par
\begin{figure}
\footnotesize
\plotone{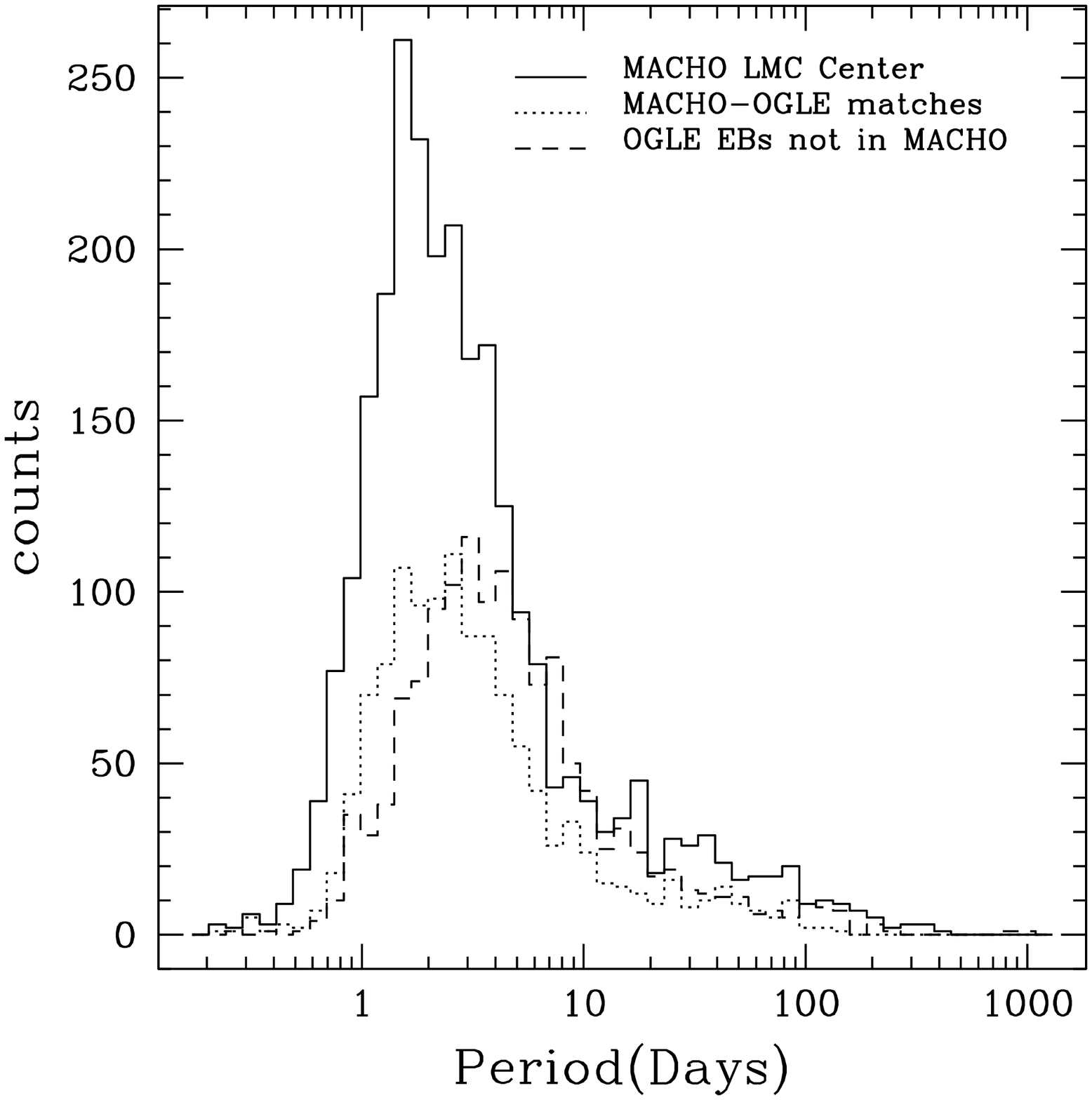}
\caption{Period distribution for $2620$ MACHO EBs in the central region of 
the LMC (continuous line), $1198$ OGLE-MACHO matches (dashed line), and $1327$ 
OGLE-II EBs without MACHO counterparts (long dashed line).
The figure shows that OGLE-II EBs with MACHO counterparts have on average 
shorter periods than the ones without.}
\label{fig:per_macho_and_ogle}
\normalsize
\end{figure}
\begin{figure}
\footnotesize
\plotone{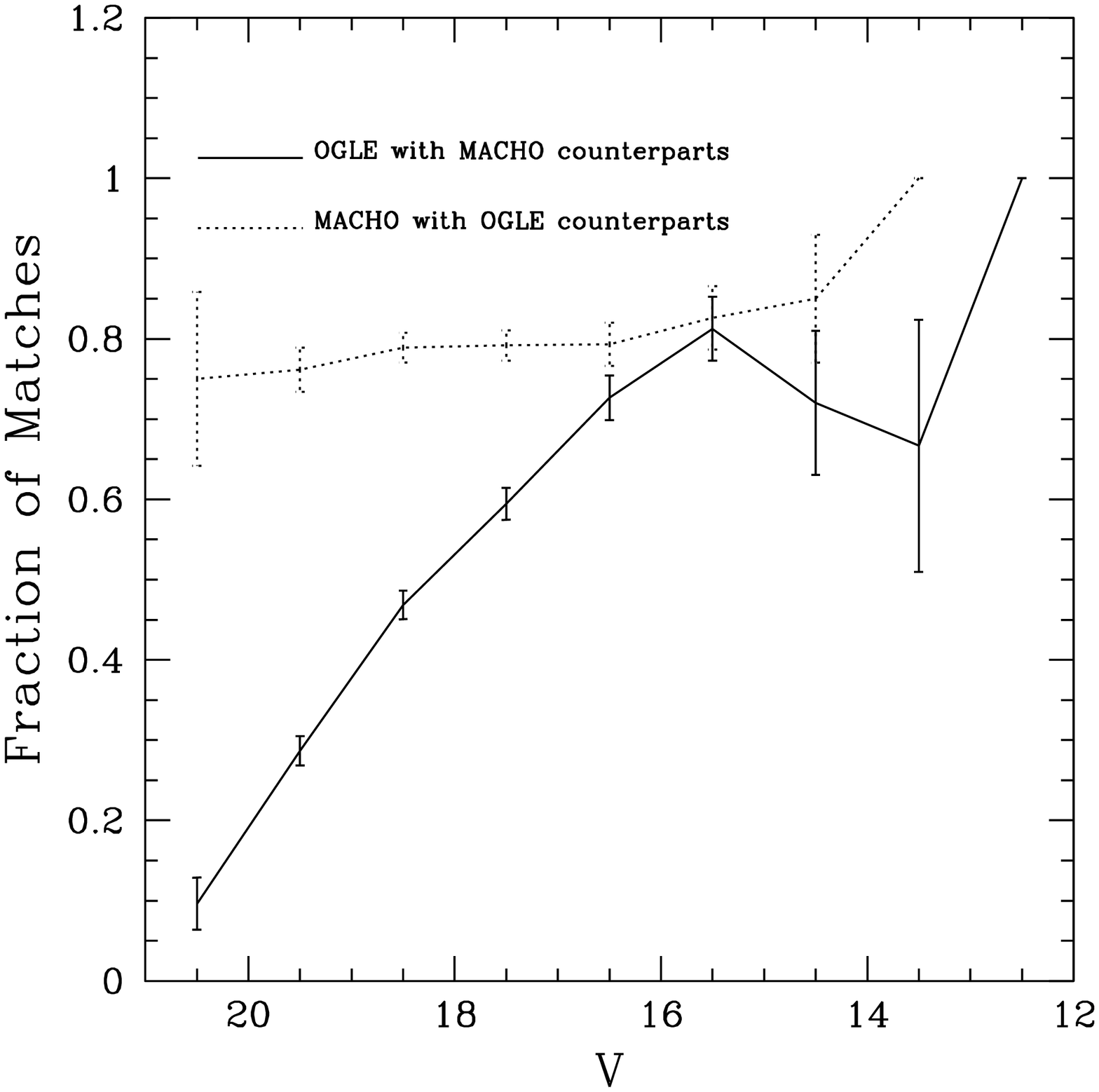}
\caption{Continuous line: $V$ distribution
of the fraction of matches for the OGLE-II LMC EBs in MACHO fields
with expected error bars.
The magnitude bins are $1~\mathrm{mag}$ wide and their centers range from 
$V=20.5~\mathrm{mag}$ to $V=12.5~\mathrm{mag}$.
Dashed line: $V$ distribution of the fraction of matches for 
the MACHO LMC EBs in OGLE-II fields.}
\label{fig:v_matches}
\normalsize
\end{figure}

We then studied the distributions of the periods: Figure 
\ref{fig:per_macho_and_ogle} shows that the OGLE-II EBs without MACHO 
counterparts have on average longer periods than MACHO EBs in the center 
of the LMC and than OGLE-II EBs with a MACHO counterpart and the difference
is statistically significant in both cases; the period distributions of
the MACHO EB of the LMC center and of the MACHO OGLE-II matches are 
statistically different as well.  
Therefore, OGLE-II finds a higher proportion of fainter objects than MACHO and this does play a role in not finding an higher number of OGLE-II counterparts to our sample.
We finally studied the distribution of the MACHO-OGLE matches as a function
of $V$.
We first counted the number of OGLE-II LMC EBs in the MACHO fields, finding
$2517$ of them, out of a total of $2580$; of these $2517$ EBs, $1236$ were 
the matches described above and $1281$ did not have a MACHO counterpart
(this last number is smaller than $1327$, the total number of OGLE EBs without
MACHO counterpart, because we are now only considering OGLE EBs in MACHO fields).
We then studied the distribution, as a function of $V$, of the
OGLE EBs in the MACHO fields that both had and did not have a MACHO counterpart
and for which the $V$ magnitude was reported: there were $1198$ of the former 
and $1267$ of the latter.
We finally performed the inverse calculation, by first counting the number of 
MACHO LMC EBs in the OGLE-II fields and finding $1551$ of them; of these 
$1225$ had an OGLE-II counterpart and $326$ did not; we studied the distribution 
of both these populations as a function of $V$.
The OGLE-II field boundaries were estimated by taking the coordinates of 
the most extreme EBs in each OGLE-II field.
These findings are summarized in Table \ref{tab:v_matches}.
\tabletypesize{\footnotesize}
\begin{deluxetable}{lc}
\tablecolumns{2}
\tablewidth{0pc}
\tablecaption{MACHO OGLE-II matches.
\label{tab:v_matches}}
\tablehead{}
\startdata
OGLE-II LMC EBs & $2580$ \\
OGLE-II LMC EBs in MACHO fields & $2517$ \\
OGLE-MACHO matches & $1236$ \\
OGLE-II LMC EBs without MACHO counterpart & $1281$ \\
OGLE-MACHO matches with reported $V$ & $1198$ \\
OGLE-II LMC EBs without MACHO counterpart with reported $V$ & $1267$ \\
MACHO EBs in OGLE fields & $1551$ \\
MACHO EBs in OGLE fields with OGLE counterpart & $1225$ \\
MACHO EBs in OGLE fields without OGLE counterpart & $326$ \\
\enddata
\end{deluxetable}
The distributions of both the OGLE-II EBs \emph{with} MACHO counterpart 
and of the MACHO EBs \emph{with} OGLE-II counterparts are shown in
Figure \ref{fig:v_matches}; the figure shows the fraction of matches in magnitude 
bins of $1~\mathrm{mag}$; the bin centers range from $V=20.5~\mathrm{mag}$ to 
$V=12.5~\mathrm{mag}$.
The error bars are estimated by assuming that the matches in each magnitude
bin follow a binomial distribution with probability $p=x/N_b$ where $x$ is 
the number of matches in each magnitude bin and $N_b$ is the total number 
of EBs; the error in the expected fraction of matches is then given by Eq 
\ref{eq:errorbar}; in both distributions of Figure. \ref{fig:v_matches} 
the error bar for the brightest magnitude bin is not shown since in both 
cases there is only one match, rendering Eq. \ref{eq:errorbar} meaningless.
The figure shows that the fraction of matches increases for brighter 
magnitudes as expected; the fall at $V=13.5~\mathrm{mag}$ in the distribution of OGLE 
matches is probably due to small number statistic as evidenced by the 
large error bar.
\begin{equation}
\label{eq:errorbar}
\sigma_{N_{b}}
=
\frac{\sqrt{p(1-p)N_b}}{N_b}
=
\sqrt{\frac{p(1-p)}{N_b}}
\approx
\frac{N_b}{N_b-1}\sqrt{\frac{\frac{x}{N_b}(1-\frac{x}{N_b})}{N_b}}.
\end{equation}
\subsection{The Small Magellanic Cloud sample}
\begin{figure}
\footnotesize
\plotone{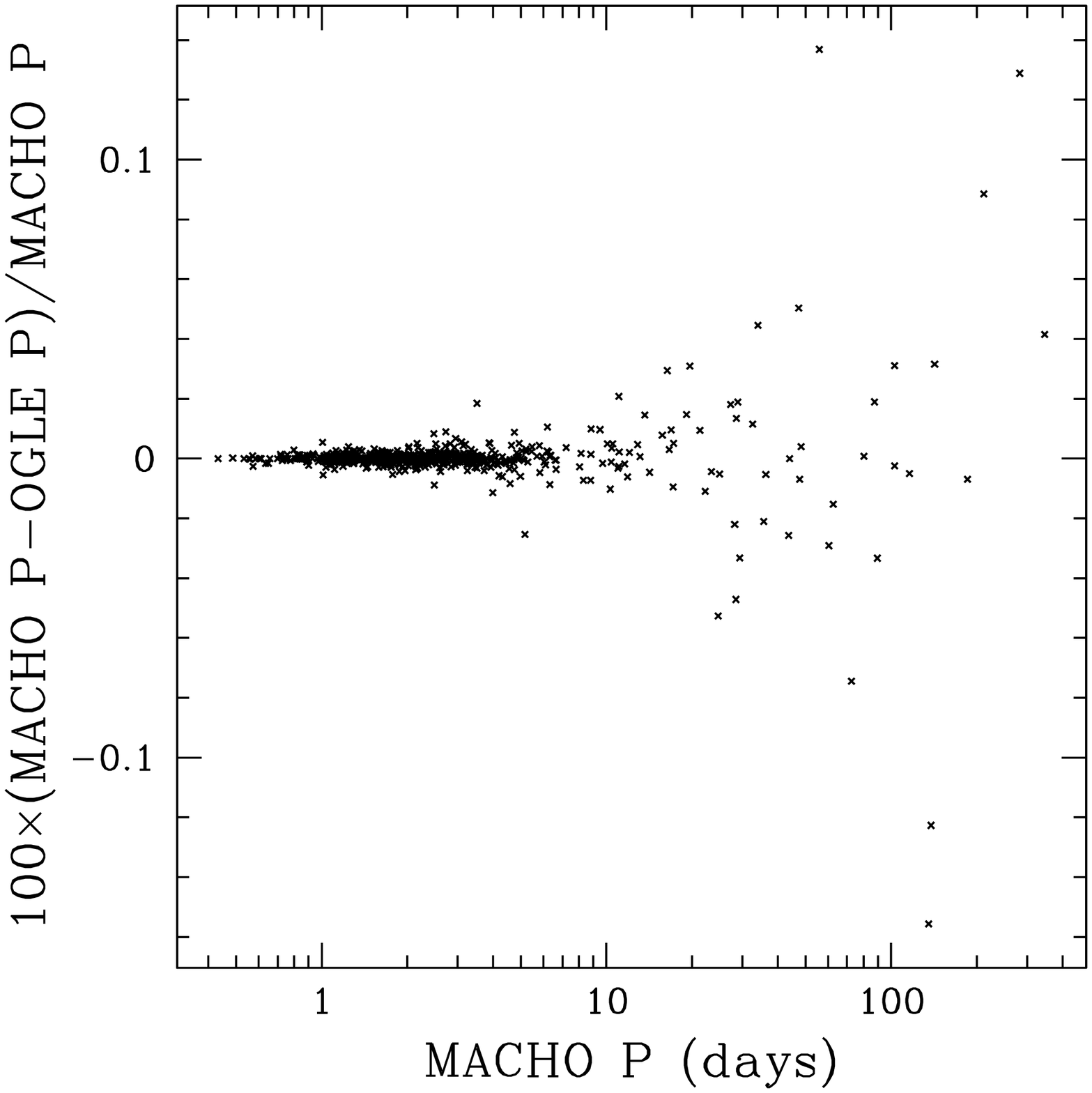}
\caption{Percentage difference for MACHO vs. OGLE-II period for the $698$ 
OGLE-II matches in the SMC sample.}
\label{fig:smccrossper}
\end{figure}

\begin{figure}
\plottwo{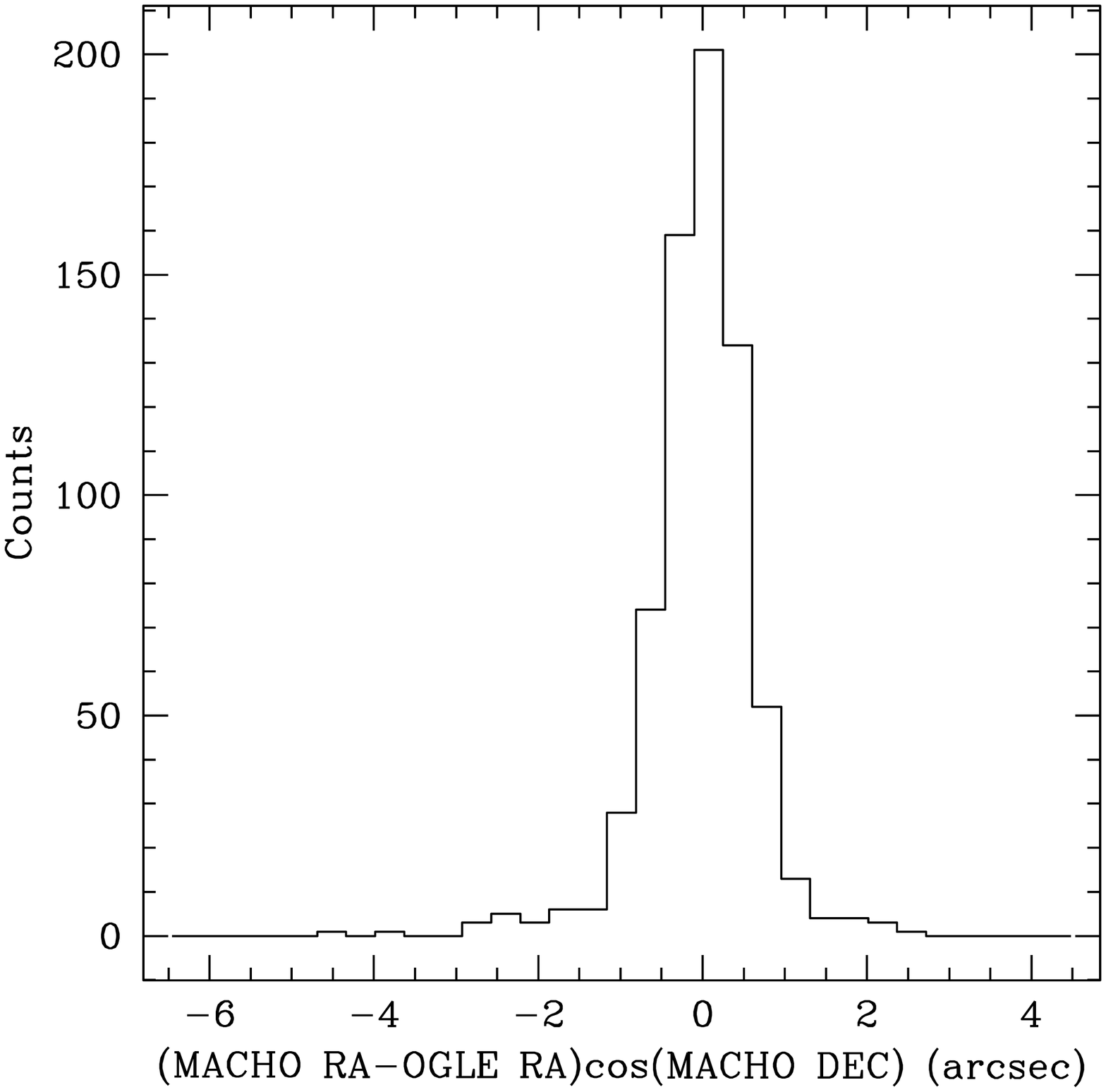}{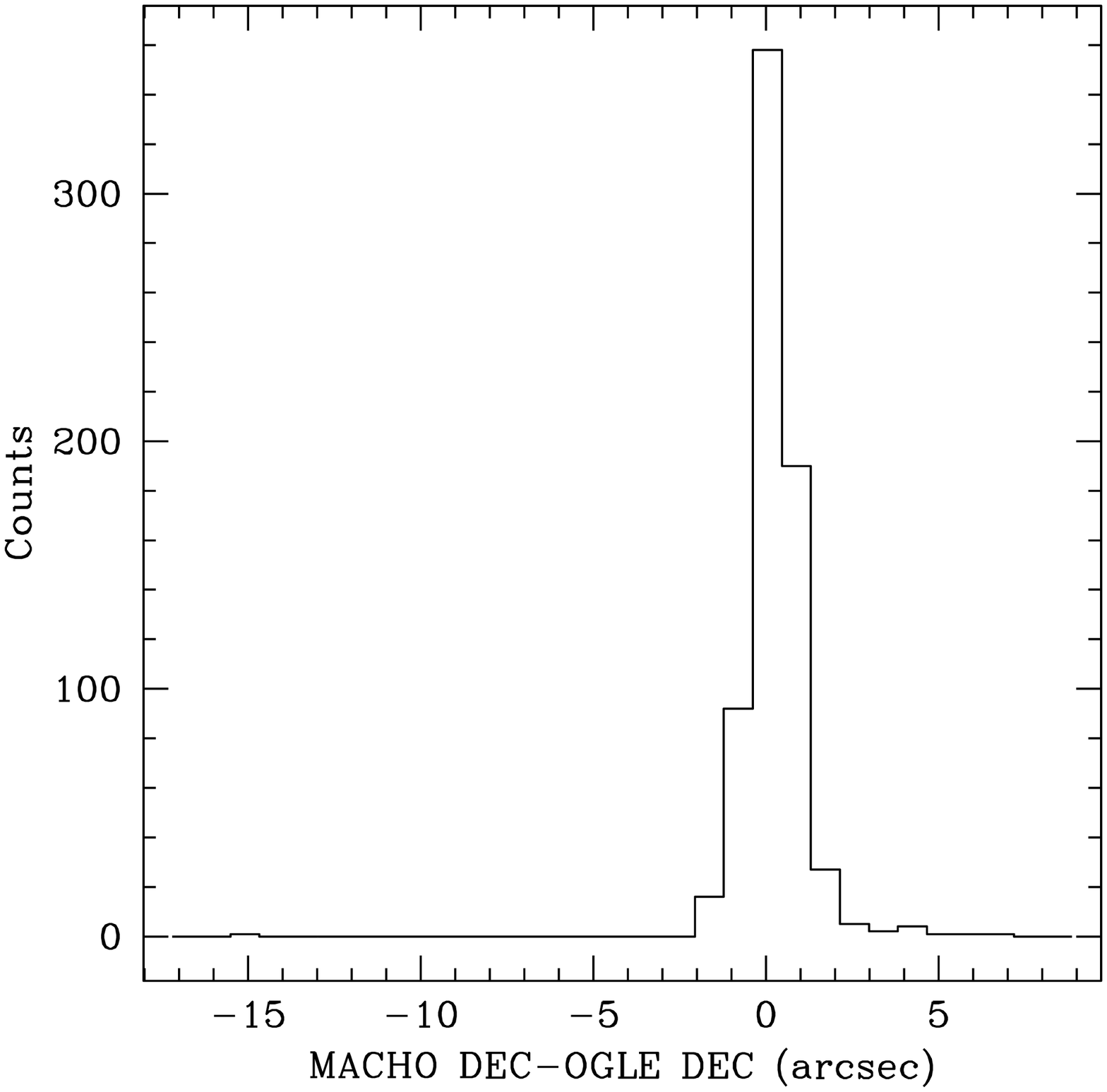}
\caption{Left Panel: histogram of the differences between Right Ascensions 
and for $698$ OGLE matches in the SMC sample.
Right Panel: histogram of the differences between declinations.
}
\label{fig:smcdiffposhist}
\end{figure}

\begin{figure}
\footnotesize
\begin{center}
\plotone{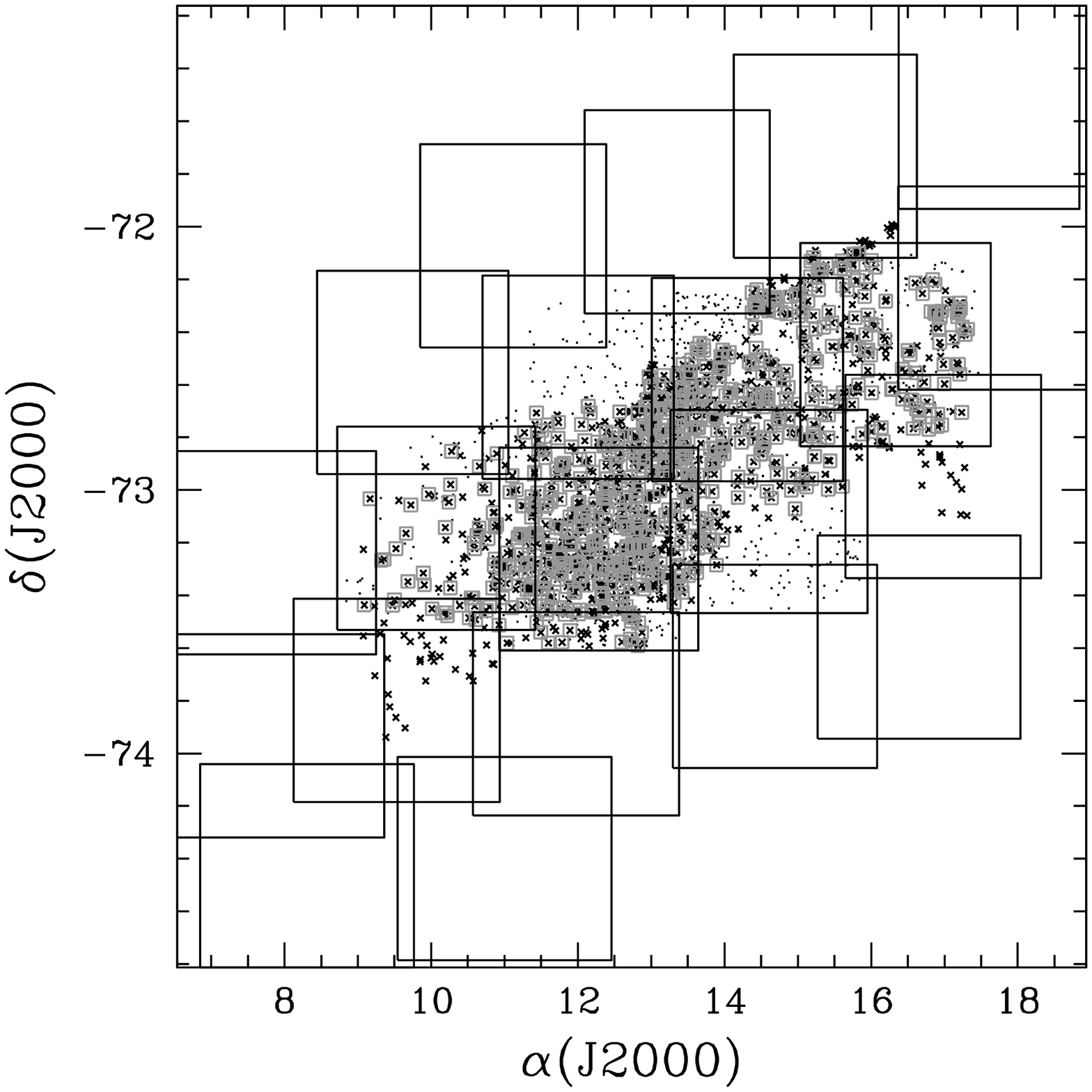}
\end{center}
\caption{Cross correlation between MACHO and OGLE-II SMC samples: points 
represent MACHO stars, crosses OGLE stars and gray empty boxes the matches.}
\label{fig:smcpos}
\normalsize
\end{figure}

\begin{figure}
\footnotesize
\plotone{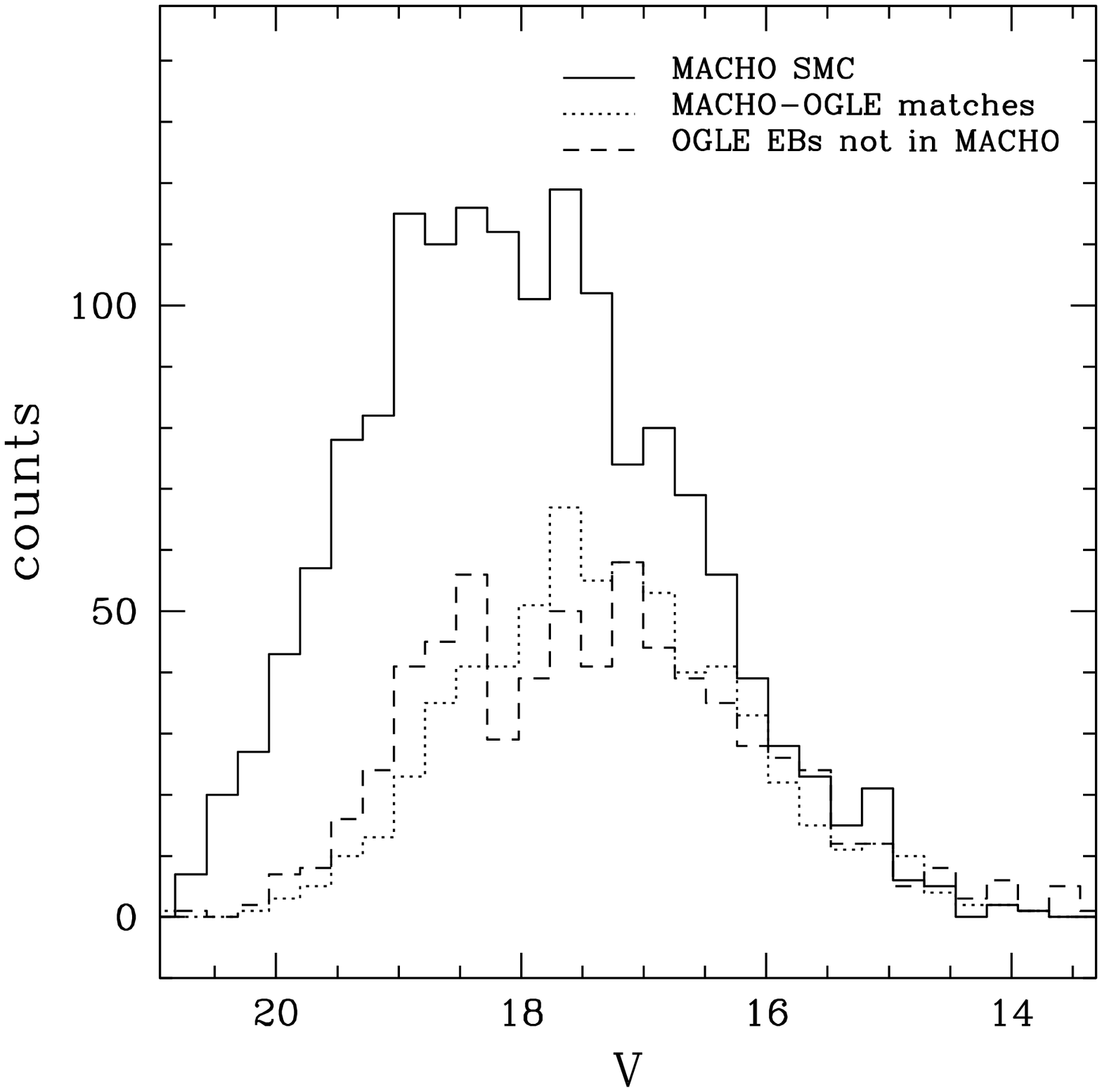}
\caption{Continuous line: $V$ distribution for $1508$ MACHO EBs in the SMC.
Dashed line: $V$ distribution for $650$ OGLE-MACHO matches.
Long dashed line: $V$ distribution for $666$ OGLE-II EBs without MACHO
counterparts.
The figure shows that OGLE-II EBs with MACHO counterparts are on 
average brighter than the ones without, and the shape of their $V$ 
distribution more closely resembles the MACHO $V$ distribution.}
\label{fig:smcv_macho_and_ogle}
\normalsize
\end{figure}

The same general considerations apply to the SMC sample: the search, 
performed with the same criteria as the LMC, produced $698$ matches.
Figure \ref{fig:smccrossper} shows the percentage difference of the MACHO 
and OGLE-II periods vs MACHO period for the matches; Figure
\ref{fig:smcdiffposhist} shows the histogram of the differences in RA and 
DEC.
\normalsize
\normalsize
Unlike the LMC, the sky coverage of the two surveys was approximately the 
same.
Figure \ref{fig:smcpos} shows the positions of the MACHO and OGLE EBs on 
the sky.
We again investigated why we did not find more matches with OGLE-II.
Looking at Figure \ref{fig:smcpos} it is evident that one of the reasons 
is the somewhat different sky coverage of the two surveys (MACHO Fields 
207, 208 and 211 are only partially covered by OGLE), but we also looked 
for other possible explanations.
Again, the most likely explanation is the ways in which the samples
were assembled, but we also considered the differences in the distributions 
of magnitudes and periods.
As for the LMC, we compared the distribution of the $V$ magnitudes for the 
$1508$ EBs in our sample which have a valid $V$ with those of $650$ matches 
and the $666$ EBs from OGLE-II without MACHO counterparts (again these two 
numbers do not add up to $1351$, the size of the OGLE-II sample, because for 
some EBs the $V$ magnitude was not reported).
The histograms of these three distributions are shown in Figure 
\ref{fig:smcv_macho_and_ogle}.
The figure shows that OGLE-II EBs with MACHO counterparts, peaking at 
$V\sim 17.5~\mathrm{mag}$ like the MACHO sample, are on average brighter 
than the ones without MACHO counterparts, which is more spread out and roughly constant between $18.5~\mathrm{mag}<V<16.5~\mathrm{mag}$.
A KS test shows that these three distributions are different at $>99\%$ 
confidence level.
The behavior at high magnitudes is different for MACHO and OGLE-II: the figure
suggests a magnitude limit of $V\sim 21~\mathrm{mag}$ for MACHO and
$V\sim 20~\mathrm{mag}$ for OGLE-II; for $V<19~\mathrm{mag}$ MACHO finds many more EBs than OGLE-II. 
\par
\begin{figure}
\footnotesize
\plotone{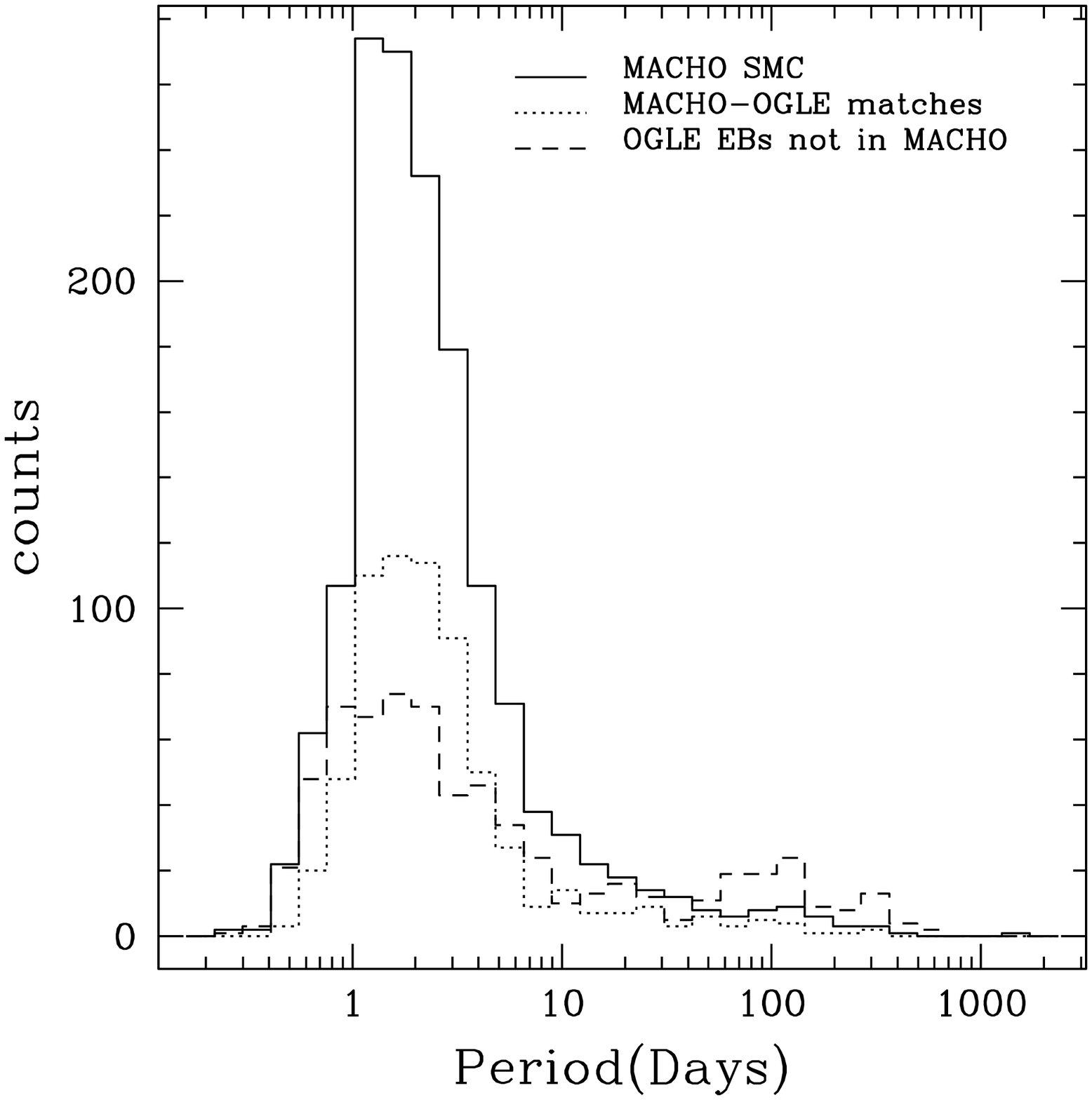}
\caption{Continuous line: Period distribution for $1508$ MACHO EBs in the SMC.
Dashed line: Period distribution for $650$ OGLE-MACHO matches.
Long dashed line: Period distribution for $666$ OGLE-II EBs without MACHO 
counterparts.
The figure shows that both MACHO and OGLE-II EBs with MACHO counterparts have periods 
that cluster more in the $1\mathrm{d}<P<10\mathrm{d}$ range.
OGLE EBs without MACHO counterparts have a larger spread 
in period and smaller ``bumps'' in
the distribution at $\sim 20\mathrm{d}$ and $100\mathrm{d}$.}
\label{fig:smcper_macho_and_ogle}
\normalsize
\end{figure}
\begin{figure}
\footnotesize
\plotone{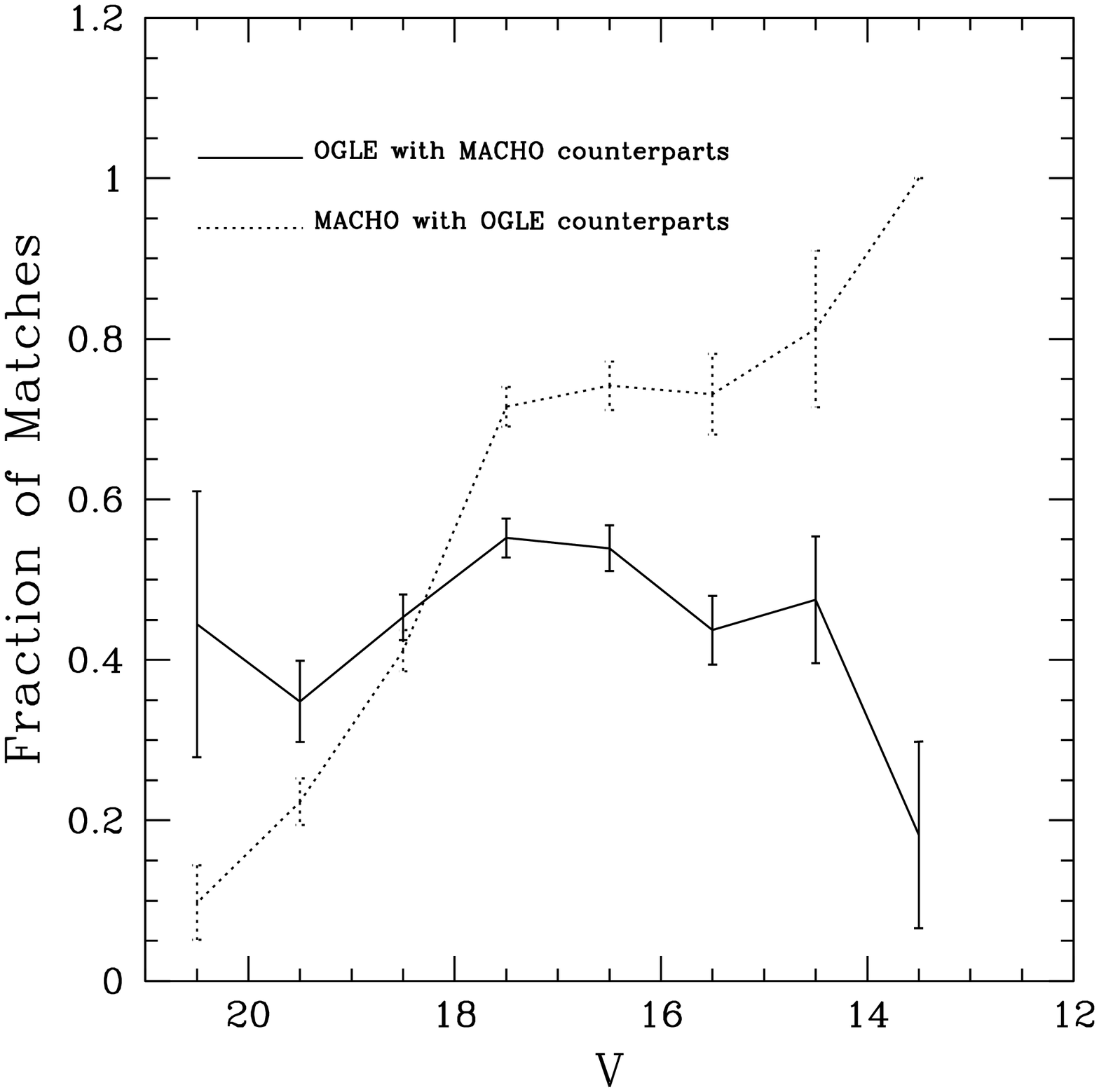}
\caption{Continuous line: $V$ distribution of the fraction of matches
of matches for the OGLE-II SMC EBs in MACHO fields with expected error bars.
The magnitude bins are $1~\mathrm{mag}$ wide and their centers range from 
$V=20.5~\mathrm{mag}$ to $V=13.5~\mathrm{mag}$.
Dashed line: $V$ distribution of the fraction of matches for the MACHO SMC EBs in OGLE-II fields.}
\label{fig:v_matches_smc}
\normalsize
\end{figure}

We then studied the distributions of the periods: Figure 
\ref{fig:smcper_macho_and_ogle} shows that OGLE-II EBs with MACHO 
counterparts have periods that cluster more in the 
$1\mathrm{d}<P<10\mathrm{d}$ range, as do the MACHO EBs and a KS test gives 
a probability $\sim 43\%$ for the two distribution to be the same.
On the other hand OGLE-II EBs without MACHO counterparts have a larger 
spread of period values, with more objects having $P>10\mathrm{d}$ and with small ``bumps'' in the distributions at $\sim 20\mathrm{d}$ and $100\mathrm{d}$
and a KS test shows it to be different from both MACHO and OGLE-II MACHO 
matches at $>99.9\%$ confidence level.
The $V$ distributions of both the OGLE-II EBs without MACHO counterpart 
and of the MACHO EBs without OGLE-II counterparts are shown if Figure 
\ref{fig:v_matches_smc}; the figure shows the fraction of matches in 
magnitude bins of $1~\mathrm{mag}$ with centers ranging from $V=20.5~\mathrm{mag}$ 
to $V=14.5~\mathrm{mag}$; the error bars are estimated again by Eq. \ref{eq:errorbar}.
\par
We conclude that differences in sky coverage and in techniques used in assembling the samples, as well as different magnitude limits and in general different behavior at high magnitude of the two surveys all play a role in not finding an higher number of OGLE-II counterparts to our sample; the fact that the SMC was less observed than the LMC by MACHO may also explain why we find fewer long period objects, while the fact that the exposure time for the SMC was 
double that of the LMC \citep{alcock99} may explain why we find many faint objects.
\section{The data on line}
\label{sec:online}
The data presented in this paper can be accessed on line at the Astronomical Journal
website\footnote{\url{http://www.journals.uchicago.edu/AJ/.}}
and are mirrored at the Harvard University Initiative in Innovative Computing 
(IIC) /Time Series Center.\footnote{\url{http://timemachine.iic.harvard.edu/;}\url{http://timemachine.iic.harvard.edu/faccioli/ebs/.}}
At both sites the data consist of a summary table for each Cloud and
light curves for all the EBs in the samples.
Light curve files contain the \emph{unfolded} data in MACHO magnitudes;
these same files can also be retrieved from the MACHO website.
Finally for the LMC, the input and output files used in the JKTEBOP
fits are provided.
\acknowledgments
This work uses public domain data from the MACHO Project whose work was 
performed under the joint auspices of the U.S. Department of Energy, 
National Nuclear Security Administration by the University of California, 
Lawrence Livermore National Laboratory under contract No. W-7405-Eng-48, 
the National Science Foundation through the Center for Particle 
Astrophysics of the University of California under cooperative agreement 
AST-8809616, and the Mount Stromlo and Siding Spring Observatory, part of 
the Australian National University.
KHC's work is performed under the auspices of the U.S. Department of Energy
by Lawrence Livermore National Laboratory in part under Contract
W-7405-Eng-48 and in part under Contract DE-AC52-07NA27344.
This work uses public domain data obtained by the OGLE Project.
We are grateful to Julia Kregenow for finding part of the eclipsing 
binary stars in the sample, to John Rice for providing the original 
reference for the Supersmoother algorithm and to Peter Eggleton for 
pointing out the difference between contact and common envelope binaries 
and Claude Lacy for providing the EBOP program.
We thank the referee for many helpful suggestions.
LF acknowledges the kind hospitality of the Institute of Geophysics and 
Planetary Physics at Lawrence Livermore National Laboratory and of the 
Harvard-Smithsonian Center for Astrophysics where part of the work was done.

\end{document}